\documentclass[aps,pre,reprint,amsmath,amssymbols,superscriptaddress]{revtex4-1}
\usepackage{graphicx}
\usepackage{bm}
\usepackage{color}
\usepackage{soul}
\usepackage{amsmath}
\usepackage{amsfonts}
\usepackage{amssymb}

\definecolor{darkBlue}{rgb}{0,0,0.6}
\definecolor{darkRed}{rgb}{0.5,0,0}
\definecolor{darkGreen}{rgb}{0,0.5,0}
\definecolor{gray}{rgb}{0.2,0.2,0.2}

\newcommand{\tz}{\tilde{z}}
\newcommand{\qv}{\mathbf{q}}
\newcommand{\uv}{\mathbf{u}}
\newcommand{\brm}[1]{\bm{{\rm #1}}}

\begin{document}
	\title{Multi-functional Twisted-Kagome lattices: Tuning by Pruning Mechanical
		Metamaterials}
	\author{Danilo B. Liarte}
	\email{liarte@cornell.edu}
	\affiliation{Cornell University, Ithaca, NY, USA}
	\author{O. Stenull}
	\author{T. C. Lubensky}
	\email{tom@physics.upenn.edu}
	\affiliation{University of Pennsylvania, Philadelphia, PA, USA}
	
	\date{\today}
	
	\begin{abstract}
This article investigates phonons and elastic response in randomly diluted lattices constructed by combining (via the addition of next-nearest bonds) a twisted kagome lattice, with bulk  modulus $B=0$ and shear modulus $G>0$, with either a generalized untwisted kagome lattice with $B>0$ and $G>0$ or with a honeycomb lattice with $B>0$ and $G=0$. These lattices exhibit jamming-like critical end-points at which $B$, $G$, or both $B$ and $G$ jump discontinuously from zero while the remaining moduli (if any) begin to grow continuously from zero.  Pairs of these jamming points are joined by lines of continuous rigidity percolation transitions at which both $B$ and $G$ begin to grow continuously from zero. The Poisson ratio and $G/B$ can be continuously tuned throughout their physical range via random dilution in a manner analogous to ``tuning by pruning" in random jammed lattices.  These lattices can be produced  with modern techniques, such as 3D printing, for constructing  metamaterials.
	\end{abstract}
	
	\maketitle

\section{Introduction}

Ball-and-spring networks provide useful and generally accurate models for the elastic properties of solids~\cite{AshcroftMer1976} from periodic crystals to disordered glasses~\cite{Thorpe1983,FengGar1985,SouslovLub2009,Binder2011,LubenskyKai2015,MaoLub2018}.  These networks undergo a transition from an elastically rigid state to a floppy one when their coordination number $z$ falls below a critical value $z_c$, usually close to the Maxwell value $z_M = 2 d$ in dimension $d$~\cite{Maxwell1864}.  Two distinct models often used to describe this behavior are (1) randomly diluted periodic lattices \cite{FengGar1985,SchwartzSen1985,GuyonCr1990} with  springs removed with probability $p$ and (2) jamming models \cite{LiuNa2010,BehringerChak2019} in which particles (usually spheres) are compressed beyond the point at which inter-particle contacts  cannot be avoided. In the  former, the transition from the floppy to the rigid state, usually called rigidity percolation (RP), both the shear modulus $G$ and the bulk modulus $B$ grow continuously from zero as $z$ increases from $z_c$.  In the latter, the transition to rigidity is characterized by a discontinuous jump in $B$ at $z_c$ and continuous growth of $G$ from zero for $z>z_c$ \cite{OhernNa2003,LiuNa2010}.  Both models can exhibit far-richer behavior depending on lattice structure and rules for removing (or adding) springs. Recent work \cite{GoodrichNAG2015,HexnerNag2018a,HexnerNag2018b,ReidPab2018} investigates a number of paths to the floppy state in a jamming model in which a network prepared by usual jamming procedures sets particle positions that are then connected pairwise by unstretched springs.  The set of springs on $B$-bonds, which most resist compression, and that on $G$-bonds, which most resist shear, are nearly independent.  If the springs are removed randomly from the entire ensemble, there is an RP transition at which both $B$ and $G$ vanish with a ratio $G/B$ that is nearly constant.  If, however, springs (on $G$-bonds) that make the largest contribution to $G$ are removed first, $G/B$ vanishes as $\Delta z \rightarrow 0$ in a jamming-like transition in which $B$ undergoes a discontinuous jump; but if springs that  make the largest contribution to $B$ (on $B$-bonds) are removed first, $G/B \rightarrow \infty$ as $\Delta z\rightarrow 0$ and $G$ undergoes a discontinuous jump. Thus by selectively removing bonds, the full range of $G/B$ from $0$ to $\infty$ and Poisson ratio from $-1$ to $1$ (in 2D) can be accessed. Reference \cite{GoodrichNAG2015} calls this process ``tuning by pruning" (TbP).

\begin{figure}
	\begin{minipage}{0.48\linewidth}
		\centering
		(a) \vspace{0.1cm} \\
		\includegraphics[width=\linewidth]{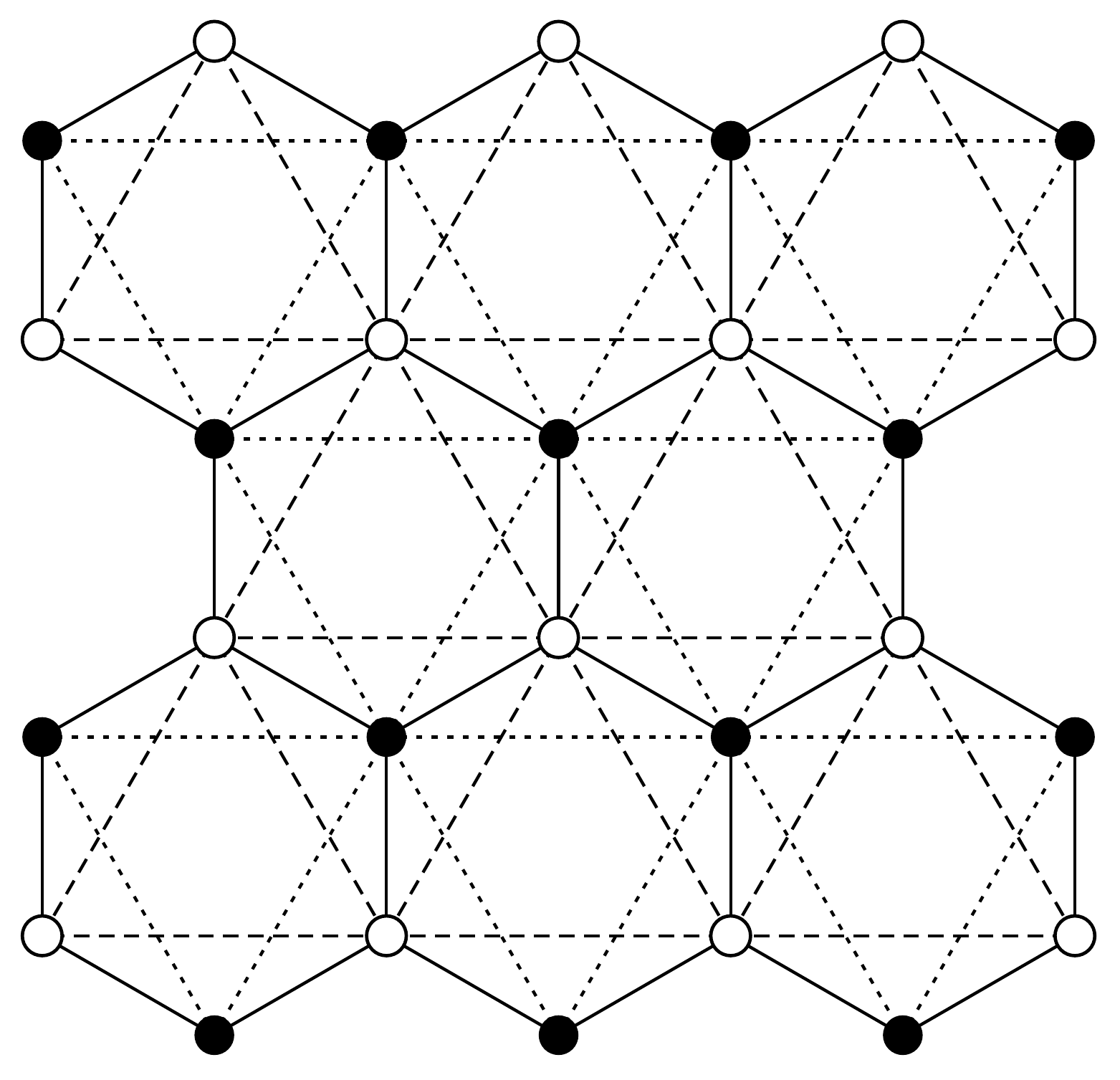}
	\end{minipage}
	\begin{minipage}{0.48\linewidth}
		\centering
		(b) \vspace{0.1cm} \\
		\includegraphics[width=\linewidth]{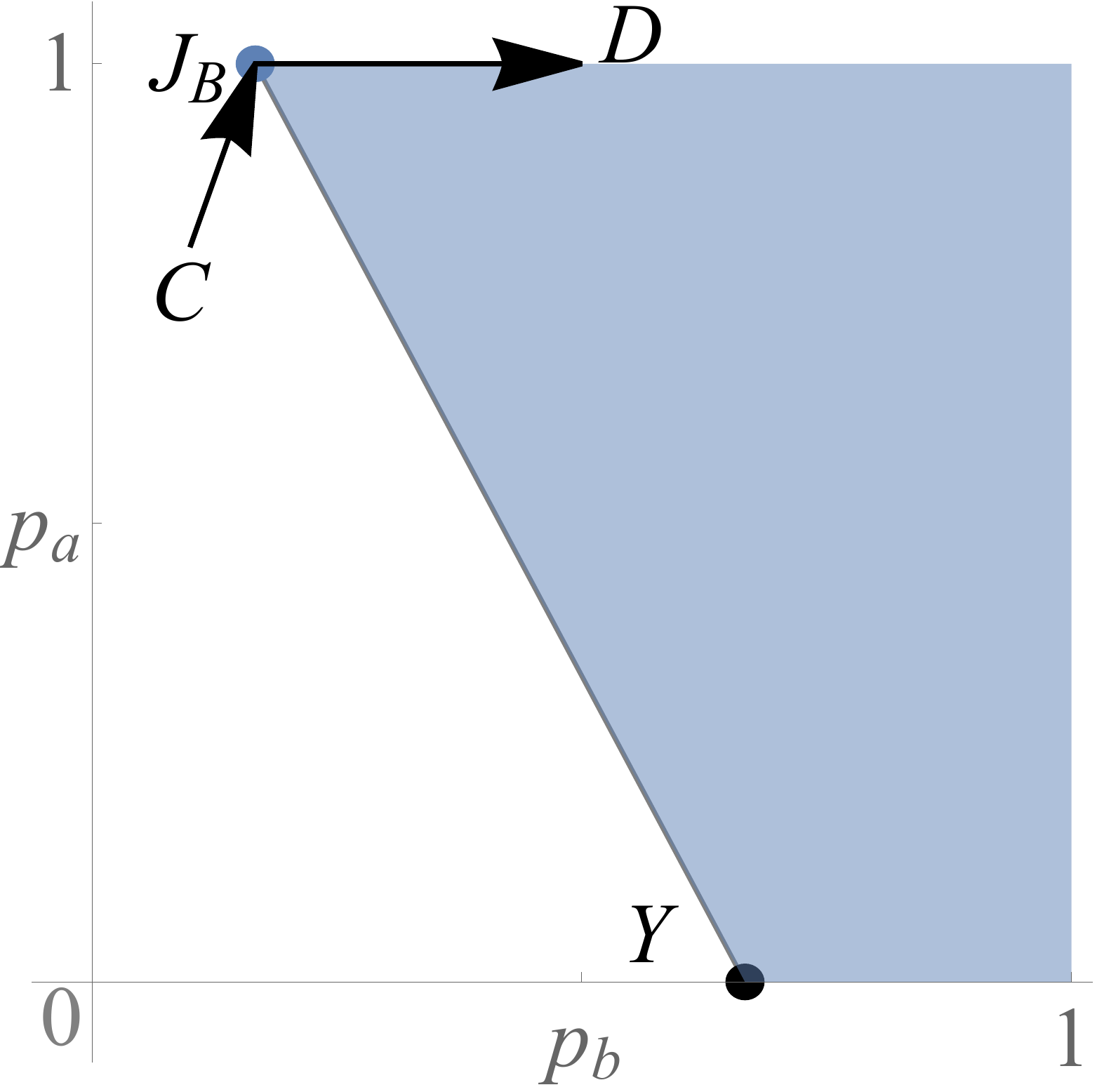}
	\end{minipage}
	\caption{(a) The  honeycomb lattice with NNN springs (dashed and dotted lines) forming two independent triangular lattices. (b) Phase diagram from Ref.~\cite{LiarteLub2019} in the $p_a-p_b$ plane showing the floppy (white region) and rigid (blue) phases, the jamming point $J_B$, the RP point $Y$ of the diluted TLs and the RP line $J_BY$. Here $p_a$ and $p_b$ correspond to the occupancy probability for each bond in the honeycomb ($a$) and triangular ($b$) sub-lattices.}
	\label{Fig:Jam-prop}
\end{figure}

Recently, we co-authored a paper \cite{LiarteLub2019} describing a model periodic lattice that exhibits both the RP and the jamming transitions and provides a range of $G/B$ analogous to the that of the TbP procedure. It consists of a honeycomb lattice (HL), which by itself has a positive $B$ even though it is under-coordinated, decorated with next-nearest-neighbor (NNN) bonds that form two independent triangular lattices (TLs) whose sites are shared by the HL [Fig.~\ref{Fig:Jam-prop}].  The bonds of HL are occupied with probability $p_a$, and those on the TLs are occupied with probability $p_b$. The connection with the TbP model is clear:  The bonds of HL are  the analog of the $B$-bonds, and the TLs are mixtures of the $B$- and $G$-bonds.  The phase diagram for this model is reproduced in Fig.~\ref{Fig:Jam-prop}.  There is a jamming critical point at $J_B= (p_a^J,p_b^J) = (1,1/6)$ and an RP line stretching from $J_B$ to $Y= (0,2/3)$. Viewed from the floppy phase, the line $(1,p_b)$ is a first-order line, and the point $J_B$ is roughly analogous to a critical endpoint in which a second-order $RP$ line ($J_B Y$) meets a first-order line \cite{ChaikinLub1995}. Both $B$ and $G$ grow with distance from the RP line, but along paths like $CJ_B D$ that pass through $J_B$, $B$ jumps discontinuously and $G$ grows continuously from zero at $J_B$ as in jamming. Paths starting at $J_B$ and ending at $Y$ cover the range of $G/B$ from $0$ to $1/2$ (or Poisson ratio from $1$ to $1/3$) without reaching any negative values.

\begin{figure}[!ht]
		\centering
		\begin{minipage}{0.48\linewidth}
			\centering
			(a) \vspace{0.1cm} \\
			\includegraphics[width=\linewidth]{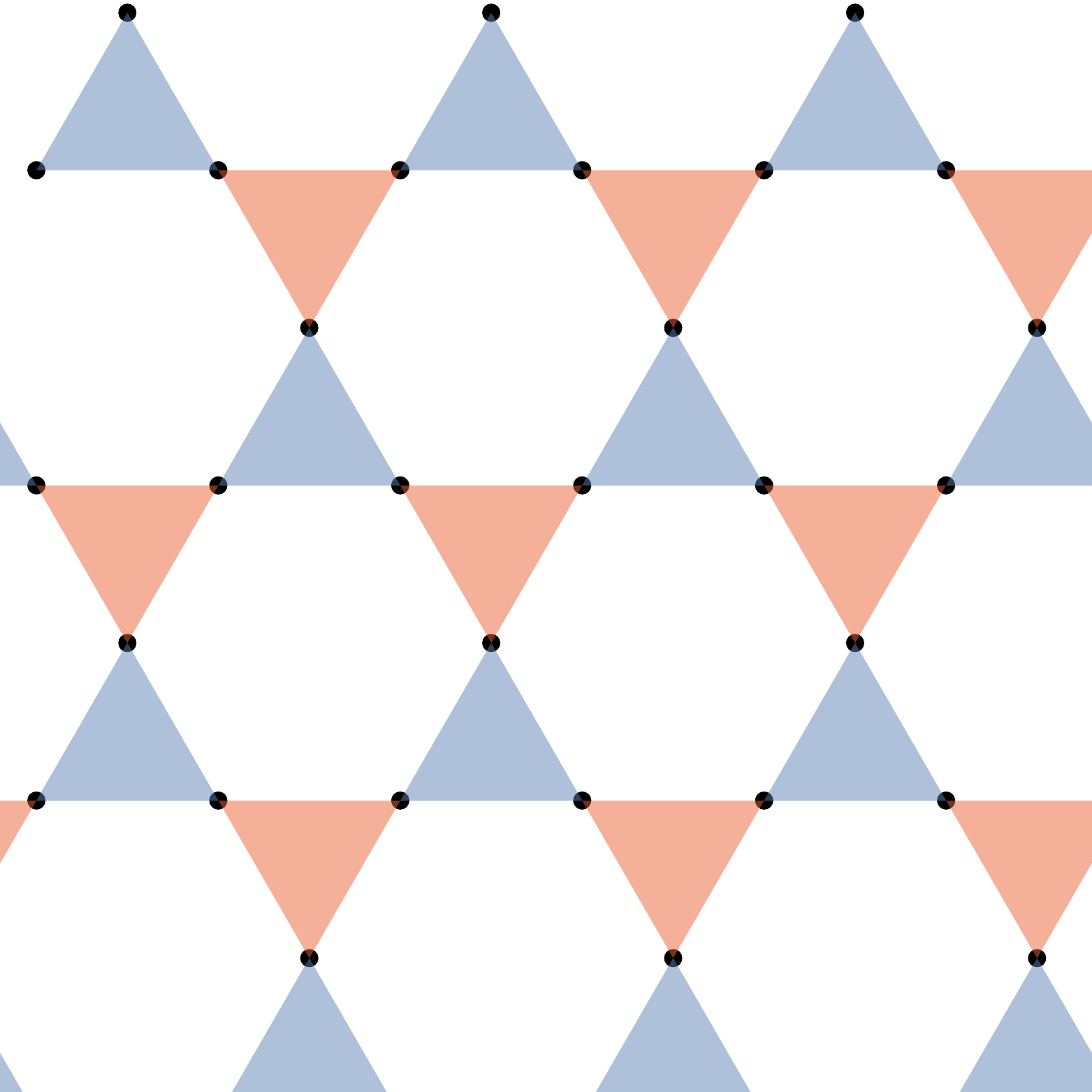}
		\end{minipage}
		\begin{minipage}{0.48\linewidth}
			\centering
			(b) \vspace{0.1cm} \\
			\includegraphics[width=\linewidth]{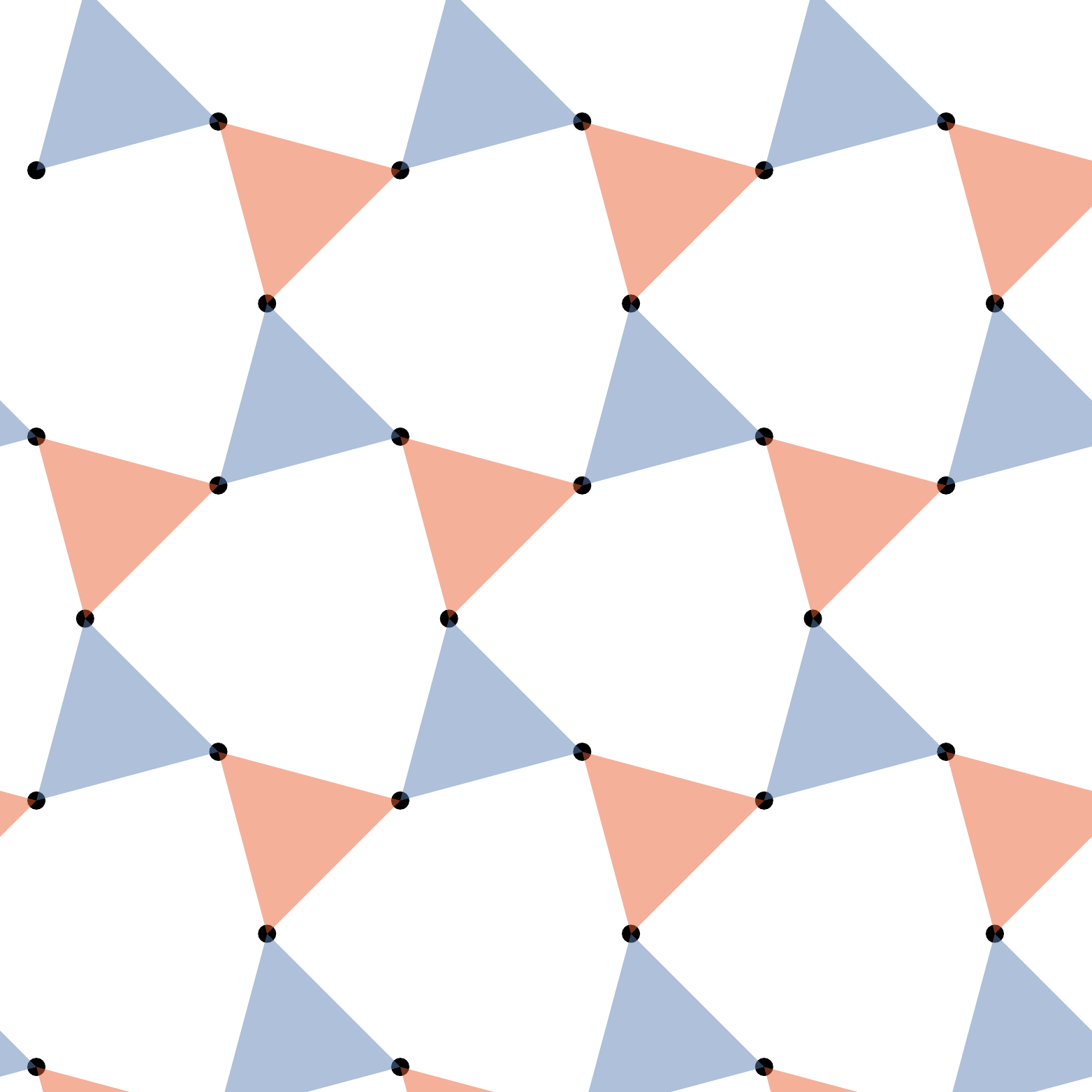}
		\end{minipage} \vspace{0.2cm} \\
		\begin{minipage}{0.48\linewidth}
			\centering
			(c) \vspace{0.1cm} \\
			\includegraphics[width=\linewidth]{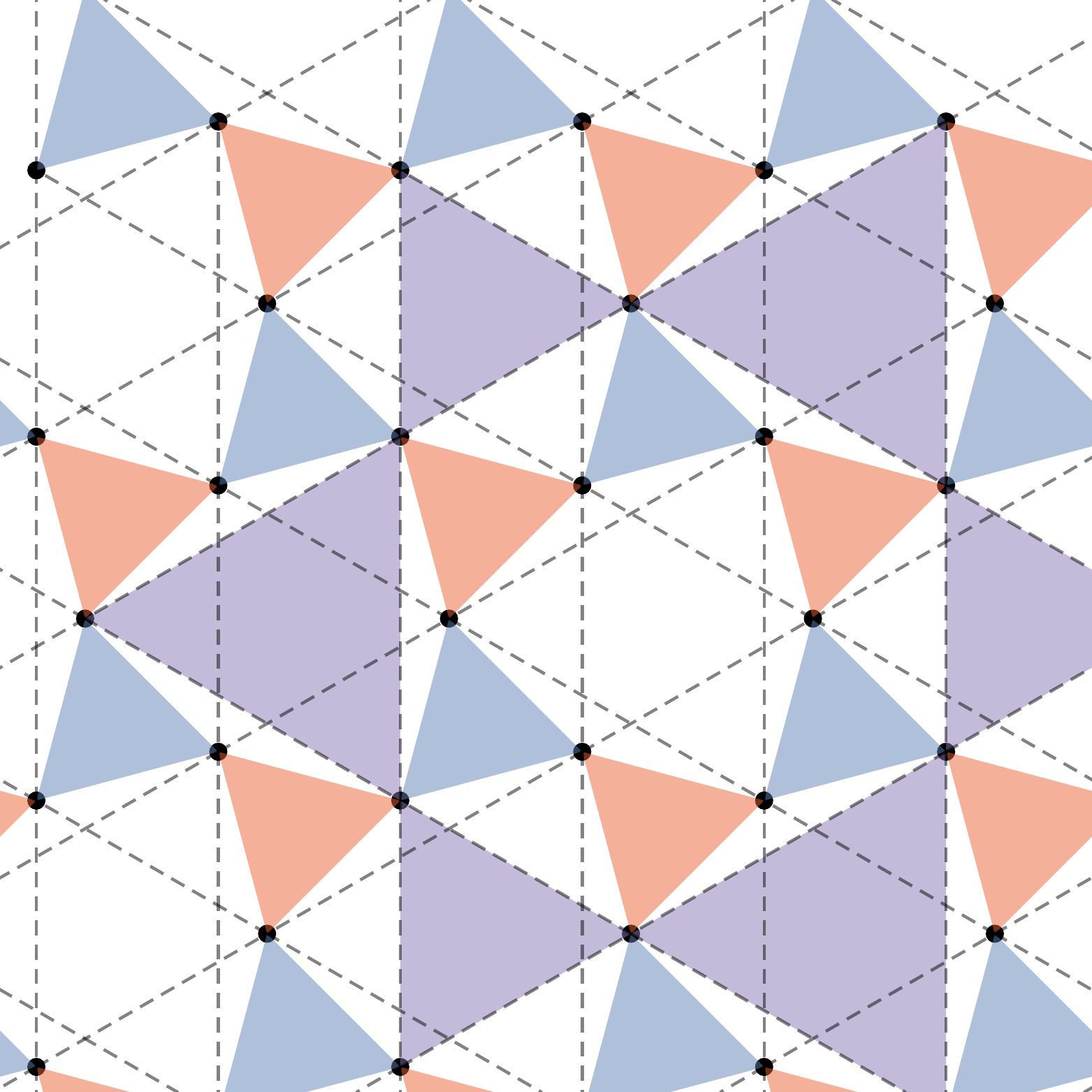}
		\end{minipage}
		\begin{minipage}{0.48\linewidth}
			\centering
			(d) \vspace{0.1cm} \\
			\includegraphics[width=\linewidth]{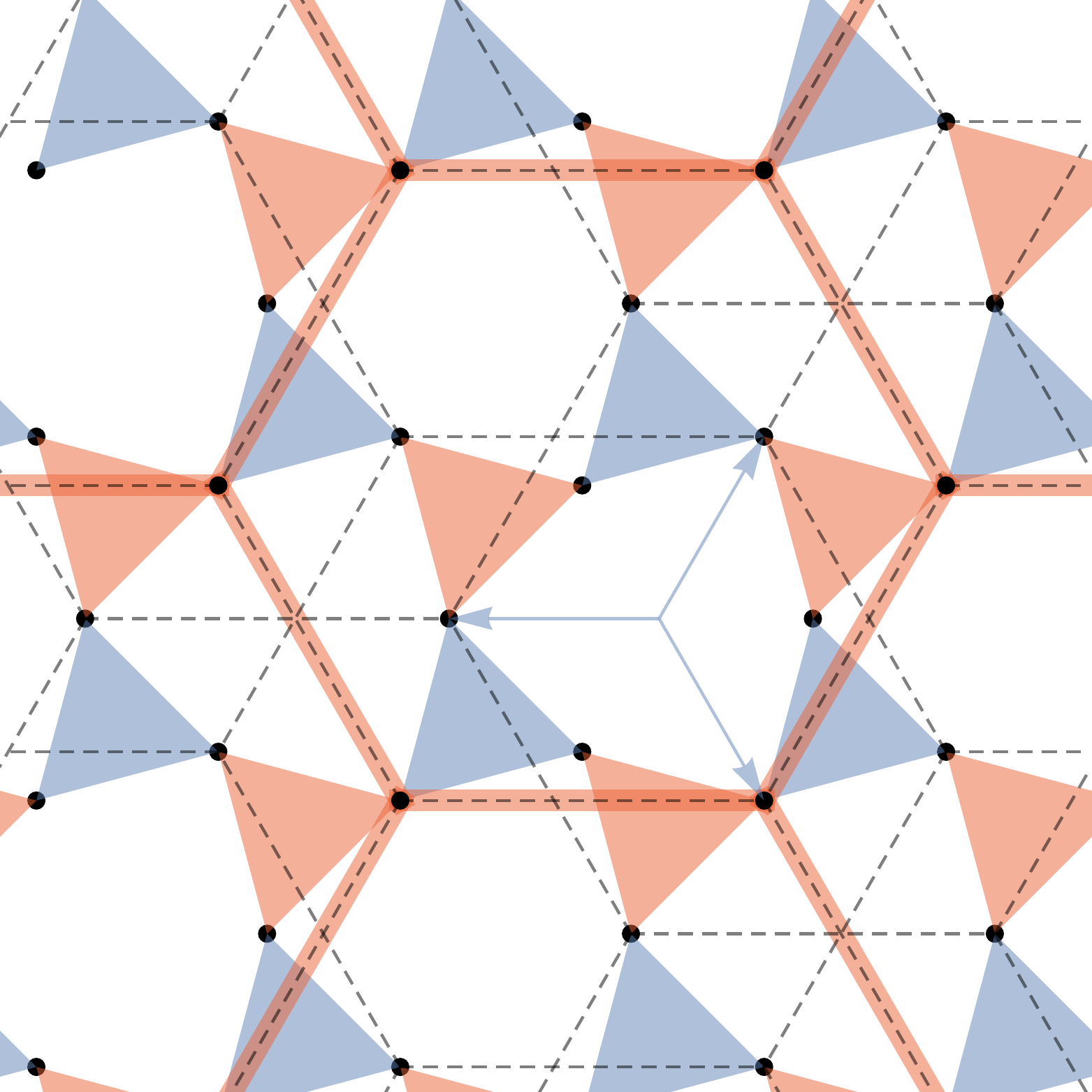}
		\end{minipage}
	\caption{(a) Untwisted Kagom\'e lattice. (b) Twisted Kagom\'e lattice with angle $\alpha>0$ between triangles equal to $\pi/12$ (see Fig.~\ref{fig:cellVectors} for an illustration of the twist angle $\alpha$). (c) TwK/GK lattice showing the GK lattices formed by NNN bonds (dashed lines, one of the GK lattices is displayed with purple triangles). Note that these lattices have triangles of two different sizes rather that the single size of the traditional Kagom\'e lattice. (d) TwK/H lattice showing honeycomb lattices (dashed lines, one of the honeycomb lattices is displayed with thick red lines). The faint blue arrows indicate the 3-fold symmetry and hence isotropic elasticity of this model.}
	\label{Fig:GK/Hfig}
	\end{figure}

\begin{figure}[!ht]
	\begin{minipage}{0.48\linewidth}
		\centering
		(a) \vspace{0.1cm} \\
		\includegraphics[height=0.9\linewidth]{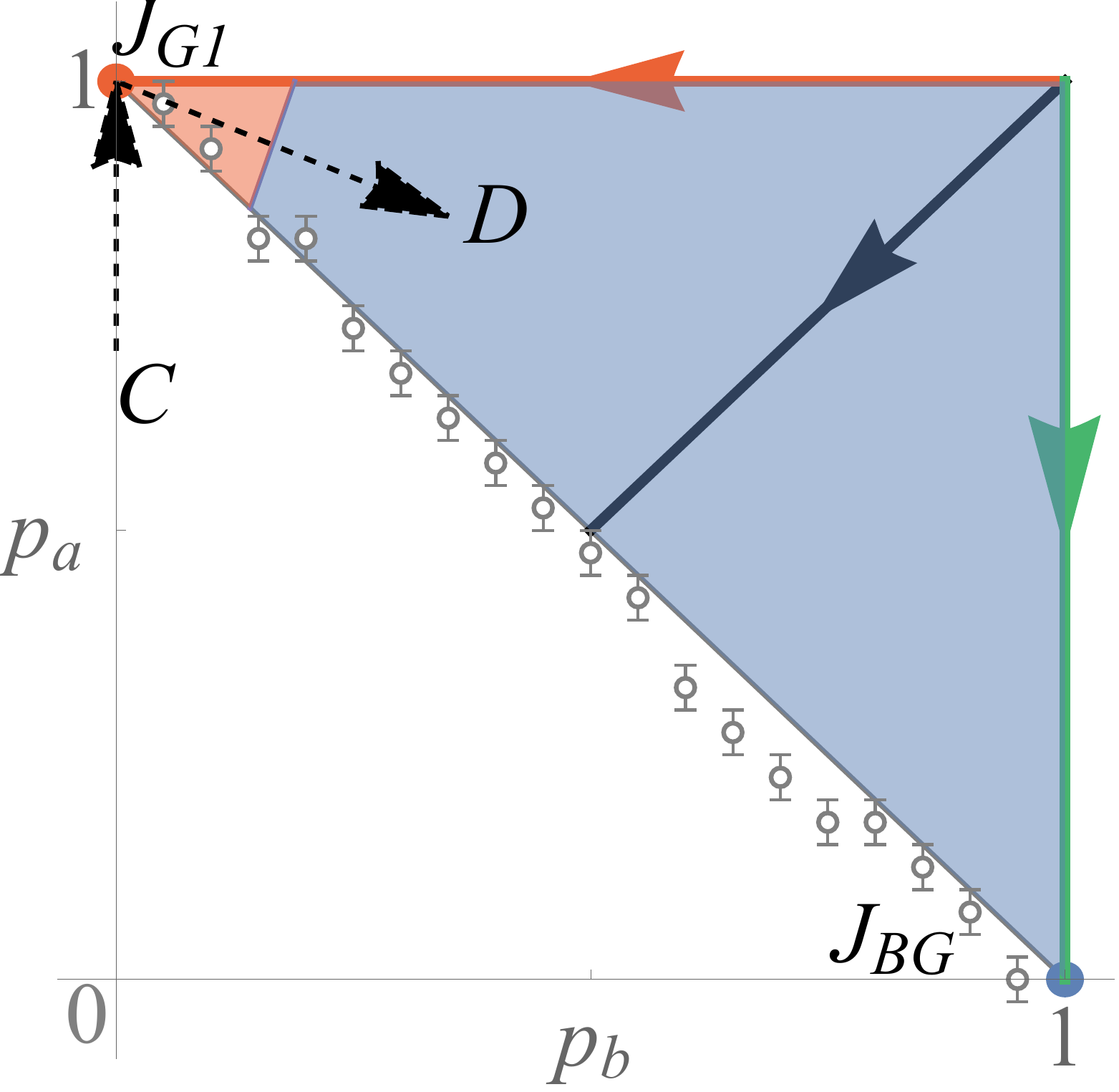}
	\end{minipage}
	\begin{minipage}{0.48\linewidth}
		\centering
		(b) \vspace{0.1cm} \\
		\includegraphics[height=0.9\linewidth]{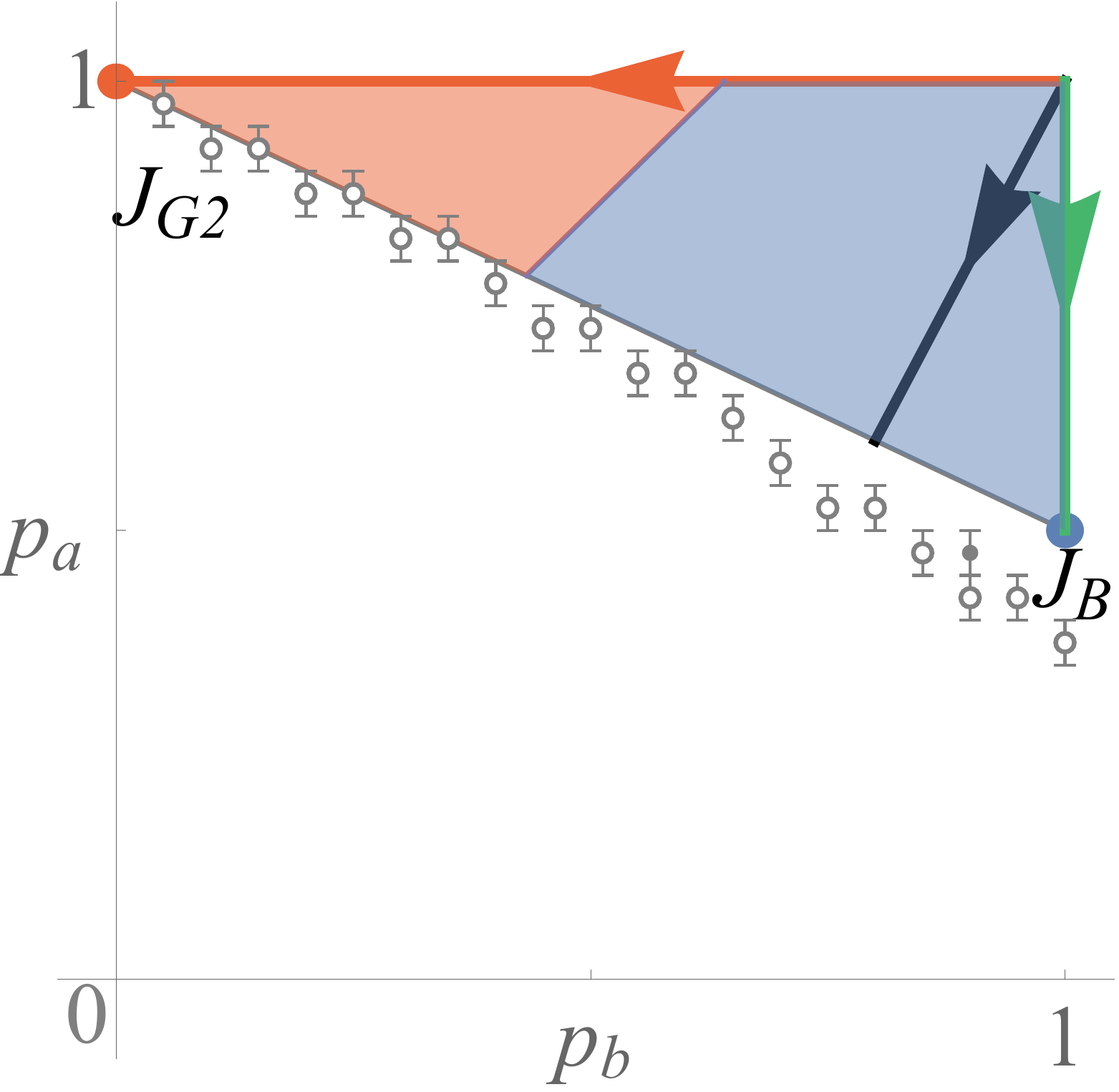}
	\end{minipage}
	\caption{Phase Diagrams of the TwK/GK (a) and TwK/H models (b) showing auxetic rigid regions on the top left (red), floppy regions at the bottom (white) and regular rigid regions with $\sigma>0$ on the top right (blue). $J_B$, $J_G$ (with $G=G1$ or $G2$) and $J_{BG}$ denote jamming, shear-jamming, and double-jamming points. The red, green, and black lines are respectively paths toward $J_G$, $J_B$ or $J_{BG}$, and RP lines along paths perpendicular to them.  The ratio $G/B$ and the Poisson ratio $\sigma$ along these lines are plotted in Fig.~\ref{Fig:Poisson}. The black-dashed arrows [$CJ_GD$ in (a)] depict a path along which the shear modulus varies discontinuously at the rigidity transition. The RP thresholds at which  $B$ and $G$ vanish are within error bars at all points except  at the value of $p_b$ two points to the left of $J_B$ in  (b).  The open circles indicate this common threshold except at the latter point at which the filled circle marks the $B$-threshold  and the open circle marks the $G$-threshold. Finally, $(1,p_b)$ and $(p_a,1)$ are the boundary $A$- and $B$-lines, respectively.
		\label{Fig:PhaseDiagrams}}
\end{figure}

This paper introduces and, using both effective medium theory (EMT) and numerical simulations, explores the elastic response of two periodic lattice models [Fig.~\ref{Fig:GK/Hfig}], both of which have average $C_3$ symmetry and macroscopic elastic energies in the isotropic class characterized by nonvanishing $B$ and $G$ with no moduli arising from anisotropy. Both models access negative values of the Poisson ratio $\sigma$, one of which accesses the full range from $\sigma = -1$ to $\sigma = +1$.  The starting point of both is the twisted kagome lattice (TwKL) [Fig.~\ref{Fig:GK/Hfig}(b)], obtained by twisting adjacent triangles in the untwisted kagome lattice (KL) [Fig.~\ref{Fig:GK/Hfig}(a)] through an angle  $\alpha$. This lattice has a nonzero shear modulus but a vanishing bulk modulus \cite{SunLub2012} and, thus, a Poisson ratio of $-1$~\footnote{Here we assume the twist angle $\alpha \in (0,\pi/3)$}. In the first model, the TwK/GK model [Fig. \ref{Fig:GK/Hfig}(c)], springs are placed on NNN bonds of the TwK with probability $p_b$. When $p_b=1$, these bonds form three independent untwisted generalized kagome lattices (GKL), composed of two different-sized rather than single-sized triangles, for which $B/G = 2$ and $\sigma = 1/3$ (see Eqs.~\eqref{eq:ModuliB} and~\eqref{eq:Moduli} in Section~\ref{sec:Energy}).  Thus points in the rigid regime cover the range of $\sigma$ from $-1$ to $1/3$.  In the second, the TwK/H model, bonds connecting a collection of NNN and third-neighbor points of the TwK lattice form three independent honeycomb lattices [Fig.~\ref{Fig:GK/Hfig}(d)].  Thus, in this lattice, bonds in the TwK lattice are the analog of $G$-bonds in the TbP model and those in HL the analog of $B$-bonds; those in the GK lattice form both $B$ and $G$ bonds. Figure \ref{Fig:PhaseDiagrams} displays the phase diagrams of these models, to be explained more fully in the next section.

\begin{figure}
	\begin{minipage}{0.48\linewidth}
		\centering
		(a) \vspace{0.1cm} \\
		\includegraphics[width=\linewidth]{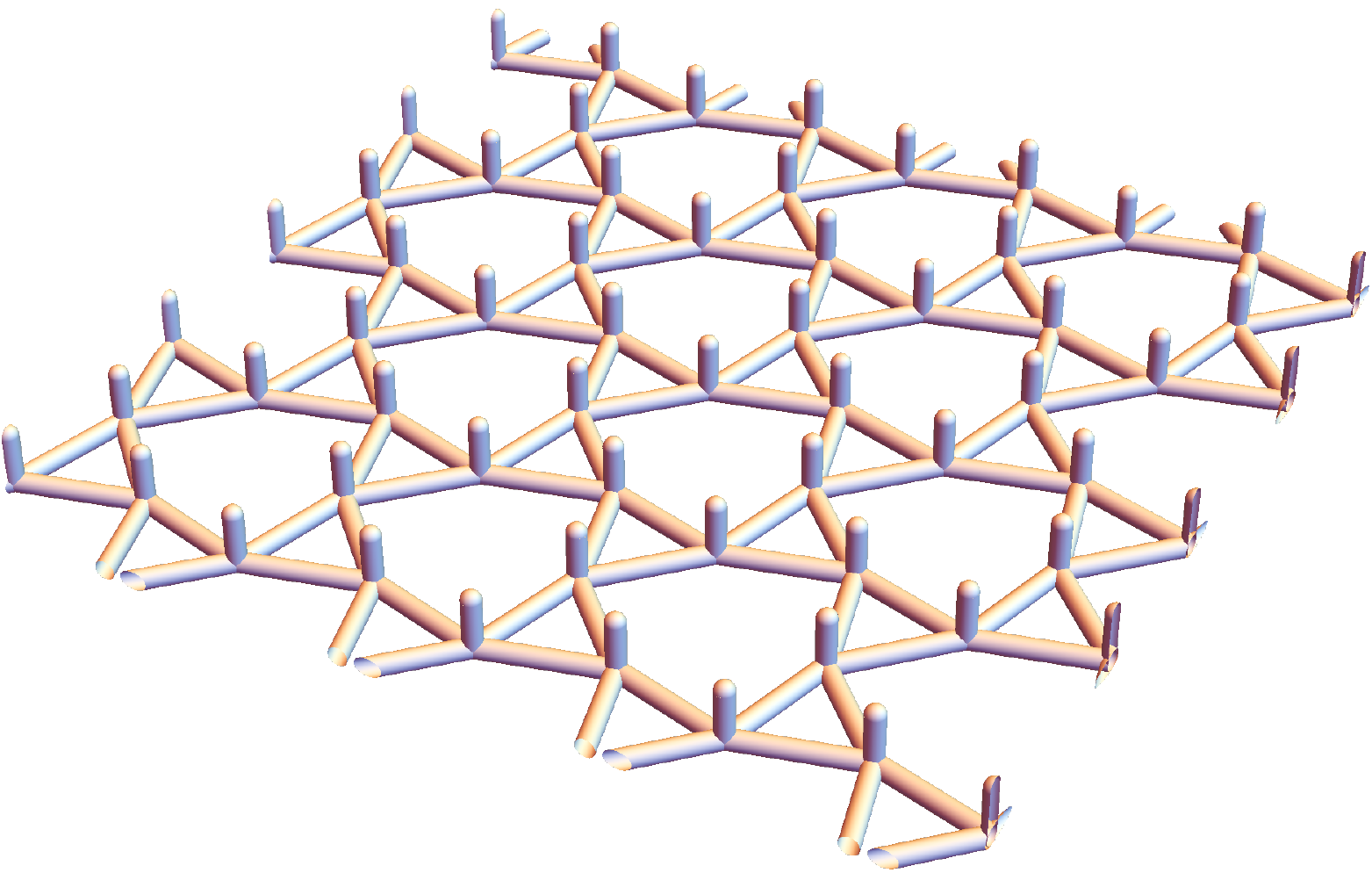}
	\end{minipage}
	\begin{minipage}{0.48\linewidth}
		\centering
		(b) \vspace{0.1cm} \\
		\includegraphics[width=\linewidth]{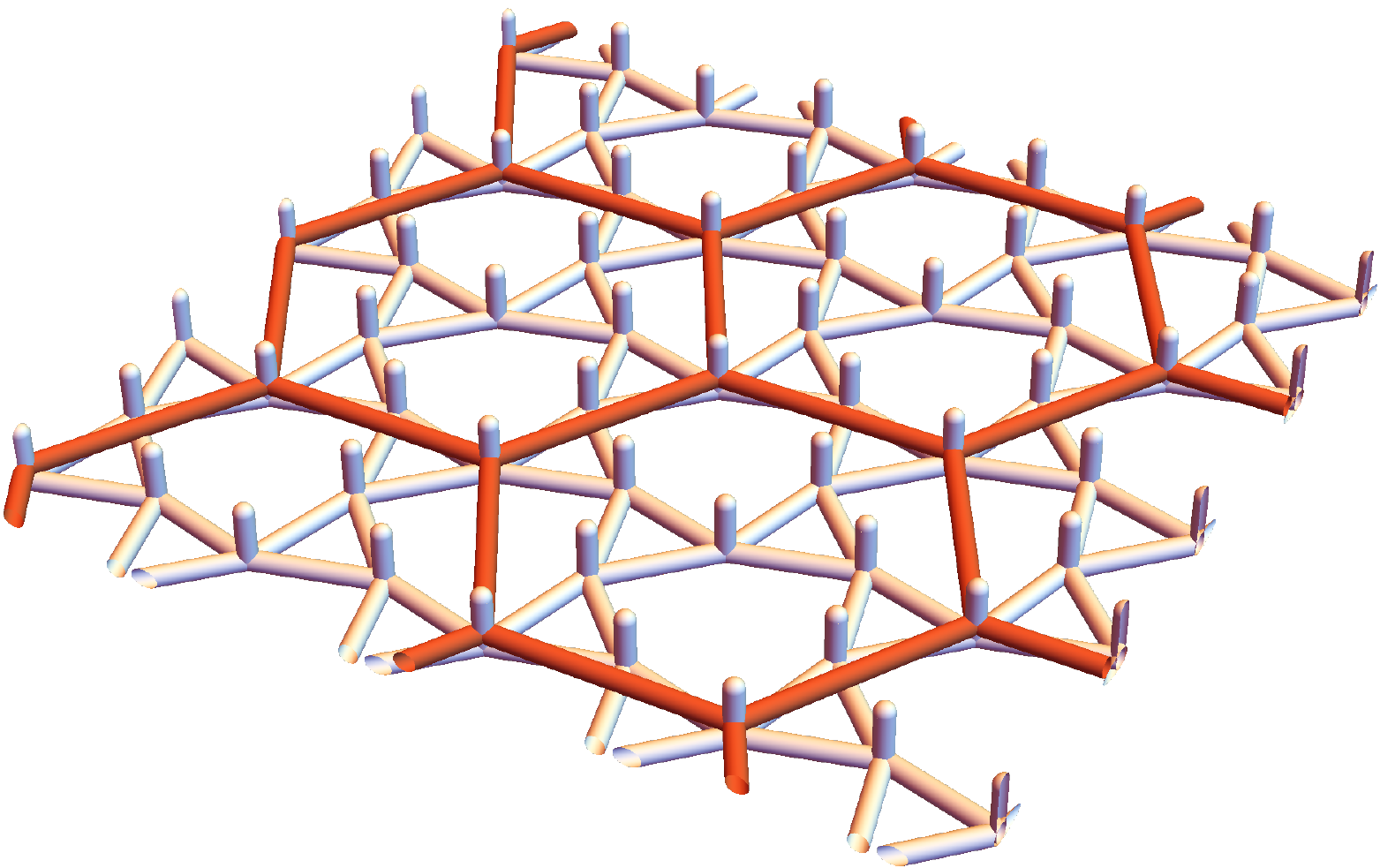}
	\end{minipage}
	\\
	\begin{minipage}{0.48\linewidth}
		\centering
		(c) \vspace{0.1cm} \\
		\includegraphics[width=\linewidth]{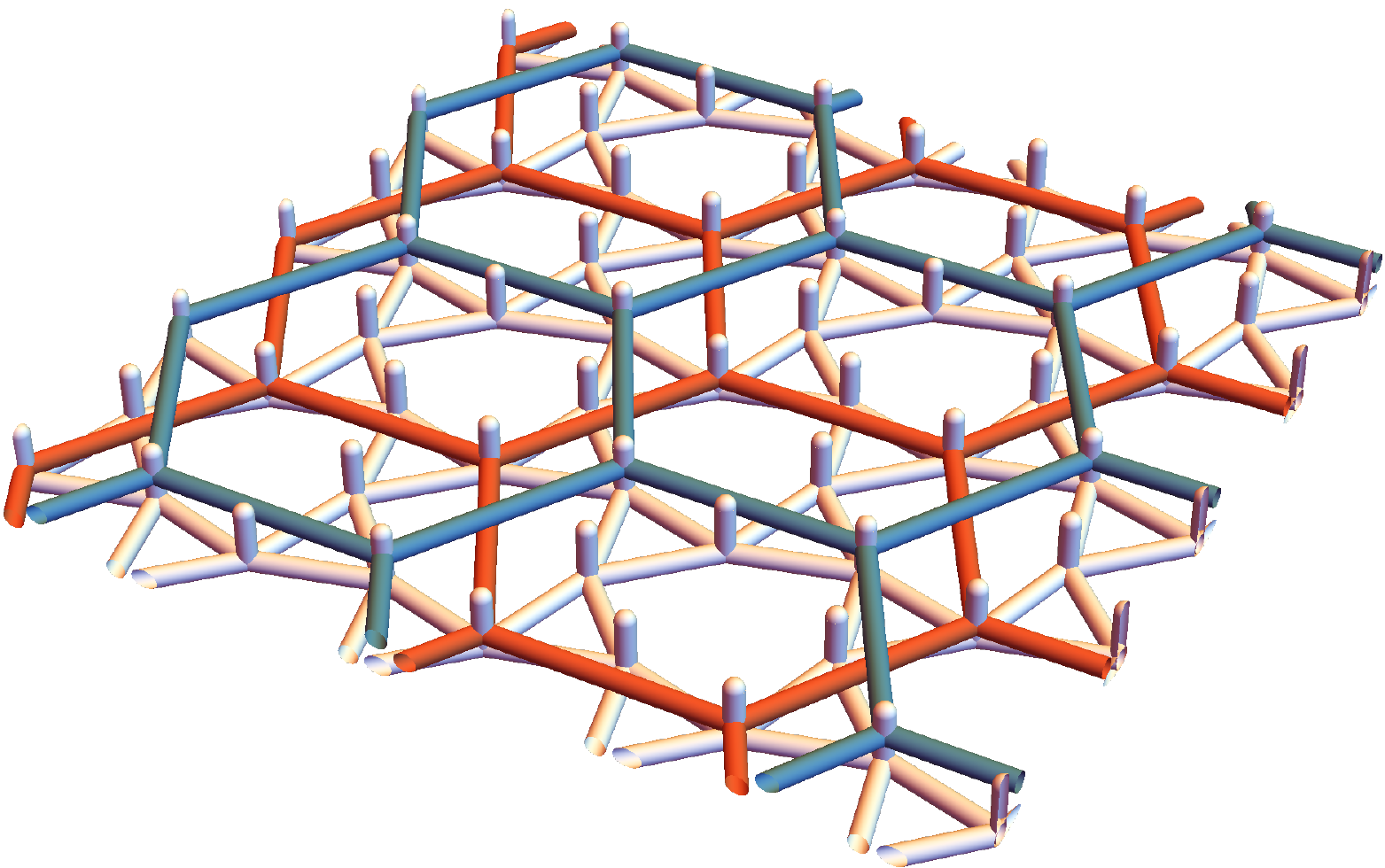}
	\end{minipage}
	\begin{minipage}{0.48\linewidth}
		\centering
		(d) \vspace{0.1cm} \\
		\includegraphics[width=\linewidth]{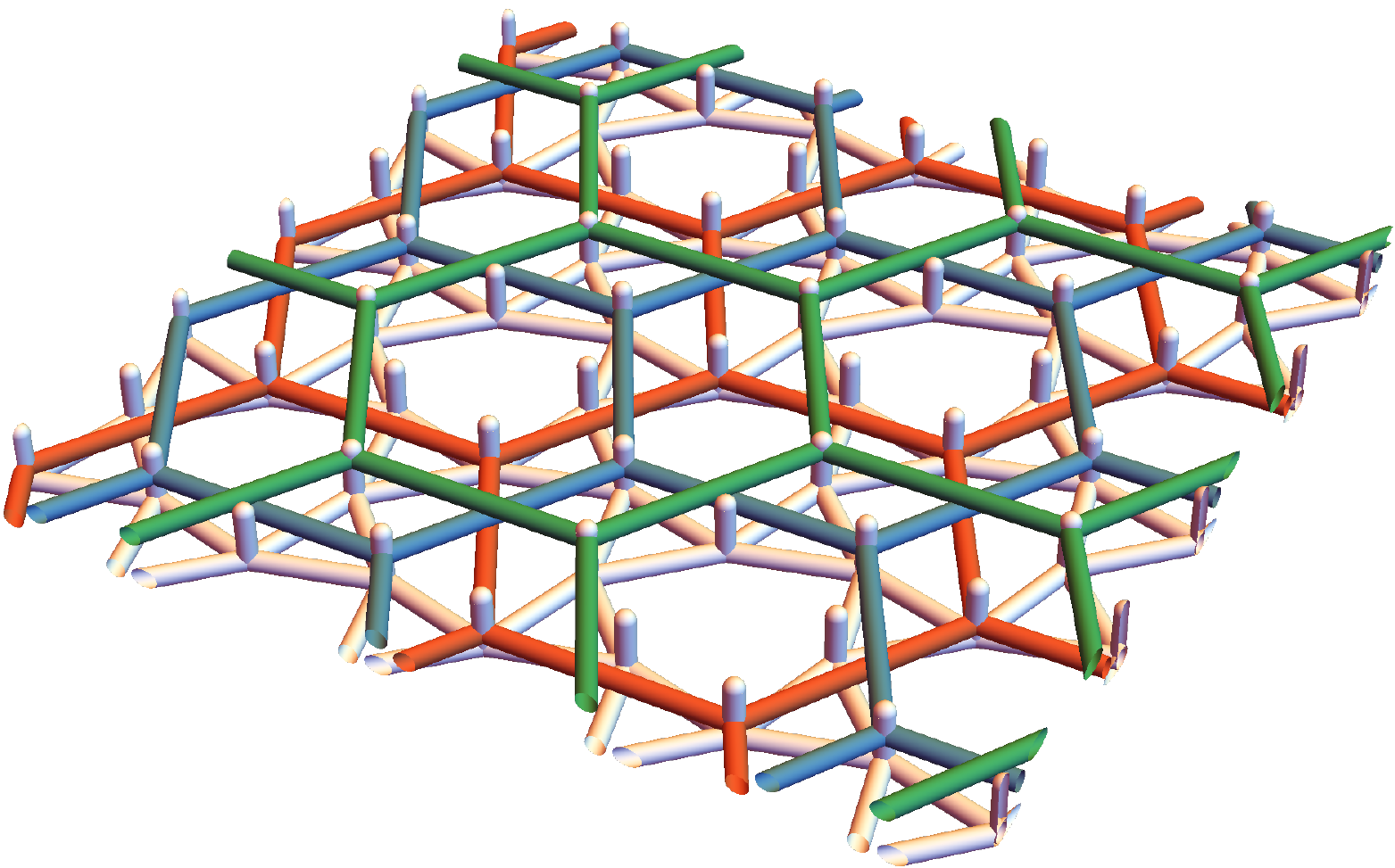}
	\end{minipage}
	\caption{One can design a 3D-printable version of the TwK/H lattice by adding pins to the sites of the twisted kagome lattice (a), and sequentially stacking the remaining three honeycomb lattices (b)-(d) on top of it.}
	\label{Fig:StackedLattices}
\end{figure}

In principle, these models provide a simple algorithm for creating, via 3D printing or related methods, physical 2D materials with  arbitrary Poisson ratios.  They do, however, suffer from a technical drawback in that the added bonds cross each other and necessarily introduce additional nodes in a purely 2D geometry.  This drawback can be addressed in two ways.  In the first, all but one of the extra, GK bond-lattices introduced in the TwK/GK lattice by the further neighbor bonds can be eliminated. Lattices constructed in this way have no bond crossings, and they have $C_3$ symmetry and thus isotropic elasticity~\footnote{On the other hand, the combination of the TwKL with one honeycomb lattice ({\it e.g.} the one formed by red bonds in Fig.~\ref{Fig:GK/Hfig}) does not have $C_3$ symmetry. We need at least three honeycomb lattices to have isotropic elasticity in the TwK/H model.}. An alternative approach is to stack different lattices formed by further-neighbor bonds in different layers connected by rigid vertical bonds between identical realizations off the original TwK lattice as shown in Fig.~\ref{Fig:StackedLattices}. 

In what follows, Sec.~\ref{sec:Results} reviews our principal results, Sec.~\ref{sec:Energy} defines our model energies and their elastic limits and auxetic response, Sec.~\ref{sec:Simulations} discusses our numerical simulations, Sec.~\ref{sec:Scaling} presents our effective medium theory (EMT) and its scaling predictions at critical points, and Sec.~\ref{sec:Conclusions} presents a summary discussion.  The appendices provide details of the lattice structures, dynamical matrices, dispersion relations, asymptotic behavior of the EMT integrals and additional three-dimensional plots of the moduli.

\section{Results}
\label{sec:Results}

Both of our models are built on the TwK lattice, whose bulk modulus is zero, but because of the different geometries imposed by the further-neighbor bonds, they have different-size unit cells (see Fig.~\ref{fig:cellVectors} at Appendix~\ref{subsec:LatticeStructures} for an illustration of the unit cells of both models).
The unit cell of the TwK/GK model has the same number of sites ($j=3$) and NN $a$-bonds ($\tz_a=6$) as the TwK lattice, and it has the same number of $b$-bonds ($\tz_b=6$) as $a$-bonds. The unit cell of the TwK/H lattice is three times as large as that of the TwK lattice with $j=9$ and $\tz_a=18$ but with $\tz_b = 9$.  [Note:  $\tz_r$ is the number of $r$-bonds in a unit cell, which is half the coordination number $z_r$ associated with these bonds]. These lattice characteristics are summarized in Table \ref{table}.
The Maxwell count, setting the number of degrees of freedom per unit cell ($jd = 2j$ in two dimensions) equal to the average number of constraints per cell ($p_a \tz_a + p_b \tz_b$), predicts the EMT RP phase boundary in the $p_a\text{--}p_b$ EMT phase diagram shown in Fig.~\ref{Fig:PhaseDiagrams} to occur at $\Delta p_{RP}=0$, where
\begin{equation}
	\Delta p_{RP} = p_a \tz_a + p_b \tz_b- jd
	\label{eq:JG1}
\end{equation}
measures the distance from the RP line along a path perpendicular to that line.
The lines of RP transitions [lines $J_G\text{--}J_{BG}$ and $J_G\text{--}J_B$] terminate at critical points at their intersections with the boundary lines $A=(p_a=1,p_b)$ and $B=(p_a, p_b = 1)$. In both cases, Eq.~(\ref{eq:JG1}) sets the  intersection with the $A$-line at
\begin{equation}
 p_b^G = (jd -\tz_a)/\tz_b = 0.
\end{equation}  
The points $J_{G}=(1,p_b^{G})$ (with $G=G1$ or $G2$) are ``shear-jamming" points~\cite{BiBeh2011,JesiSet2017,BehringerChak2019}, at which $G$ jumps discontinuously from zero in paths (such as $CJ_{G1}D$ in Fig.~\ref{Fig:PhaseDiagrams}(a)) from the floppy region~\footnote{Our use of shear-jamming, which is defined by a discontinuous jump in $G$ at the jamming point, differs from that of Refs.~\cite{BiBeh2011,JesiSet2017,BehringerChak2019}, which refers to jamming induced by shear.}. The second intersection at $p_b^{RP} = 1$ occurs at $p_a^{BG} = 0$, i.e., at $J_{BG} = (0,1)$ in the TwK/GK model and at $p_a^{B} =(jd - \tz_b)/z_a = (2\times 9 - 9)/18=1/2$, i.e., at $J_{BG}=(1/2,1)$ in the TwK/H model. The HL of the TwK/H model at $J_B$ is fully formed and resists compression, but the system is still on the RP-line along which $G=0$.  Thus $J_B$ is a jamming point at which $B$ jumps discontinuously. At the point $J_{BG} = (0,1)$, only the three GKLs survive, each consisting of three grids of sample-traversing lines of parallel bonds with two rather than a single spacing between lines.  These lines provide states of self-stress that lead to both $B$ and $G$ being positive \cite{LubenskyKai2015}.  As a consequence, $J_{BG}$ is a double ``jamming" point at which both $B$ and $G$ jump continuously from zero. 

\begin{table}
	\begin{tabular}{|l|c|c|c|c|} \hline
		& $\tz_a$ &$\tz_b$ &$j$ & $s$ \\ \hline
		TwK/GK & $6$ & $6$ & $3$ & $6$ \\ \hline
		TwK/H & $18$ & $9$ & $9$ & $9$ \\ \hline
	\end{tabular}
	\caption{Table of basic parameters of the TwK/GK and TwK/H lattices.}
	\label{table}
\end{table}

Figure \ref{Fig:PhaseDiagrams} also shows data simulation points that indicate an RP-transition line that lies mostly below, but close to, the EMT RP-line and terminates within numerical error at the EMT points $J_G$ and $J_{BG}$. The difference between the EMT RP-lines is greatest in the TwK/H model near $J_B$.
We set the twist angle $\alpha=\pi/12$ for all numerical results presented in this paper, since our conclusions do not vary with $\alpha$ (even for the self-dual case of $\alpha=\pi/4$~\cite{FruchartVit2020}).

In general, effective-medium theory yields a more faithful representation of the disordered network in the limit of weak lattice dilution.
As shown in previous studies (see {\it e.g.} Refs.~\cite{FengGar1985,SchwartzSen1985,JacobsTho1995}), EMT generally provides accurate but not exact estimates of elastic moduli and phase boundaries, largely because it fails to deal with redundant bonds~\cite{JacobsTho1995} that lead to over- and under-constrained regions in randomly diluted samples.
Our results here and in our previous work~\cite{LiarteLub2019} further support these studies. At the points $J_{G1}$ and $J_{G2}$, the lattices are pure TwK and both the EMT and simulations correctly find that the rigidity transition occurs exactly at these points.

We simulated $32^2$ and $64^2$ unit cells (3072 and 12288 sites) for the TwK/GK model, and $16^2$ and $32^2$ unit cells (2304 and 9216 sites) for the TwK/H model. For both lattices, the difference in the data when we compared the two system sizes was negligible, which reassured us that our system sizes were large enough for finite size effects to be weak.

In the vicinity of the ``jamming" critical points in the EMT, all of the elastic moduli $K$ that undergo a discontinuous jump and satisfy the simple scaling equation,
\begin{equation}
	\frac{K}{K_0}=\frac{\Delta p_{RP}}{\Delta p_{RP} + \mathcal{C}_M \Delta p_M} = \left( 1+ \mathcal{C}_M \frac{\Delta p_{M}}{\Delta p_{RP}} \right)^{-1} ,
\label{eq:K-K0}
\end{equation}
where $\mathcal{C}_M$ is a numerical constant that depends on the jamming point~\footnote{The meaning of the index $M$ (from `majority') will become clear when we introduce the concept of a `majority' lattice in Sec.~\ref{sec:Scaling}; Specific numbers for $\mathcal{C}_M$ can be obtained for each jamming point using Eqs.~\eqref{eq:kalphan0},~\eqref{eq:kalphan1} and Table~\ref{tab:JammingPoints}.}, $\Delta p_M$ equals $1-p_a$ for the two $J_G$ points, $1-p_b$ for the $J_B$, and $J_{BG}$ points and where $\Delta p_{RP}$ is defined in Eq.~(\ref{eq:JG1}). This scaling form predicts $K=K_0$ when $\Delta p_M = 0$ for any $\Delta p_{RP} \geq 0$.  Thus, for example $G$ undergoes a discontinuous jump at the point $J_G$ along a path such as $CJ_{G1} D$ in Fig.~\ref{Fig:PhaseDiagrams}(a).  Away from the jamming points and near the RP line, all moduli grow linearly with $\Delta p_{RP}$ with a coefficient that changes with distance along the RP line. This behavior is clearly indicted in Eq.~(\ref{eq:K-K0}).

\begin{figure}[!ht]
		\centering
		\begin{minipage}{0.48\linewidth}
			\centering
			(a) \vspace{0.1cm} \\
			\includegraphics[height=0.9\linewidth]{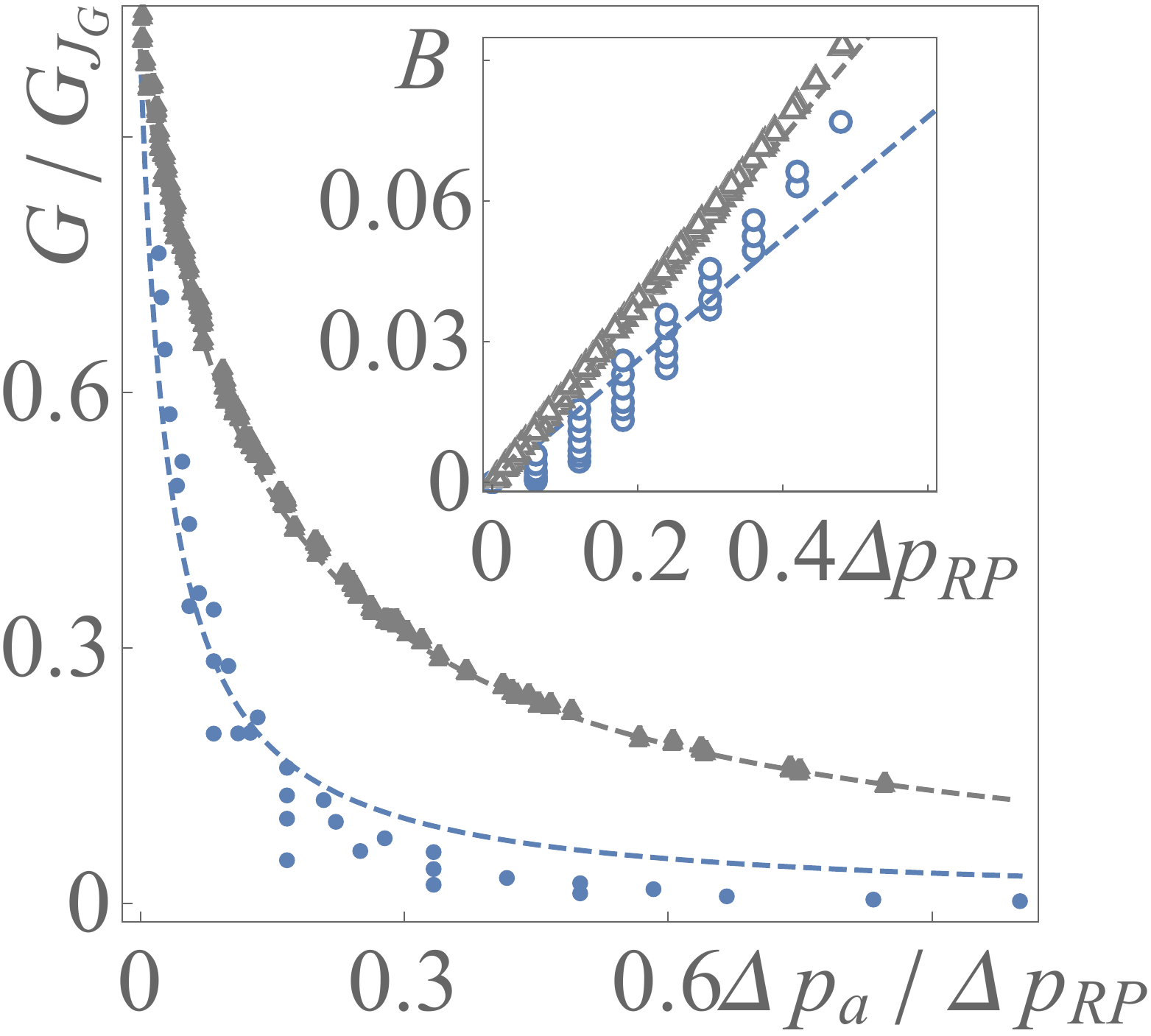}
		\end{minipage}
		\begin{minipage}{0.48\linewidth}
			\centering
			(b) \vspace{0.1cm} \\
			\includegraphics[height=0.9\linewidth]{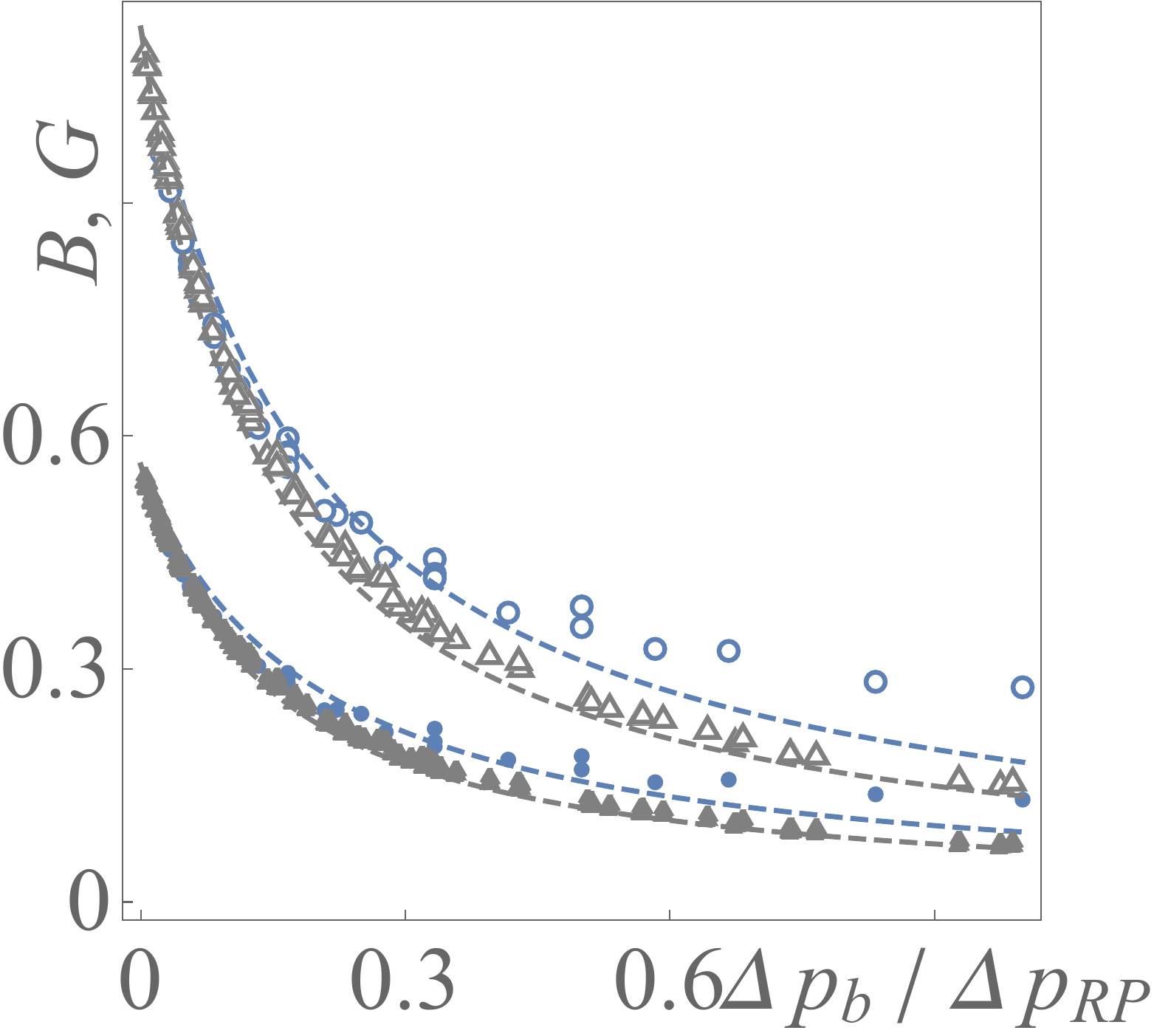}
		\end{minipage} \vspace{0.2cm} \\
		\begin{minipage}{0.48\linewidth}
			\centering
			(c) \vspace{0.1cm} \\
			\includegraphics[width=\linewidth]{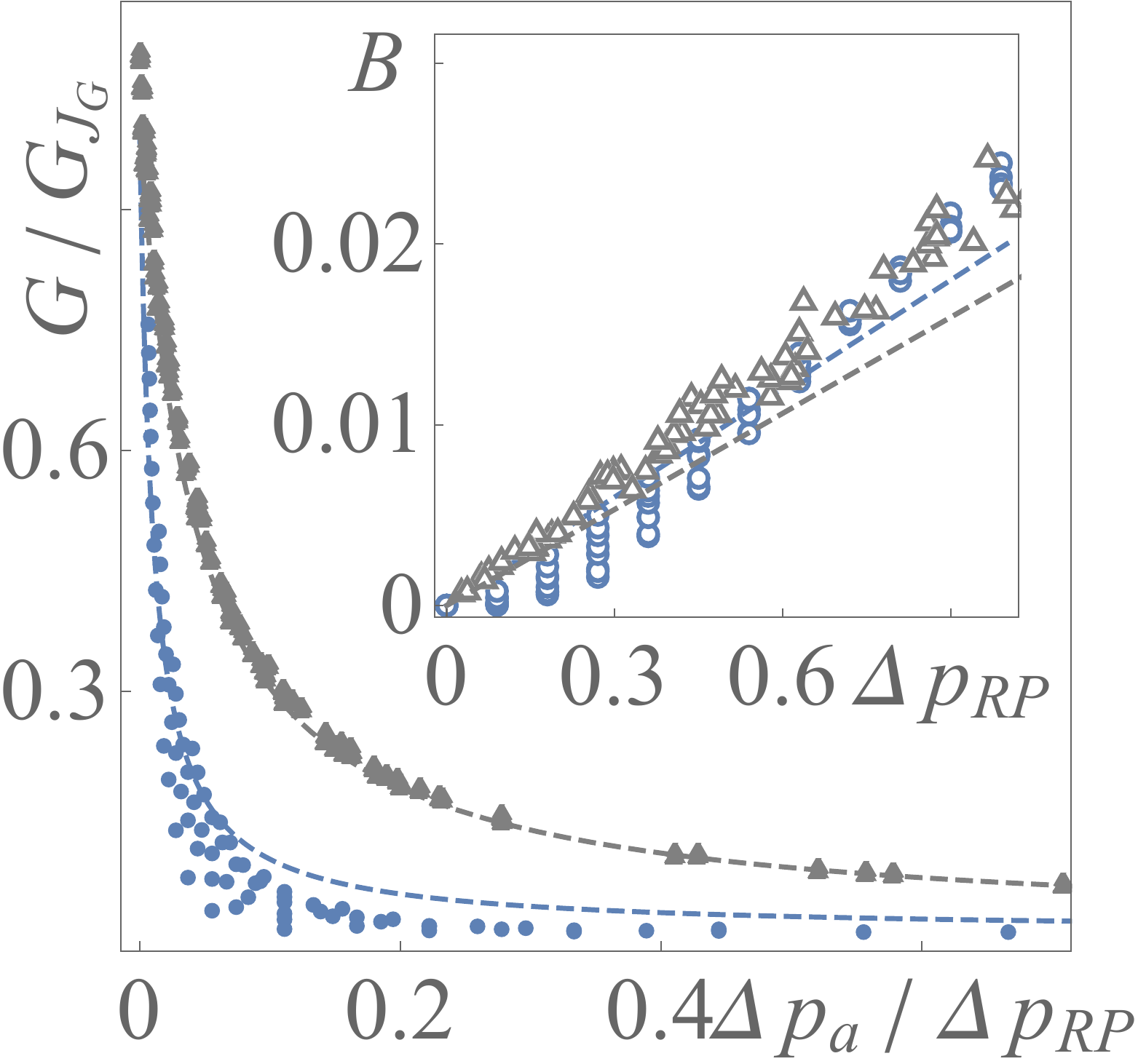}
		\end{minipage}
		\begin{minipage}{0.48\linewidth}
			\centering
			(d) \vspace{0.1cm} \\
			\includegraphics[width=\linewidth]{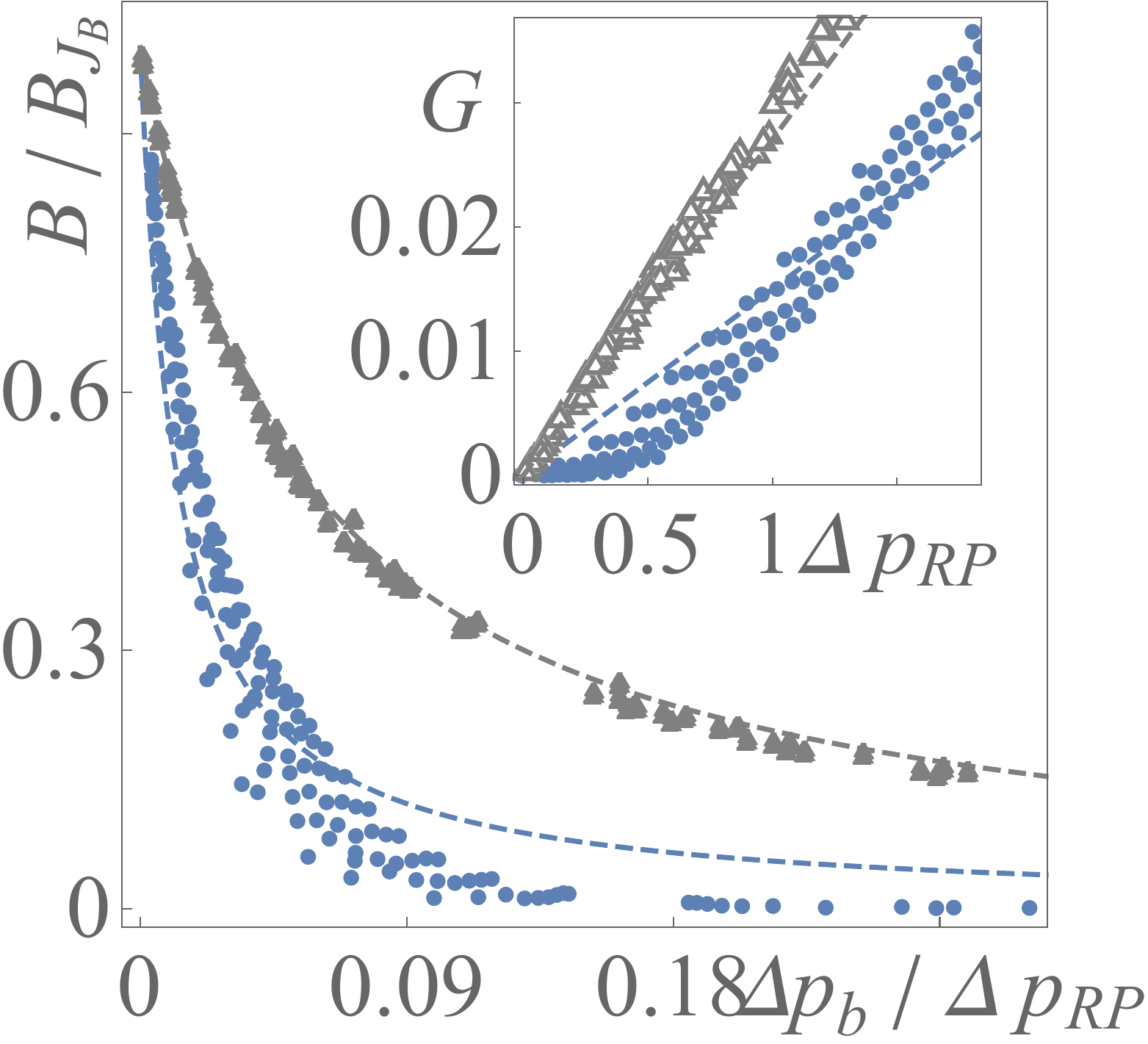}
		\end{minipage}
	\caption{Scaling behavior of the TwK/GK ((a) and (b)) and TwK/H ((c) and (d)) models. Filled and open circles represent the shear and bulk moduli, respectively. Gray triangles and blue circles correspond, respectively, to full EMT solutions and to numerical simulations for a set of points in a rigid region in the neighborhood of $J_G$ ((a) and (c)), $J_B$ (d), and $J_{BG}$ (c). The dashed lines correspond to our analytical predictions (Eq.~\ref{eq:K-K0} normalized near the critical points).
	\label{Fig:Scaling}}
	\end{figure}

\begin{figure}[!ht]
		\centering
		\begin{minipage}{0.48\linewidth}
			\centering
			(a) \vspace{0.1cm} \\
			\includegraphics[height=0.9\linewidth]{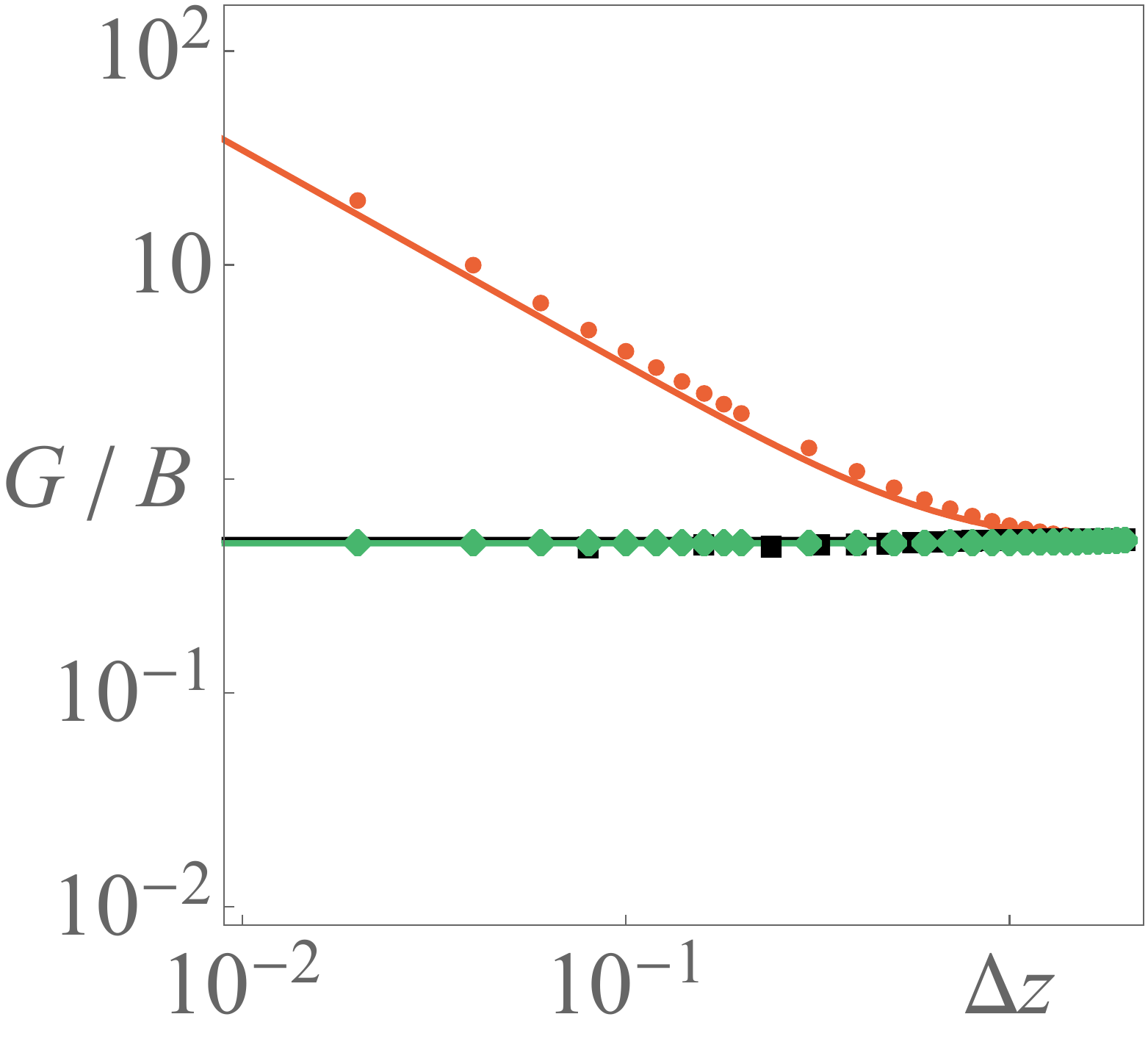}
		\end{minipage}
		\begin{minipage}{0.48\linewidth}
			\centering
			(b) \vspace{0.1cm} \\
			\includegraphics[height=0.9\linewidth]{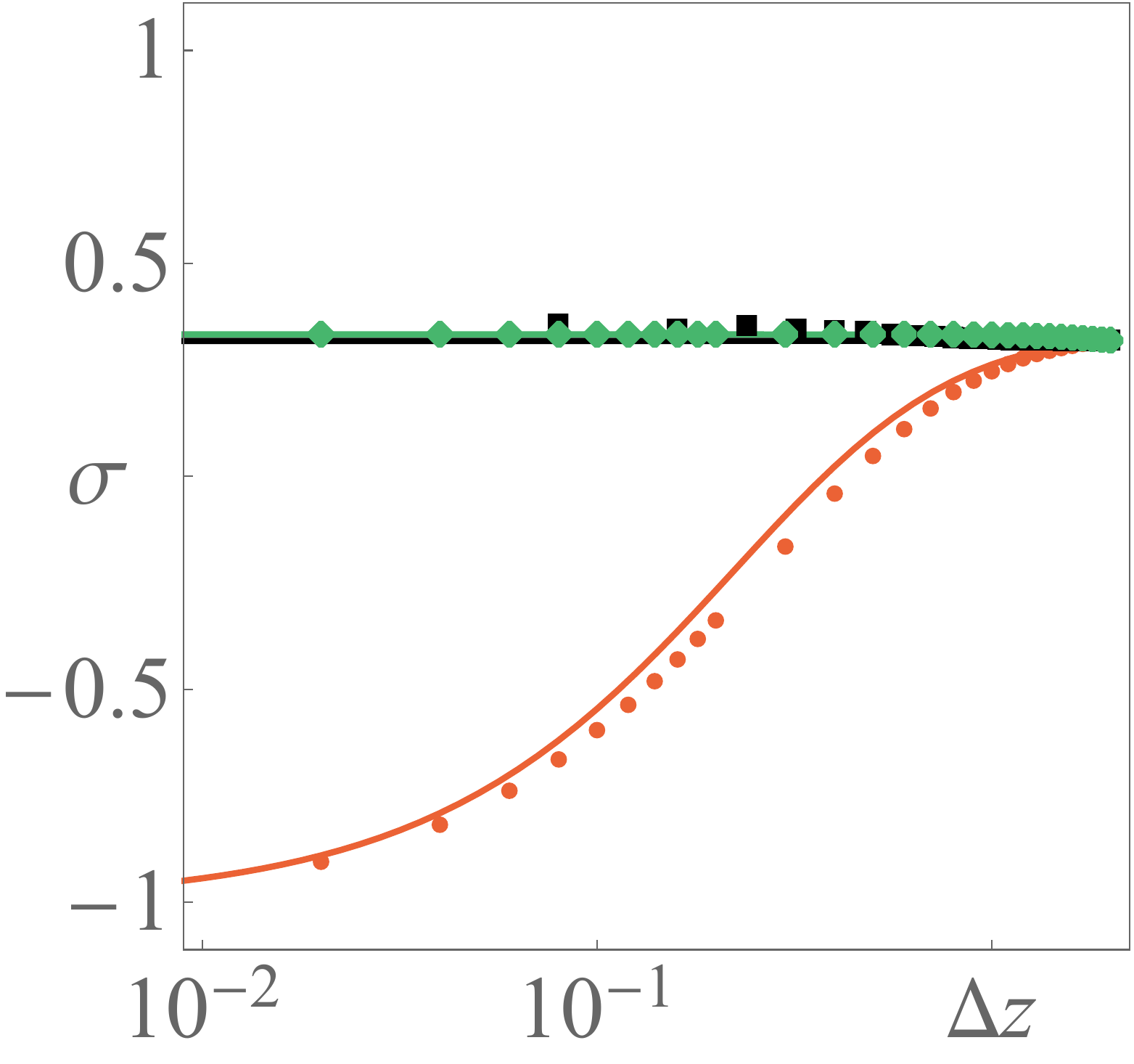}
		\end{minipage} \vspace{0.2cm} \\
		\begin{minipage}{0.48\linewidth}
			\centering
			(c) \vspace{0.1cm} \\
			\includegraphics[height=0.9\linewidth]{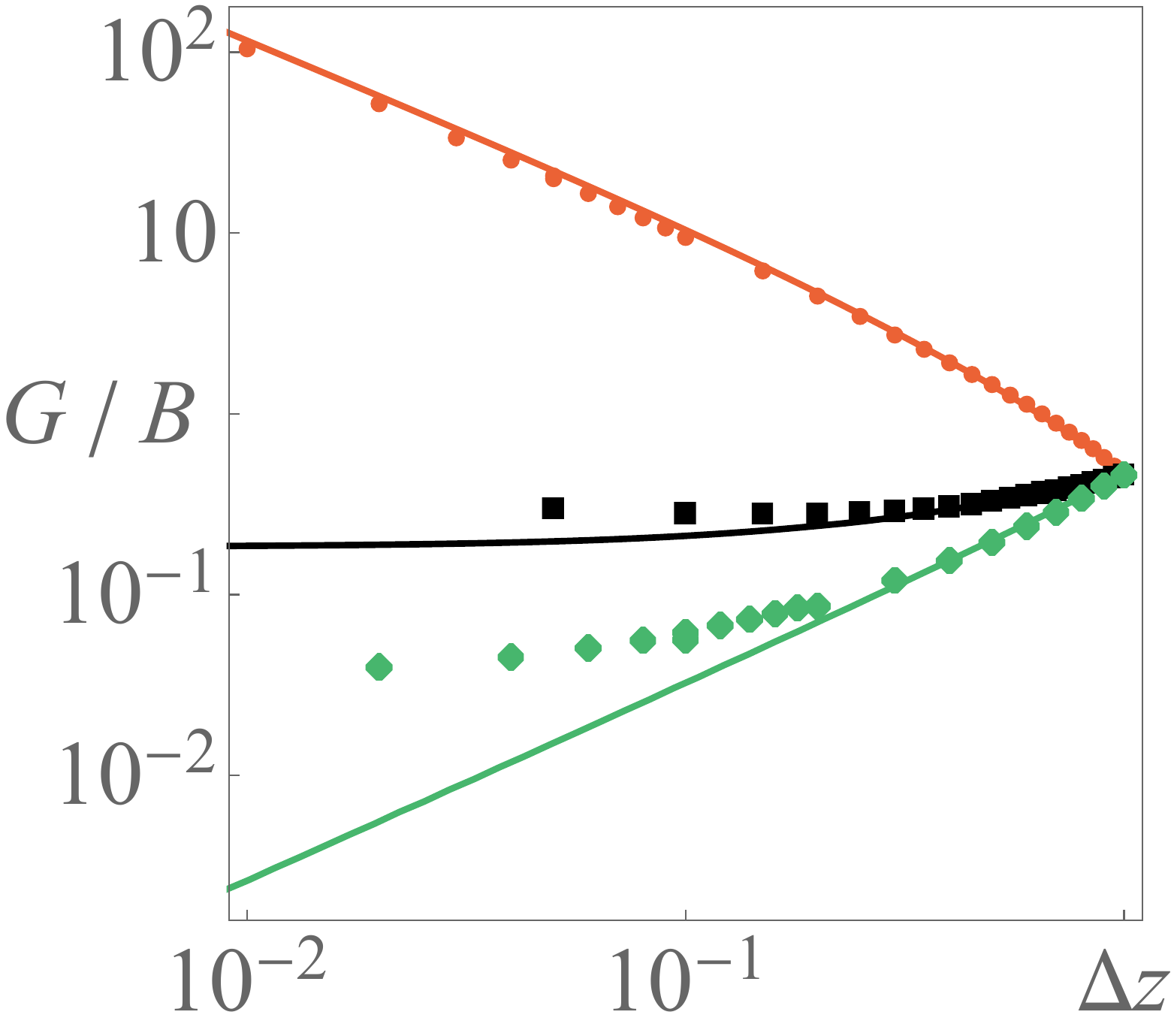}
		\end{minipage}
		\begin{minipage}{0.48\linewidth}
			\centering
			(d) \vspace{0.1cm} \\
			\includegraphics[height=0.9\linewidth]{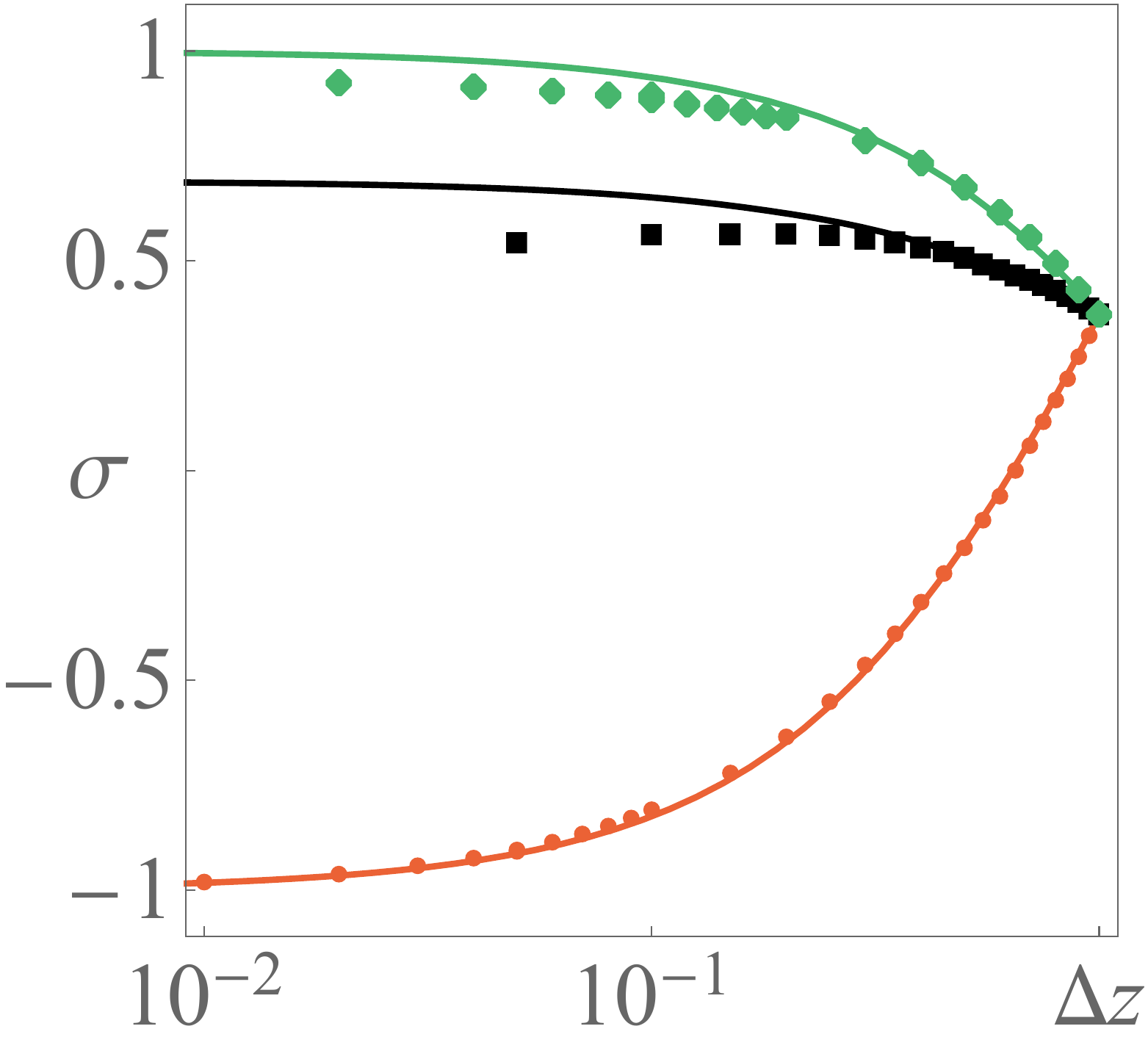}
		\end{minipage} 
	\caption{Simulation (symbols) and EMT (lines) results for $G/B$ ((a) and (c)) and Poisson ratio ((b) and (d)) as a function of $\Delta z \equiv 2 \Delta p_{RP} / (j\,d)$ for the TwK/GK ((a) and (b)) and the TwK/H ((c) and (d)) models, along paths towards $J_G$ (red circles), $J_{BG}$ or $J_{B}$ (green diamonds), and the RP line (black squares), as depicted in Figure~\ref{Fig:PhaseDiagrams}. The discrepancy between simulation and EMT at low $\Delta z$ for some paths is largely due to the discrepancy for the value of the phase boundary $z_c$, which is largest near $J_{B}$ of the TwK/H model.
	\label{Fig:Poisson}}
	\end{figure}

Figure \ref{Fig:Scaling} shows numerical evaluation of the full EMT equations in the vicinity of jamming points collapse onto the analytical form of Eq.~(\ref{eq:K-K0}), with the coefficient $c_J$ depending on the jamming point.  The simulation data collapses onto a modification of the Eq.~(\ref{eq:K-K0}) that takes into account of the fact that the RP transition-line lies below the RP line~\footnote{More precisely, we modify Eq.~\eqref{eq:K-K0} in two ways to collapse simulation data. First, we consider a constant $\mathcal{C}_M$ that is different from the numerical value found in EMT. Second, we change the definition of $\Delta p_{RP}$ from $p_a \tz_a + p_b \tz_b- jd$ to $\Delta p_{RP} = (k_1 p_a + k_2) \tz_a + p_b \tz_b- jd$, and choose $k_1$ and $k_2$ so that $\Delta p_{RP}\approx 0$ when $\min (B,G) < 10^{-10}$ in the simulation data}.
As required, the numerical solution to the EMT equations also show linear growth of the bulk modulus $B$ near the $J_G$ points of both the TwK/GK and TwK/H models and of the shear modulus $G$ at the $J_B$ point of the TwK/H model.  The simulation data are consistent with linear growth of B near the $J_G$ points but are more consistent with quadratic behavior, which may be due to  finite-size effects, of $G$ very near the $J_B$ point of the TwK/H model. Figure \ref{Fig:Poisson} shows the variation of $G/B$ and the Poisson ratio along the paths shown in Fig.~\ref{Fig:PhaseDiagrams}. Note that dilution of our lattices induces changes in the network geometry and hence strongly affects the Poisson ratio, in agreement with the results of Ref.~\cite{HanifpourZap2018}. Three-dimensional plots of $B$ and $G$ obtained both from our EMT and our numerical simulations are shown in Appendix~\ref{sec:ExtraPlots}. 
 
\section{Model energies and elastic energies}
\label{sec:Energy}

	We consider the harmonic interaction energy arising from central force springs:
\begin{equation}
E
= \sum_{\alpha \in \{a,b\}} \frac{k_\alpha}{2} \sum_{\{i,j\} \in C_\alpha}
g_{ij}^\alpha\left[ (\bm{u}_j-\bm{u}_i) \cdot \hat{\bm{r}}_{ij} \right]^2,
\label{eq:EnergyRSpace}
\end{equation}
where $\bm{u}_i$ is a displacement vector,
$\hat{\bm{r}}_{ij}=(\bm{r}_i-\bm{r}_j)/|\bm{r}_i-\bm{r}_j|$, with $\bm{r}_i$ giving the
position of site $i$ in the reference lattice, and $C_\alpha$ is a set of neighbor pairs of sites for sub-lattice $\alpha$.
In EMT, $g_{ij}^\alpha = 1, \forall \, i, j$, and bonds in lattices $a$ and $b$ are populated with springs with spring constants $k_a$ and $k_b$ satisfying a set of self-consistent equations depending on probabilities $p_a$ and $p_b$.
In the simulations, $k_\alpha=1$ and $g_{ij}^{\alpha}$ is a bimodal random variable equal to one with probability $p_\alpha$ and zero with complementary probability $1-p_\alpha$.
In Appendix~\ref{sec:Lattice}, we provide details about the lattice structures, dynamical matrices of our models, and details of phonon dispersion relations of the TwK/GK model.

In the long-wavelength limit, Eq.~\eqref{eq:EnergyRSpace} reduces to the elastic isotropic limit
\begin{equation}
\frac{E}{V}
= \frac{B}{2} \left(u_{xx} + u_{yy }\right)^2 + 2 \, G\left[ {u_{xy}}^2 + \frac{1}{4}
\left(u_{xx} -u_{yy} \right)^2 \right],
\label{eq:IsotropicEnergy}
\end{equation}
where $u_{ij}$ are components of the linearized strain tensor, $B$ and $G$ are the bulk and shear moduli, respectively, and $V$ is the volume.

For the TwK/GK model, analytical expressions for $B$ and $G$ in terms of $k_a$, $k_b$ and $\alpha$ can easily be derived:
\begin{align}
& B
= \frac{3}{4} \, k_b \, \frac{2k_a +3 k_b -k_a \cos 2 \alpha}{k_a +2 k_b -k_a \cos 2 \alpha}, \label{eq:ModuliB}\\
& G
= \frac{3}{16} \left( k_a +3 k_b\right),
\label{eq:Moduli}
\end{align}
where $\alpha$ is the twist angle of the TwK lattice.
Note that $B \rightarrow 0$ and $G>0$ for $k_b \rightarrow 0$ and $k_a>0$, except at $\alpha=0$, where $B \rightarrow (3/8) k_a >0$. The Poisson ratio,
\begin{equation}
\sigma
= \frac{B-G}{B+G},
\label{eq:Poisson}
\end{equation}
is negative (auxetic structure) for
\begin{equation}
\sin^2 \alpha > \frac{k_b (k_a + 3k_b)}{k_a(k_a-k_b)}.
\label{eq:AuxeticCondition}
\end{equation}
The phase diagram of Fig.~\ref{Fig:PhaseDiagrams} shows auxetic regions in red and and non-auxetic regions in blue for $\alpha = \pi/12$.

Calculation of the moduli for the TwK/H model poses a greater challenge than it does for the TwK/GK model, and we present only numerical solutions for $\alpha=\pi/12$.
Let $U_\text{aff}$ be the $N_b$-dimensional vector of affine bond deformations and $\hat{\text{t}}_\alpha$ be the $\alpha$th orthonormal basis vector of ker($Q$) (state of self-stress), where $N_b$ is the number of bonds in a unit cell and $Q$ is the equilibrium matrix~\cite{LubenskyKai2015}.
To evaluate $B$ and $G$ for arbitrary numerical values of $k_a$ and $k_b$, we first project affine deformations into the states of self stress of our lattice model: $U_\text{aff}^\alpha = U_\text{aff} \cdot \hat{\text{t}}_\alpha$.
We then use Eq.~(3.10) of~\cite{LubenskyKai2015}, which analytically includes the effects of nonaffine distortions, to express the elastic free energy as a quadratic form in terms of strain components~\footnote{Note that our matrix of spring constants $\bm{k}$ is not (in general) proportional to the identity matrix, so we cannot use the limiting form $(k/2V) \sum_\alpha (\text{U}_\text{aff} \cdot \hat{\text{t}}_\alpha)^2$ on the right of Eq.~(3.10) of Ref.~\cite{LubenskyKai2015}.}.
To extract $B$ and $G$, we compare the resulting free energy with the isotropic elastic energy given by Eq.~\eqref{eq:IsotropicEnergy}.
As expected, our numerical evaluations show that $B \rightarrow 0$ and $G>0$ when $k_b \rightarrow 0$ and $k_a>0$ (as in the region near $J_G$ in Fig.~\ref{Fig:PhaseDiagrams}b), whereas $G \rightarrow 0$ and $B>0$ for $k_a\rightarrow 0$ and $k_b>0$ (as in the region near $J_B$ in Fig.~\ref{Fig:PhaseDiagrams}b).
To find the threshold for auxetic behavior, we numerically solve the equation $B=G$ (corresponding to $\sigma=0$) for $\eta \equiv k_b / k_a$, and find that the poisson ratio is negative (auxetic structure) for $k_b \lessapprox 0.37 k_a$.

\section{Numerical Simulations}
\label{sec:Simulations}
In this section, we briefly describe our numerical simulations. As a first step, we generate supercells composed of $N_{\text{cell}} = L \times L$ unit cells of our two model lattices. For the TwK/GK we use $L=64$, and for TwK/H with its unit cell three times larger, we use $L=32$. The resulting number of sites per supercell is $12,288$ for the TwK/GK and $9,216$  and the TwK/H. Next, we randomly remove $a$ and $b$ bonds from the supercells with probability $1 - p_a$ and $1 - p_b$, respectively. Care is taken that the removal of the exterior bonds is consistent with periodic boundary conditions. 

To calculate the elastic moduli of the resulting diluted supercells, we apply affine deformations via multiplying the site positions with the deformation tensor,
\begin{equation}
	\label{Lambda}
	\brm{ \Lambda}_{\text{bulk}} =  \left( 
	\begin{array}{cc}
		1+\frac{\chi}{2} & 0\\
		0 & 1+\frac{\chi}{2}
	\end{array} 
	\right)
	\quad \mbox{or} \quad
	\brm{ \Lambda}_{\text{shear}} =  \left( 
	\begin{array}{cc}
		1 & \frac{\chi}{2} \\
		\frac{\chi}{2} & 1
	\end{array} 
	\right) ,
\end{equation} 
for bulk and pure shear deformation, respectively ({\it i.e.} the displacement $\bm{u}_i = \Lambda \cdot \bm{x}_i$, where $\bm{x}_i$ is the equilibrium position of site $i$ in the absence of any applied deformation). We set $\chi$, specifying the magnitude of the deformation, to $0.01$. In addition to the affine deformation, the displacement $\uv_i$ is given a non-affine component $\delta \uv_i$, $\uv_i \to \uv_i + \delta \uv_i$ to allow for a relaxation of the supercell. Then, we minimize the resulting energy as given in Eq.~(4) over the $\delta \uv_i$ using a conjugate gradient algorithm adapted from Numerical Recipes~\cite{Recipies-C}. Denoting the minima of the elastic energy density $f = E/V$ [{\it cf.} Eq.~\eqref{eq:IsotropicEnergy}] with respect to the two applied deformations by $f^{\text{min}}_{\text{bulk}}$ and $f^{\text{min}}_{\text{shear}}$, the bulk and shear moduli of the TwK/GK are then obtained as
\begin{align}
	B = \frac{2 f^{\text{min}}_{\text{bulk}}}{\chi^2}
	\quad \mbox{and} \quad
	G = \frac{f^{\text{min}}_{\text{shear}}}{2 \chi^2} .
\end{align}
For the TwK/H, we divide the right hand sides by an extra factor of $3$ to compensate for fact that the unit cell is three time larger than that of the TwK/GK. Finally, the so-obtained moduli are averaged over a number (usually ten) lattice realizations for any fixed given pair of $p_a$ and $p_b$.

\section{EMT and critical scaling}
\label{sec:Scaling}

This section provides details of our EMT calculations and their results. We assign occupancy probabilities $p_a$ and $p_b$ for bonds on sub-lattices $a$ (the TwK sub-lattice) and $b$ (the GK sub-lattice in the TwK/GK model and the H sub-lattice in the TwK/H model, respectively).
The effective spring constants $k_a$ and $k_b$ satisfy a set of self-consistent equations given by the EMT~\cite{FengGar1985,MaoLub2011,MaoLub2013,LiarteLub2016,LiarteLub2019}:
\begin{equation}
k_a
= \frac{p_a-h_a}{1-h_a}, \quad
k_b
= \frac{p_b-h_b}{1-h_b},
\label{eq:CPA}
\end{equation}
where
\begin{equation}
h_\alpha
= \frac{k_\alpha}{\tilde{z}_\alpha N_c} \sum_{\bm{q}} \text{Tr} \left[K_\alpha (\bm{q})
\cdot D^{-1} (\bm{q}) \right], \quad \alpha = a,b,
\label{eq:Integral}
\end{equation}
where $\tilde{z}_\alpha$ is the number of $\alpha$-bonds per unit cell [See Table~\ref{table}], and $N_c$ is the number of unit cells.
$K_\alpha$ is the normalized stiffness matrix, $D=k_a K_a + k_b K_b$ is the dynamical matrix, and the trace is taken over $j d$-dimensional matrices (see Appendix~\ref{sec:Lattice} for details).
The integrals $h_\alpha$ satisfy the index summation rule \cite{FengGar1985,LiarteLub2019}:
\begin{equation}
\tilde{z}_a h_a + \tilde{z}_b h_b = j d ,
\label{eq:SumRule}
\end{equation}
which  establishes that $h_a$ and $h_b$ are not independent. 

The functions $h_a$ and $h_b$ depend upon which lattice they are associated with.  They can be evaluated numerically for any $k_a$ and $k_b$, and we provide graphs of them in Appendix~\ref{sec:ExtraPlots}. Here we derive analytical expressions for these functions in the vicinity  of each of the jamming points.  Before proceeding, however, it is useful to introduce the concept of majority and minority lattices associated with these critical points.  The \textit{majority} lattice is the one whose bond occupation probability is exactly one at the jamming point in question, and the minority  lattice is the one whose bond occupation probability is less that one at the same point. 

\vspace{6pt}
\noindent
\textbf{Jamming  points $J_{G1}$ and $J_{G2}$} (see Fig.~\ref{Fig:PhaseDiagrams}):  In both cases, the majority lattice is the TwKL, whose stiffness matrix $K_a\equiv K_M$  is fully gapped, and thus invertible, for all $\qv$ except $\qv=0$. The subscript $M$ refers to the majority lattice.  The evaluation of the expansion of $K_M$ in powers of $k_m/k_M=k_b/k_a$, where $m$ refers to the minority lattice proceeds as follows:
\begin{subequations}  
\begin{align}
h_M & = h_a   = \frac{1}{\tz_M N_c} 
\sum_{\bm{q}} \text{Tr} \left[ K_M (\qv)
\cdot \left(K_M+\frac{k_m}{k_M} K_m \right)^{-1} \right] \label{eq:h_M}\\
& = \frac{1}{\tz_M N_c} \sum_{\bm{q}} \text{Tr}[K_M\cdot K_M^{-1} - (k_m/k_M) K_M^{-1} K_m  \cdots ] \\
& = \frac{j d}{\tz_M} - \frac{1}{c_M \tz_M}\frac{k_m}{k_M} =1-\frac{1}{c_M \tz_M}\frac{k_m}{k_M} \equiv 1- \Delta h_M,
\label{eq:DelhM1}
\end{align}
\end{subequations}
where $\Delta h_M = 1- h_M$ and
\begin{equation}
c_M=\left[\frac{1}{N_c} \sum_{\bm{q}} \text{Tr} \left( K_M^{-1}\cdot K_m \right) \right]^{-1}
\label{eq:cMdef}
\end{equation}
with the numerical constant $c_M$ (see Table~\ref{tab:JammingPoints}) depending on the jamming point. Note that in  both cases, $h_M \rightarrow 1$ as $k_m/k_M \rightarrow 0$. The value  of $h_m$, the  minority field then follows  directly from Eq.~(\ref{eq:SumRule}):
\begin{equation}
h_m = h_b = \frac{1}{\tz_m}(jd -\tz_M h_M)= \frac{\tz_M}{\tz_m} \Delta h_M ,
\end{equation}
because $j d - \tz_M = 0$ for the $J_{G}$ points of both models.

\begin{table}
\begin{tabular}{|l|c|c|c|c|} \hline
	& $J_{G1}$ & $J_{G2}$ &	$J_B$ & $J_{BG}$ \\ \hline
	$B$ & 0 & 0 & $0.75$ & $9/8$ \\ \hline
	$G$ & $3/16$ & $0.1875$ & 0 & $9/16$ \\ \hline
	$\sigma$ & $-1$ & $-1$ & $+1$ & $1/3$ \\ \hline
	$c_M$ & 0.035 & 0.030 & 0.037 & 0.035 \\ \hline  
\end{tabular}
	\caption{Values of $B$, $G$, $\sigma$ and the parameter $c_M$ in the vicinity of jamming points. $J_{G1}$ is the shear jamming point of the TwK/GK lattice and $J_{G2}$ that of the TwK/H lattice.}
\label{tab:JammingPoints}
\end{table}

\vspace{6pt}
\noindent
\textbf{Jamming point $J_{BG}$:} In this case, the majority lattice is the $b$-lattice, which consists of three distinct GKLs that decouple from each other and from the minority TwKL or $a$-lattice. The stiffness matrix $K_M=K_b$ has two zero modes for each wavenumber $\qv$ along the symmetry  lines  $\Gamma K$ and $KM$ in the Brillouin zone. The result is that the calculation of $h_a$ and $h_b$ is considerably more complicated that it is at the $J_G$ points.  Fortunately, the ``heavy lifting" for this calculation has already been done in Ref.~\cite{MaoLub2013}  with the result
\begin{align}
h_{mBG} & = 1 - \frac{1}{\tz_a}\left(\frac{1}{c_M}\frac{k_a}{k_b} \right)^{1/2} \\
h_{MBG} & =  \frac{jd - \tz_a h_m}{\tz_M} =1-\frac{1}{z_b}\left(\frac{1}{c_M}\frac{k_a}{k_b}\right)^{1/2} .
\label{eq:hMGB}
\end{align}
We reemphasize at this point that a nonzero $k_a$ at $J_{BG}$ produces both a nonzero $B$ and a nonzero $G$, and both undergo a discontinuous jump.
Also note that the constants $c_M$ appearing in Eq.~\eqref{eq:hMGB} and later in Eq.~\eqref{eq:hmB} are numerically estimated using the definition of the $h$ integrals; they cannot be evaluated using Equation~\eqref{eq:cMdef}.

\vspace{6pt}
\noindent
\textbf{Jamming point $J_B$:} The majority lattice is again the $b$-lattice and the minority lattice the $a$-lattice. Now $K_M$ has several zero modes for each wavenumber in the Brillouin zone and is thus non-invertible, which considerably complicates the calculation of the $h$'s. The count of zero modes in $K_M$ is obtained as follows:  When $k_m=k_a=0$, there are three  sites per unit cell (or equivalently per wavenumber) that are unattached to the network and unconstrained in their motion.  This gives $3 \times 2 = 6$ zero modes per wavevector $\qv$. In addition when $k_a=0$, the three H lattices are not attached to each other nor to the TwK lattice, and each of the three H lattices has one zero mode per $\qv$ for a total of $d_{M0}= 9$ zero modes per $\qv$. In Eq.~(\ref{eq:h_M}),  $[K_M+(k_m/k_M) K_m ]$ is projected onto the range of $K_M$ whose dimension is  $d_R = j d- d_{M0} = 2 \times 9 - 9 = 9$.  
The  limit of $k_m \rightarrow 0$ gives $h_M=h_b = d_R/\tz_b=1$. In addition though it may not be immediately obvious, $h_M$ has a well-behaved power series in $k_m/k_M$.  As a result, $h_M$ has the same functional form as it has in the vicinity of the $J_G$ points. $h_m$, however is different in that its value $k_m \rightarrow 0$ is not zero, as follows from the application of Eq.~(\ref{eq:SumRule}):
\begin{align}
h_{mB} & = \frac{j d - \tz_b h_b}{\tz_a}=\frac{j d - \tz_b}{\tz_a}+\frac{\tz_b}{\tz_a} \Delta h_M \nonumber \\
& =\frac{1}{2}+\frac{1}{c_M \tz_M}\frac{k_a}{k_b}.
\label{eq:hmB}
\end{align}

\vspace{6pt}
We are now ready to calculate the effective spring constants near all of the jamming points. Following Eqs.~\eqref{eq:CPA},~\eqref{eq:JG1} and~\eqref{eq:SumRule}, we can express $k_M$ and $k_m$ as
	\begin{align}
& k_M
= \frac{\tz_M \Delta h_M -\tz_M \Delta p_M}{z_M
	\Delta h_M},
\label{eq:CPA2a} \\
& k_m
= \frac{\Delta p_{RP} + \tz_M \Delta p_M -\tz_M \Delta h_M}{s
	-\tz_M \Delta h_M},
\label{eq:CPA2b}
\end{align}
where
\begin{equation}
s= \tz_a +\tz_b - j d .
\end{equation}
Taking the ratio of $k_m$ to $k_M$ and using Eqs.(\ref{eq:DelhM1}) and (\ref{eq:hMGB}), we obtain
\begin{align}
c_M  \left(\tz_M \Delta h_M \right)^{n}
& \approx \frac{\Delta p_{RP} +\tz_M \Delta p_M -\tz_M \Delta h_M}{s
	-\tz_M \Delta h_M} \nonumber \\
& \quad \times \frac{1}{\tz_M\Delta h_M
	-\tz_M \Delta p_M}  ,
\label{eq:DeltahEq}
\end{align}
where $n=0$ applies to the $J_G$ and $J_B$ points and $n=1$ applies to the $J_{BG}$ point. Solving this equation for $\Delta h_M$ when $n=0$, we obtain
\begin{equation}
\tz_M \Delta h_M - \tz_M \Delta p_M
\approx \frac{\Delta p_{RP}}{1 + s \, c_M},
\label{eq:Deltahn0}
\end{equation}
and then from Eqs.~(\ref{eq:CPA2a}) and (\ref{eq:CPA2b}),
\begin{equation}
k_M
\approx \frac{\Delta p_{RP}}{\Delta p_{RP} + (1+s\,c_M) \tz_M \Delta p_M},
\label{eq:kalphan0}
\end{equation}
and
\begin{equation}
k_{m}
\approx \frac{c_M \, \Delta p_{RP}}{1 + s \, c_M} .
\label{eq:kalphabarn0}
\end{equation}

Finally when $n=1$ ($J_{BG}$), the equation for $\Delta h_M$ is quadratic rather than linear with a solution to second order in $\Delta p_{RP}$ and $\Delta p_M$ of
\begin{equation}
z_M \Delta h_M
\approx \left(\Delta p_{RP} + \tz_M \Delta p_M \right)
\left(1-s\, c_M \Delta p_{RP}\right),
\label{eq:Deltahn0}
\end{equation}
\begin{equation}
k_M
\approx \frac{\Delta p_{RP}}{\Delta p_{RP} + \tz_M \Delta p_M},
\label{eq:kalphan1}
\end{equation}
\begin{equation}
k_m
\approx c_M \Delta p_{RP} \left(\Delta p_{RP} + \tz_M \Delta p_M \right).
\label{eq:kalphabarn1}
\end{equation}

\section{Review and Future Questions}
\label{sec:Conclusions}

This paper has presented an analysis, via Effective-Medium Theory (EMT) and numerical simulations, of the varied elastic and phonon properties of model lattices of central-force harmonic springs that tune continuously from a twisted kagome lattice with $B=0$ and $G>0$ to either a honeycomb lattice with $B>0$ and $G=0$ or to a generalized untwisted kagome lattice with both $B$ and $G$ greater than zero. In each case the two extreme lattices share the same lattice sites but have a different and mutually exclusive set of bonds, which can be occupied with springs with probabilities $p_a$ and $p_b$.  The phase diagrams in the $2D$ $p_a-p_b$ space [Fig.~\ref{Fig:PhaseDiagrams}] exhibit jamming critical-end-points, at which one of or both $B$ and  $G$ jump discontinuously from zero, that terminate lines of second-order rigidity-percolation transitions separating the rigid from the floppy regime. EMT provides a semi-quantitative picture, verified by simulations, of the various transitions and, in particular, an analytic representation of elastic moduli in the vicinity of the jamming points.
	
The values of $G/B$ and the Poisson ratio $\sigma$ vary continuously with $p_a$ and $p_b$, which can be tuned to reach arbitrarily close to physical limits such as $\sigma=\pm 1$. Our algorithm for reaching these limits is less complicated than ``tuning by pruning" (TbP) \cite{GoodrichNAG2015,HexnerNag2018a} in that it involves only the variation of $p_a$ and $p_b$  rather than the testing of the effects of removing each individual spring in the lattice.  On the other hand, our algorithm only calculates the average effect of dilution.  For a given average coordination number $z$ after dilution, there are certainly specific spring configurations that get closer to physical limits than does the average configuration.  By construction TbP takes the  system as close as possible to a given goal such as the maximum value of $G/B$ or $\sigma$.  This presumably explains why references \cite{GoodrichNAG2015,HexnerNag2018a} access more extreme values of $G/B$ or $\sigma$ for a given $z$ than does our approach.  It would be interesting to investigate in more detail the statistical distributions of $G/B$ and $\sigma$ arising from random dilution, or to apply the TbP to our system.

It would also be interesting to create laboratory versions of our lattices, which can certainly be done using  modern fabrication techniques like $3D$ printing, and to measure their elastic and mechanical properties.  These synthetic lattices will necessarily have bending forces that favor particular angles between bonds and thereby  increase their rigidity relative to that of simple central-force models.  The effect of these bending forces has yet to be studied in detail. Their effect on surfaces states of topological mechanical lattices and on auxetic transitions have been studied in Refs.~\cite{StenullLub2019} and~\cite{RensLer2019}, respectively.

\begin{acknowledgments}	
	We benefited from useful conversations with Andrea Liu, Xiaoming Mao and James Sethna.
	This work was supported in part by NSF MRSEC/DMR-1720530 (TCL and OS), NSF DMR-1719490 (DBL). TCL's work on this research was supported in part by the Isaac  Newton  Institute  for  Mathematical Sciences during the  program  ``Soft Matter Materials - Mathematical Design Innovations”  and by the International Centre for Theoretical Sciences (ICTS) during the program - ''Bangalore School on Statistical Physics - X (Code: ICTS/bssp2019/06)."  
\end{acknowledgments}

\appendix

\section{Lattice structures, dynamical matrices, and dispersion relations}
	\label{sec:Lattice}

In this section we provide additional information relating to the lattice structures and dynamical matrices 
of both the TwK/GK and TwK/H models, as well as dispersion relations for the TwK/GK model.

\subsection{Lattice structures}
\label{subsec:LatticeStructures}

\begin{figure}[!ht]
			\centering
			\begin{minipage}{0.49\linewidth}
				\centering
				(a) \vspace{0.1cm} \\
				\includegraphics[height=\linewidth]{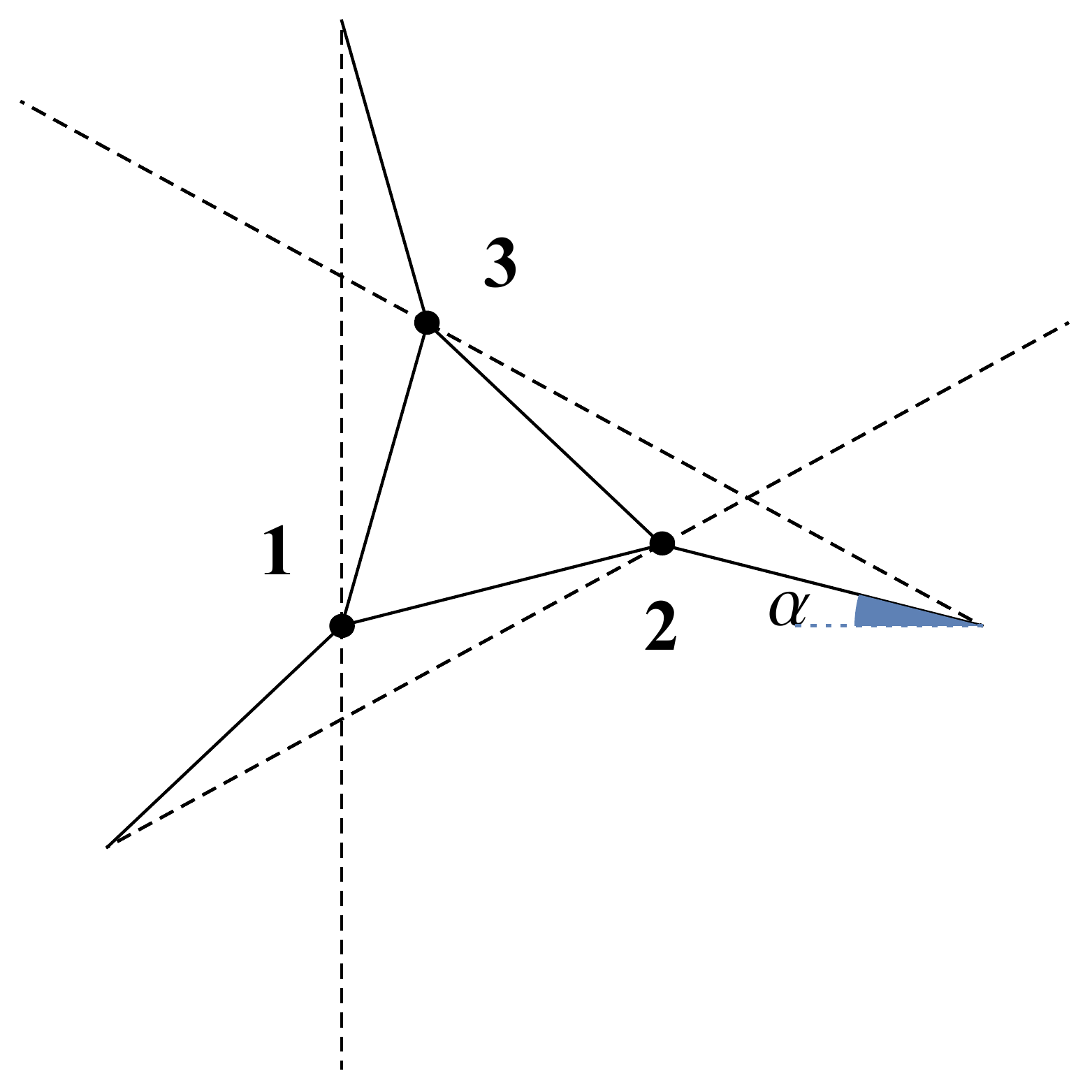}
			\end{minipage}
			\begin{minipage}{0.49\linewidth}
				\centering
				(b) \vspace{0.1cm} \\
				\includegraphics[height=\linewidth]{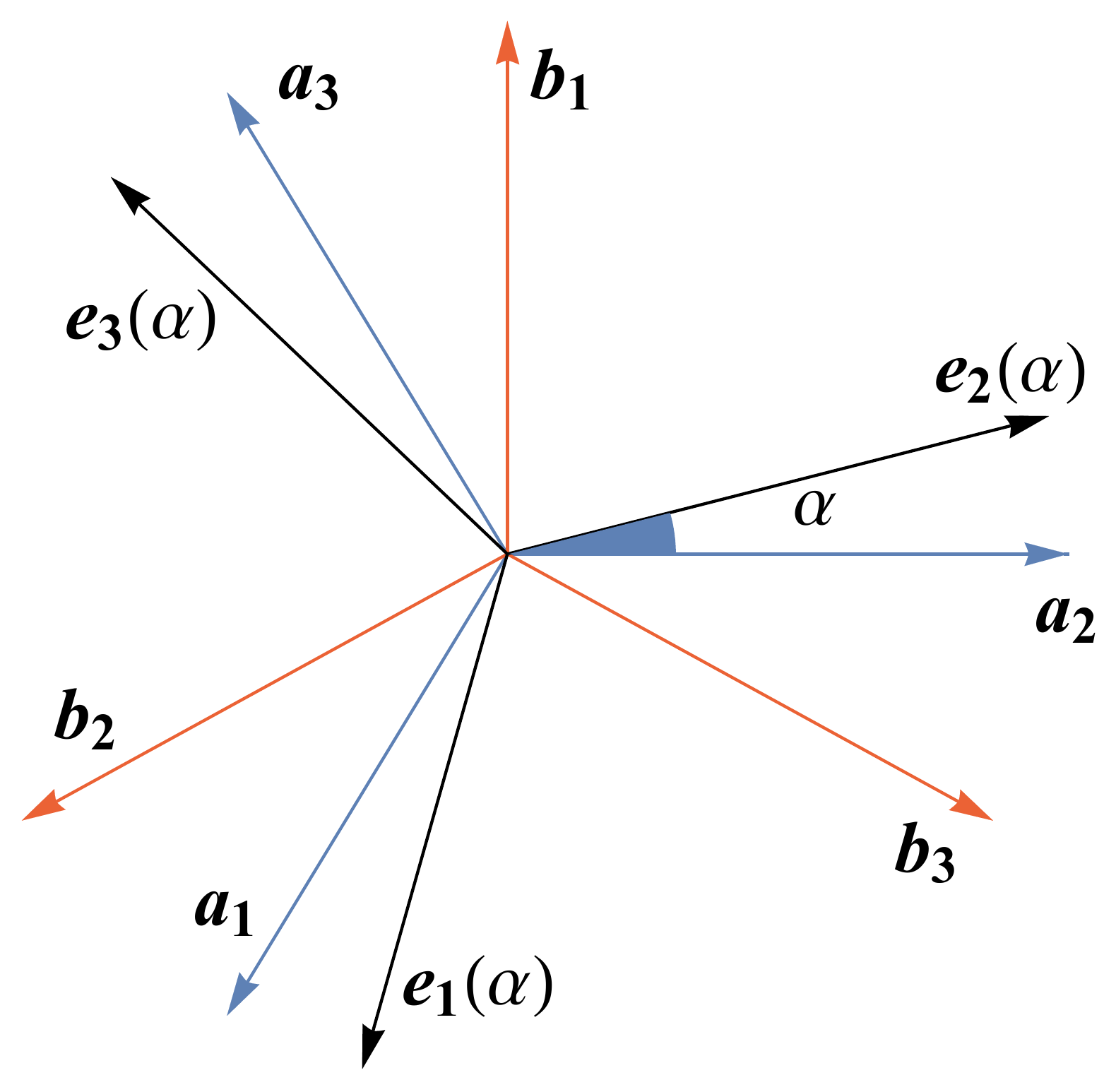}
			\end{minipage} \\
			\centering
			(c) \vspace{0.1cm} \\
			\centering
			\includegraphics[width=0.8\linewidth]{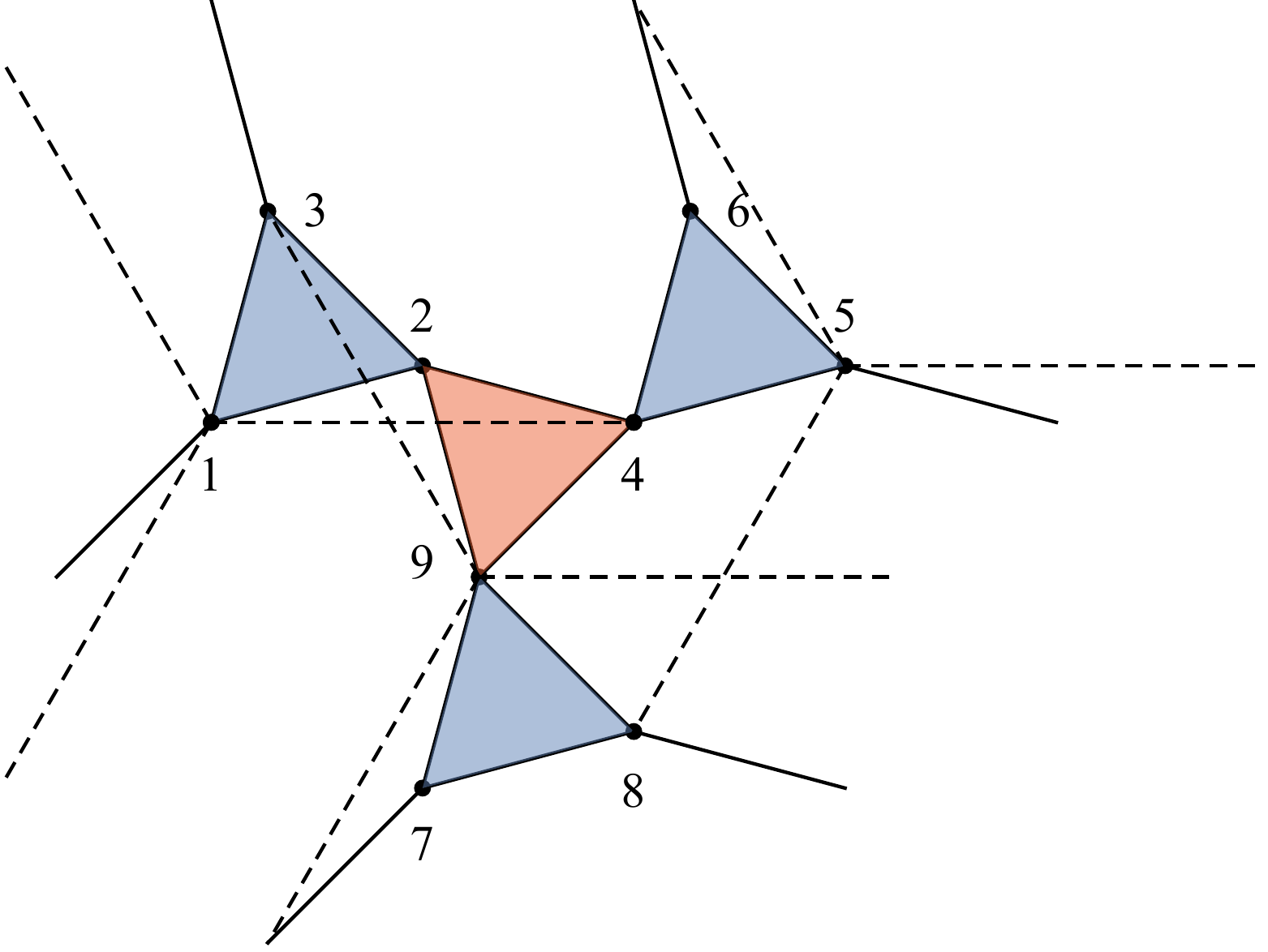}
			\caption{(a) Unit cell of the TwK/GK lattice showing its three-points
			basis (1, 2 and 3), six bonds connecting nearest neighbors (solid lines) and
			six bonds connecting next-nearest neighbors (dashed).
			(b) Sets of unit vectors used in our calculations.
			(c) Unit cell of the TwK/H model showing its nine-point basis, eighteen bonds connecting
			nearest neighbors (solid lines) and nine bonds of the $b$ sub-lattice (dashed lines).
			}
			\label{fig:cellVectors}
		\end{figure}

		Figure~\ref{fig:cellVectors}(a) shows the unit cell of the TwK/GK lattice, its three-point basis
		and a set of unit vectors used in our calculations.
		We set the origin of each cell at the position of the first atom of the unit cell, so that atoms of the
		three-point basis are located at $c_1(\alpha) = (1 / \cos \alpha) \, R(\alpha) \cdot (0,0)$,
		$c_2(\alpha) = (1/\cos \alpha) \, R(\alpha) \cdot (1/2,0)$ and
		$c_3 (\alpha) = (1/\cos \alpha) \, R(\alpha) \cdot (1/4, \sqrt{3}/4)$, where
		\begin{equation}
			R(\alpha)
				= \left(
					\begin{array}{cc}
						\cos \alpha & - \sin \alpha \\
						\sin \alpha & \cos \alpha
					\end{array}
				\right)
		\end{equation}
		is a rigid rotation matrix and the $\cos \alpha$ factor in the denominator ensures that the cell size
		does not change with twist angle.
		The lattice translation vectors are given by $\bm{a}_1 = (-1/2,-\sqrt{3}/2)$, $\bm{a}_2 = (1,0)$ and
		$\bm{a}_3 = (-1/2,\sqrt{3}/2)$.
		The vectors $\bm{e}_i (\alpha) = R(\alpha) \cdot \bm{a}_i$ determine the directions of bonds for
		the $a$-sublattice.
		The vectors $\bm{b}_1 = (0,1)$, $\bm{b}_2 = (-\sqrt{3}/2,-1/2)$ and $\bm{b}_3 = (\sqrt{3}/2,-1/2)$
		are perpendicular to $\bm{a}_2$, $\bm{a}_3$ and  $\bm{a}_1$, respectively, and determine the
		directions of bonds of the $b$-sublattice.
		See Fig.~\ref{fig:cellVectors}(b) for an illustration of the $\bm{a}$, $\bm{b}$ and $\bm{e}$ vectors.
		Figure~\ref{fig:cellVectors}(c) shows the unit cell of the TwK/H lattice with its nine-point basis.
		Note that here $\bm{b}_i$ are lattice translation vectors for the TwK/H model.

\subsection{Dynamical matrices}
\label{subsec:DynamicalMatrices}

		Equation~\eqref{eq:EnergyRSpace} can be written in Fourier space as
		\begin{equation}
			E 
				= \frac{1}{2{N_c}^2}\sum_{\bm{q},\bm{q}^\prime} \bm{u}(\bm{q}) \cdot
					D(-\bm{q}, \bm{q}^\prime) \cdot \bm{u} (\bm{q}^\prime),
		\end{equation}
		where $N_c$ is the number of cells, $\bm{u} (\bm{q})$ is the Fourier transform of
		$\bm{u} (\bm{r}) = (1/N_c) \sum_q \bm{u} (\bm{q}) e^{i \bm{q} \cdot \bm{r}}$,
		and the dynamical matrix is given by,
		\begin{equation}
			D(-\bm{q}, \bm{q}^\prime)
				= N_c \delta_{\bm{q},\bm{q}^\prime} D(\bm{q})
		\end{equation}
		with
		\begin{equation}
			D(\bm{q})
				= \sum_{\alpha \in \{a,b\}} k_\alpha K_\alpha (\bm{q}),
		\end{equation}
		where $K_\alpha$ is the stiffness matrix,
		\begin{equation}
			K_\alpha
			= \sum_{n=1}^{\tilde{z}_\alpha} \bm{B}_n^\alpha (\bm{q}) \otimes \bm{B}_n^\alpha
				(-\bm{q}),
		\end{equation}
		where $\otimes$ denotes an outer product between two vectors, and
		$\tilde{z}_\alpha$ is the number of bonds per unit cell of sub-lattice $\alpha$.
		For the TwK/GK model, the $B$-vectors are given by:
		\begin{align}
			B_1^a (\bm{q})
				& = \left( \bm{e}_1 (\alpha), \bm{0}, -\bm{e}_1 (\alpha) \right), \nonumber \\
			B_2^a (\bm{q})
				& = \left( -\bm{e}_2 (\alpha), \bm{e}_2 (\alpha), \bm{0} \right), \nonumber \\
			B_3^a (\bm{q})
				& = \left( \bm{0}, - \bm{e}_3 (\alpha), \bm{e}_3 (\alpha) \right), \nonumber \\
			B_4^a (\bm{q})
				& = \left( -\bm{e}_1 ( - \alpha), \bm{0}, e^{-i\bm{q} \cdot \bm{a}_1} \, \bm{e}_1 (-\alpha) \right), \nonumber \\
			B_5^a (\bm{q})
				& = \left( e^{-i\bm{q} \cdot \bm{a}_2} \, \bm{e}_2 (-\alpha), -\bm{e}_2 (-\alpha), \bm{0} \right), \nonumber \\
			B_6^a (\bm{q})
				& = \left( \bm{0}, e^{-i\bm{q} \cdot \bm{a}_3} \, \bm{e}_3 (-\alpha), -\bm{e}_3 (-\alpha) \right), \nonumber
		\end{align}
		with $\bm{0}$ denoting a two-dimensional null vector, and
		\begin{align}
			B_1^b (\bm{q})
				& = \left( -\bm{b}_1, e^{-i \bm{q} \cdot \bm{a}_3} \bm{b}_1, \bm{0} \right), \nonumber \\
			B_2^b (\bm{q})
				& = \left( \bm{0}, -\bm{b}_2, e^{-i \bm{q} \cdot \bm{a}_1} \bm{b}_2 \right), \nonumber \\
			B_3^b (\bm{q})
				& = \left( e^{-i \bm{q} \cdot \bm{a}_2} \bm{b}_3, \bm{0}, -\bm{b}_3 \right), \nonumber \\
			B_4^b (\bm{q})
				& = \left( -\bm{b}_1, e^{-i \bm{q} \cdot \bm{a}_3} \bm{b}_1, \bm{0} \right), \nonumber \\
			B_5^b (\bm{q})
				& = \left( \bm{0}, -\bm{b}_2, e^{-i \bm{q} \cdot \bm{a}_1} \bm{b}_2 \right), \nonumber \\
			B_6^b (\bm{q})
				& = \left( e^{-i \bm{q} \cdot \bm{a}_2} \bm{b}_3, \bm{0}, -\bm{b}_3 \right). \nonumber
		\end{align}
		For the TwK/H model, the B-vectors are given by:
		\begin{align}
			B_1^a (\bm{q})
				& = \left( \bm{e}_1 (\alpha), \bm{0}_{1}, -\bm{e}_1 (\alpha) , \bm{0}_{6}\right), \nonumber \\
			B_2^a (\bm{q})
				& = \left( -\bm{e}_2 (\alpha), \bm{e}_2 (\alpha) , \bm{0}_{7}\right), \nonumber \\
			B_3^a (\bm{q})
				& = \left( \bm{0}_{1}, -\bm{e}_3 (\alpha) , \bm{e}_3 (\alpha), \bm{0}_{6}\right), \nonumber \\
			B_4^a (\bm{q})
				& = \left( -\bm{e}_1 (-\alpha), \bm{0}_{4}, e^{-i \bm{q}\cdot \bm{b}_2} \bm{e}_1 (-\alpha) ,
				\bm{0}_{3}\right), \nonumber \\
			B_5^a (\bm{q})
				& = \left( \bm{0}_{1}, -\bm{e}_2 (-\alpha), \bm{0}_{1}, \bm{e}_2 (-\alpha) , \bm{0}_{5}\right), \nonumber \\
			B_6^a (\bm{q})
				& = \left( \bm{0}_{2}, -\bm{e}_3 (-\alpha), \bm{0}_{1}, e^{i \bm{q}\cdot \bm{b}_3} \bm{e}_3 (-\alpha) ,
				\bm{0}_{4}\right), \nonumber \\
			B_7^a (\bm{q})
				& = \left(\bm{0}_{3}, \bm{e}_1 (\alpha), \bm{0}_{1}, -\bm{e}_1 (\alpha) , \bm{0}_{3}\right), \nonumber \\
			B_8^a (\bm{q})
				& = \left(\bm{0}_{3}, -\bm{e}_2 (\alpha), \bm{e}_2 (\alpha) , \bm{0}_{4}\right), \nonumber \\
			B_9^a (\bm{q})
				& = \left( \bm{0}_{4}, -\bm{e}_3 (\alpha) , \bm{e}_3 (\alpha), \bm{0}_{3}\right), \nonumber \\
			B_{10}^a (\bm{q})
				& = \left(\bm{0}_{3}, -\bm{e}_1 (-\alpha), \bm{0}_{4}, \bm{e}_1 (-\alpha) \right), \nonumber \\
			B_{11}^a (\bm{q})
				& = \left( \bm{0}_{4}, -\bm{e}_2 (-\alpha), \bm{0}_{1},  e^{i \bm{q}\cdot \bm{b}_2} \bm{e}_2 (-\alpha),
				\bm{0}_{2}\right), \nonumber \\
			B_{12}^a (\bm{q})
				& = \left( \bm{0}_{5}, -\bm{e}_3 (-\alpha), \bm{0}_{1}, e^{-i \bm{q}\cdot \bm{b}_1} \bm{e}_3 (-\alpha) ,
				\bm{0}_{1}\right), \nonumber \\
			B_{13}^a (\bm{q})
				& = \left(\bm{0}_{6}, \bm{e}_1 (\alpha), \bm{0}_{1}, -\bm{e}_1 (\alpha)\right), \nonumber \\
			B_{14}^a (\bm{q})
				& = \left(\bm{0}_{6}, -\bm{e}_2 (\alpha), \bm{e}_2 (\alpha) , \bm{0}_{1}\right), \nonumber \\
			B_{15}^a (\bm{q})
				& = \left( \bm{0}_{7}, -\bm{e}_3 (\alpha) , \bm{e}_3 (\alpha)\right), \nonumber \\
			B_{16}^a (\bm{q})
				& = \left(\bm{0}_{2}, e^{i \bm{q}\cdot \bm{b}_1} \bm{e}_1 (-\alpha), \bm{0}_{3}, -\bm{e}_1 (-\alpha),
				\bm{0}_2 \right), \nonumber \\
			B_{17}^a (\bm{q})
				& = \left( e^{-i \bm{q}\cdot \bm{b}_3} \bm{e}_2 (-\alpha), \bm{0}_{6},  -\bm{e}_2 (-\alpha),
				\bm{0}_{1}\right), \nonumber \\
			B_{18}^a (\bm{q})
				& = \left( \bm{0}_{1}, \bm{e}_3 (-\alpha), \bm{0}_{6}, -\bm{e}_3 (-\alpha) \right),
		\end{align}
		with $\bm{0}_n$ denoting a ($2n$)-dimensional null vector, and
		\begin{align}
			B_1^b (\bm{q})
				& = \left( -\bm{a}_1,  \bm{0}_2, e^{-i \bm{q} \cdot \bm{b}_2} \bm{a}_1, \bm{0}_5 \right), \nonumber \\
			B_2^b (\bm{q})
				& = \left( -\bm{a}_2,  \bm{0}_2, \bm{a}_2, \bm{0}_5 \right), \nonumber \\
			B_3^b (\bm{q})
				& = \left( -\bm{a}_3,  \bm{0}_2, e^{-i \bm{q} \cdot \bm{b}_3} \bm{a}_3, \bm{0}_5 \right), \nonumber \\
			B_4^b (\bm{q})
				& = \left(\bm{0}_4, -\bm{a}_1, \bm{0}_2, \bm{a}_1, \bm{0}_1 \right), \nonumber \\
			B_5^b (\bm{q})
				& = \left(\bm{0}_4, -\bm{a}_2, \bm{0}_2, e^{i \bm{q} \cdot \bm{b}_2} \bm{a}_2, \bm{0}_1 \right), \nonumber \\
			B_6^b (\bm{q})
				& = \left(\bm{0}_4, -\bm{a}_3, \bm{0}_2, e^{-i \bm{q} \cdot \bm{b}_1} \bm{a}_3, \bm{0}_1 \right), \nonumber \\
			B_7^b (\bm{q})
				& = \left(\bm{0}_2, e^{i \bm{q} \cdot \bm{b}_1} \bm{a}_1, \bm{0}_5, -\bm{a}_1 \right), \nonumber \\
			B_8^b (\bm{q})
				& = \left(\bm{0}_2, e^{-i \bm{q} \cdot \bm{b}_3} \bm{a}_2, \bm{0}_5, -\bm{a}_2 \right), \nonumber \\
			B_9^b (\bm{q})
				& = \left(\bm{0}_2, \bm{a}_3, \bm{0}_5, -\bm{a}_3 \right).
		\end{align}
		All vectors ($\bm{b}_i$, $\bm{e}_i$ and $\bm{a}_i$) are defined in Section~\ref{subsec:LatticeStructures}.

\subsection{Dispersion curves}
\label{subsec:DispersionCurves}

\begin{figure}[!ht]
	\centering
	\begin{minipage}{0.7\linewidth}
		\centering
		(a) \vspace{0.1cm} \\
		\includegraphics[height=4cm]{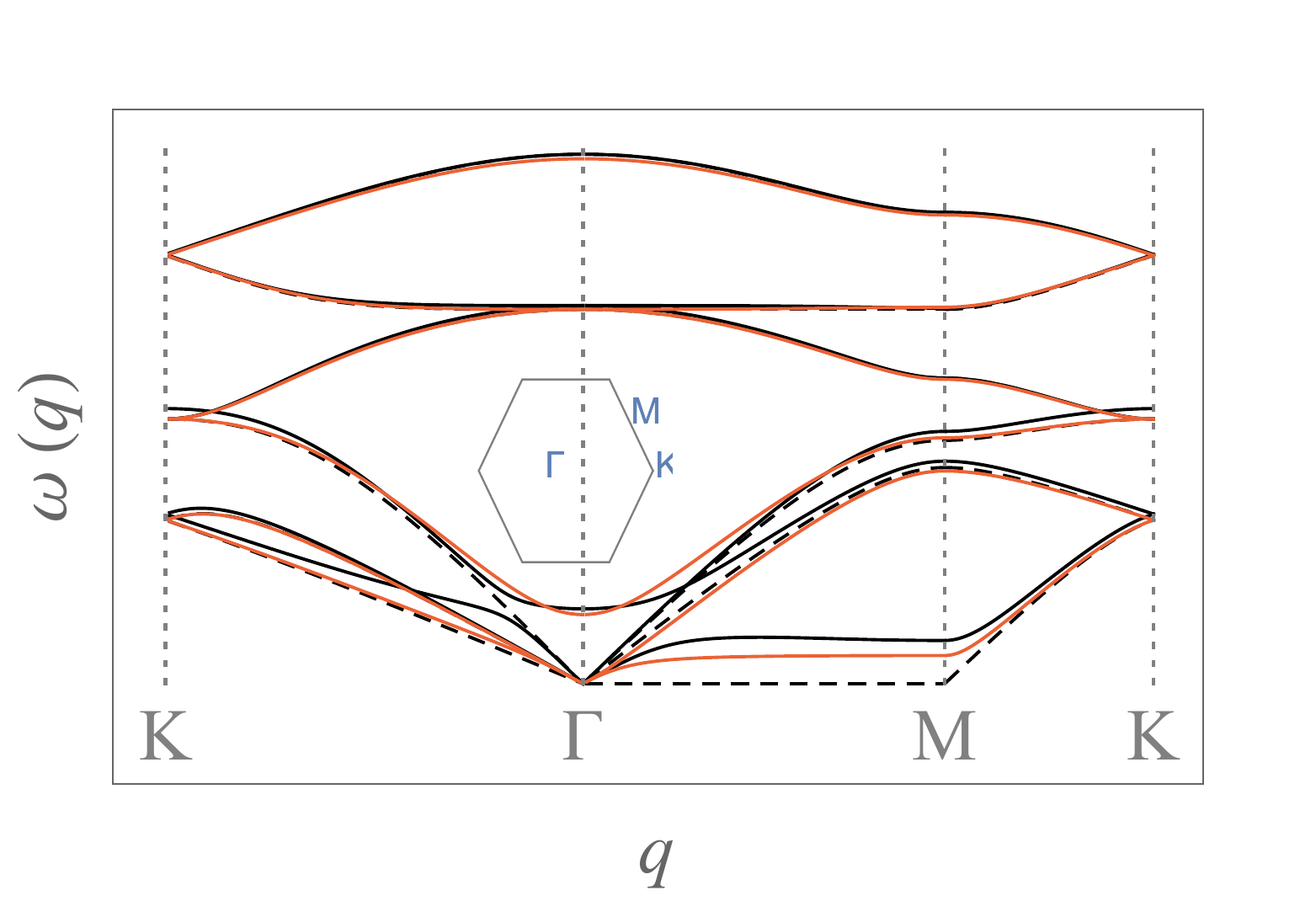}
	\end{minipage}
	\begin{minipage}{0.28\linewidth}
		\centering
		(b) \vspace{0.1cm} \\
		\includegraphics[height=4cm]{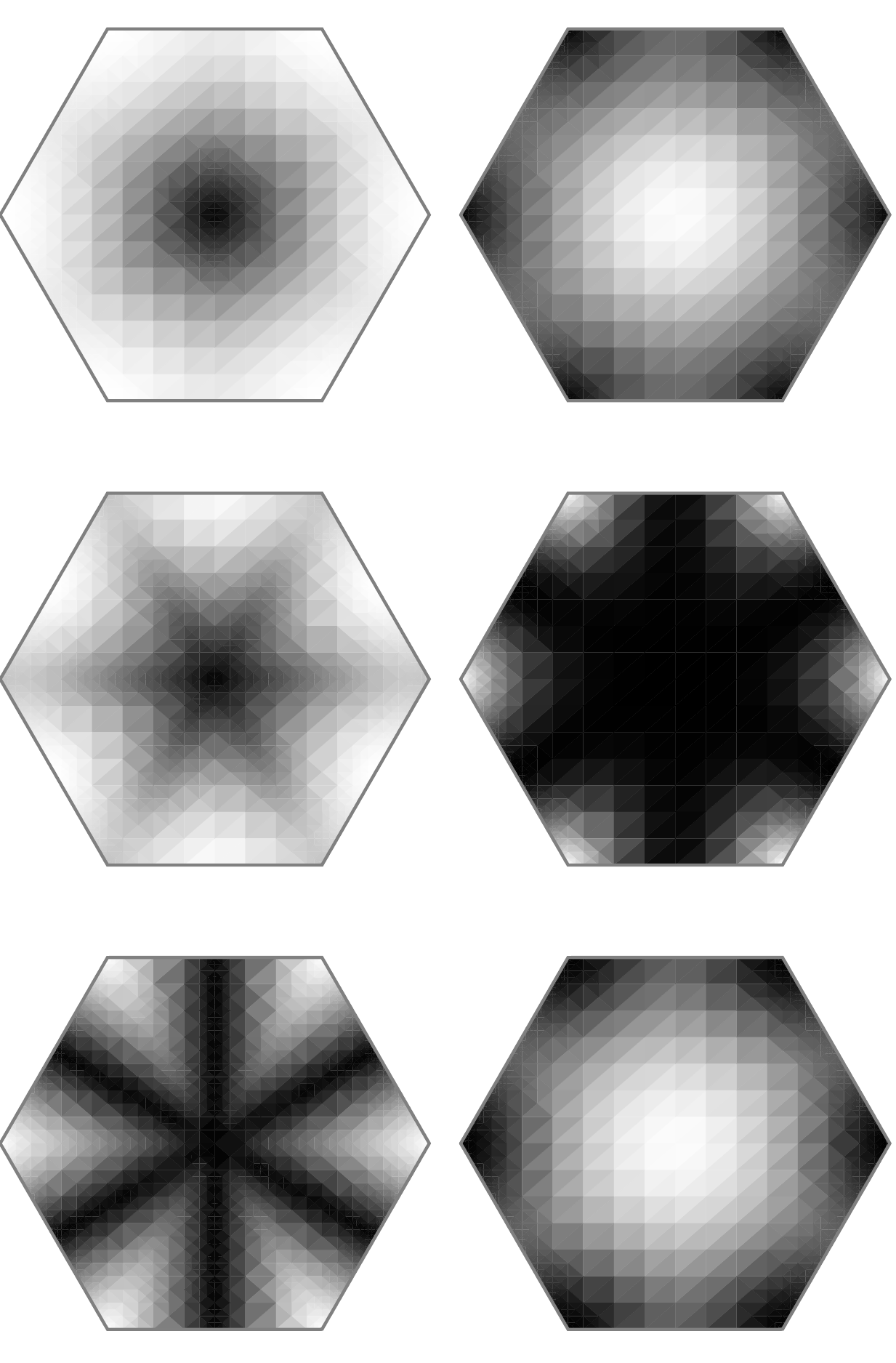}
	\end{minipage} \vspace{0.2cm} \\
	\begin{minipage}{0.7\linewidth}
		\centering
		(c) \vspace{0.1cm} \\
		\includegraphics[height=4cm]{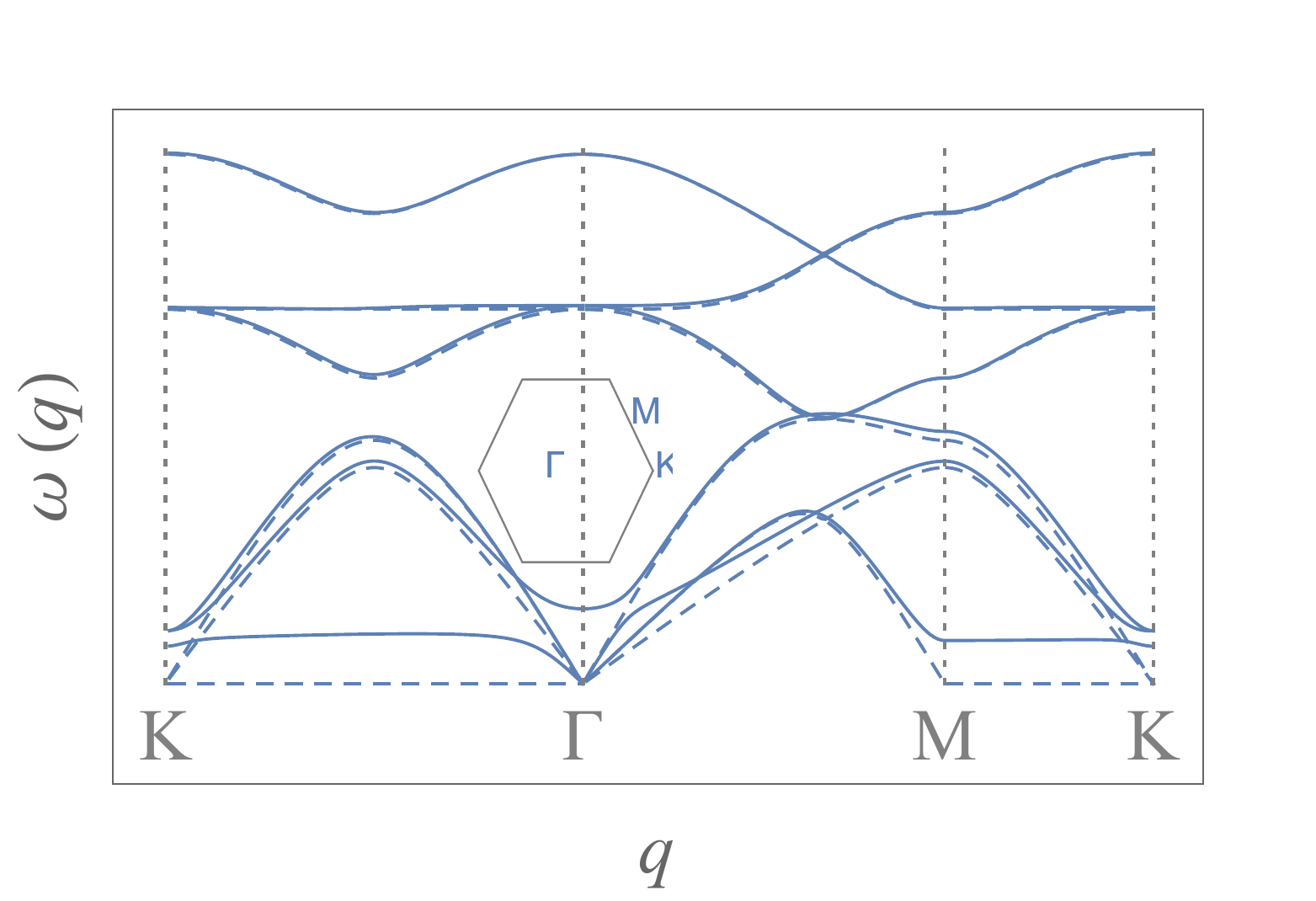}
	\end{minipage}
	\begin{minipage}{0.28\linewidth}
		\centering
		(d) \vspace{0.1cm} \\
		\includegraphics[height=4cm]{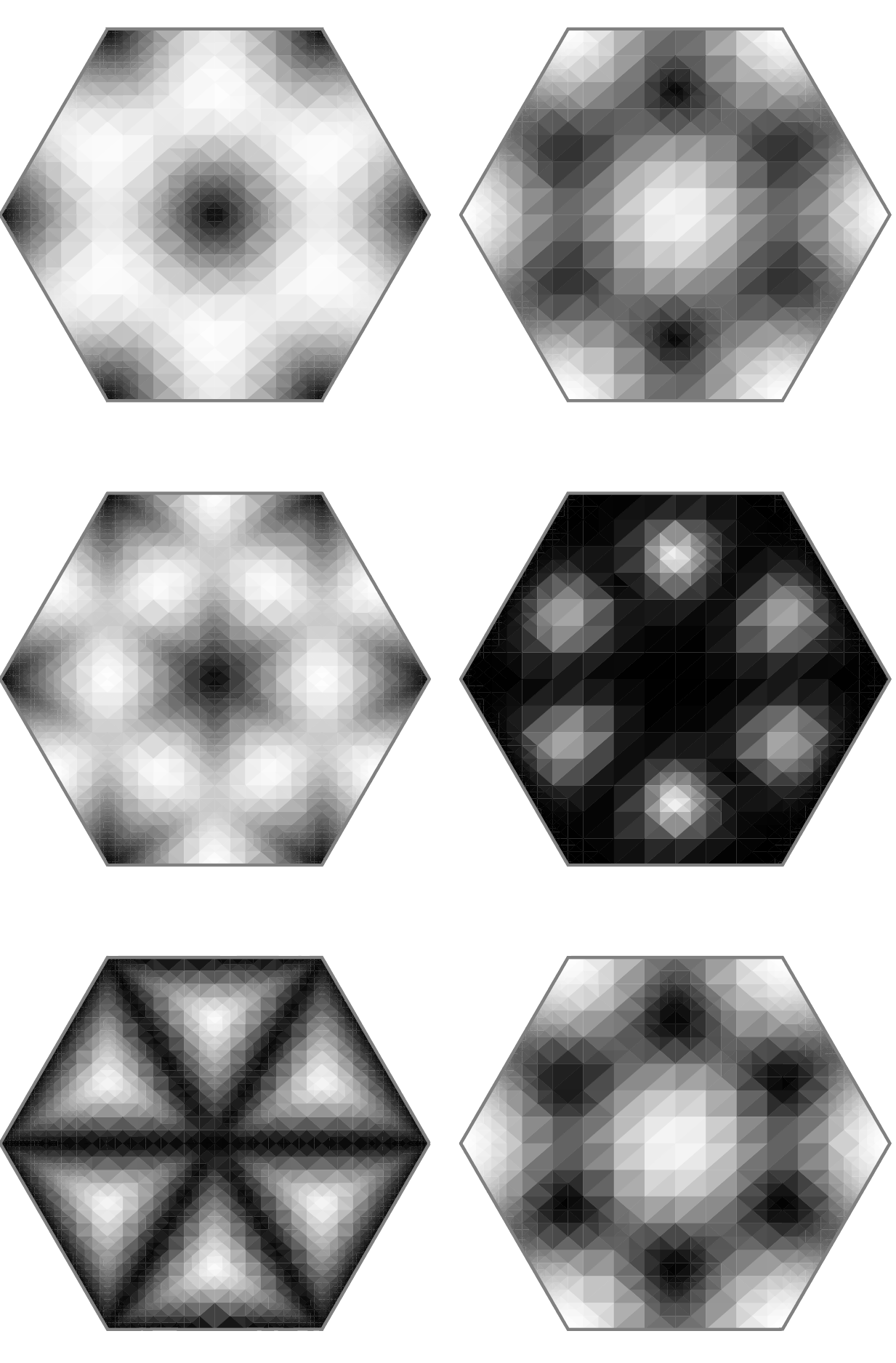}
	\end{minipage}
	\caption{(a) Dispersion curves of the KL, $k_a=1$, $\alpha = 0$ and $k_b = 0$ (black dashed), the K/GK lattice, $k_a = 1$, $\alpha = 0$, and $k_b = 0.02$ (black solid), and the TwKL, $k_a = 1$, $\alpha = \pi /12$, and $k_b = 0$ (red). (b) Density plot of the six eigenmodes for the KL. (c) Dispersion curves for the GKL, $k_a = 0$, $\alpha = 0$, and $k_b = 1$ (solid blue), and K/GK lattice, $k_a = 0.02$, $\alpha = 0$, and $k_b = 1$ (solid blue). (d) Density plot of the six eigenmodes for the GKL. In all cases, the addition of NNN bonds to the K lattices raises all zero-frequency eigenmodes to the NN lattices to nonzero frequency.}
%
	\label{fig:dispersion}
\end{figure}

Figure~\ref{fig:dispersion} shows dispersion curves ($\omega_i (\bm{q})$ is the square root of the $i$-th Eigenvalue of $D(\bm{q})$) of the TwK/GK and K/GK ($\alpha=0$) lattices along symmetry lines [(a) and (c)] and dispersion densities over the first Brillouin zone [(b) and (d)] for $k_a$, $k_b$ and $\alpha$ corresponding to regions in the phase diagram near $J_G$ [(a) and (b)] and $J_{BG}$ [(c) and (d)].
Notice that the Kagome lattices have modes that vanish along lines in the Brillouin zone
($\Gamma M$ for the TwKL; $\Gamma K$ and $KM$ for the GKL; recall that
the GKL has an orientation that is rotated  by $\pi/6$ with respect to the untwisted KL).
As discussed in Refs.~\cite{MaoLub2018,SunLub2012}, the untwisted K and GK lattices have straight lines of bonds, whose number scales as the lattice perimeter ($\propto N_{\text{cell}}^{1/2}$), which support states of self stress that by the Calladine Index Theorem \cite{Calladine1978,PellegrinoCal1986} then require an equal number of zero modes, one for each wavevector on the lines $\Gamma M$ in the Brillouin Zone.
Twisting these lattices eliminates the straight lines of bonds, their states of self stress, and associated zero-frequency modes.
Adding NNN bonds also eliminates zero modes.
On the other hand, we cannot separate the effects of twist angle and TwKL bonds on the dispersion curves of the TwK/GK model near $J_{BG}$ [(c) and (d) plots]. If we set $k_a = 0$, then $\alpha > 0$ has no effect on the dispersion curves. For the purposes of this paper, increasing $\alpha$ from zero when $k_a > 0$ and $k_b \approx 1$ does not lead to qualitative changes in the dispersion curves.
The phonon dispersion curves of the TwK/H model is more complicated (with 18 modes per
wavevector $\bm{q}$.
They do not offer additional insight, and we do not show them here.

\section{Asymptotic limit of the EMT Integrals and global behavior of the elastic moduli}
	\label{sec:ExtraPlots}

Here we discuss plots showing the asymptotic behavior of the EMT integrals $h_\alpha$ near the jamming
points, for both the TwK/GK and the TwK/H models.
We also show 3D plots of the moduli as a function of $p_a$ and $p_b$ for both models.

	\begin{figure}[!ht]
		\centering
		\begin{minipage}{0.49\linewidth}
			\centering
			(a) \vspace{0.1cm} \\
			\includegraphics[height=\linewidth]{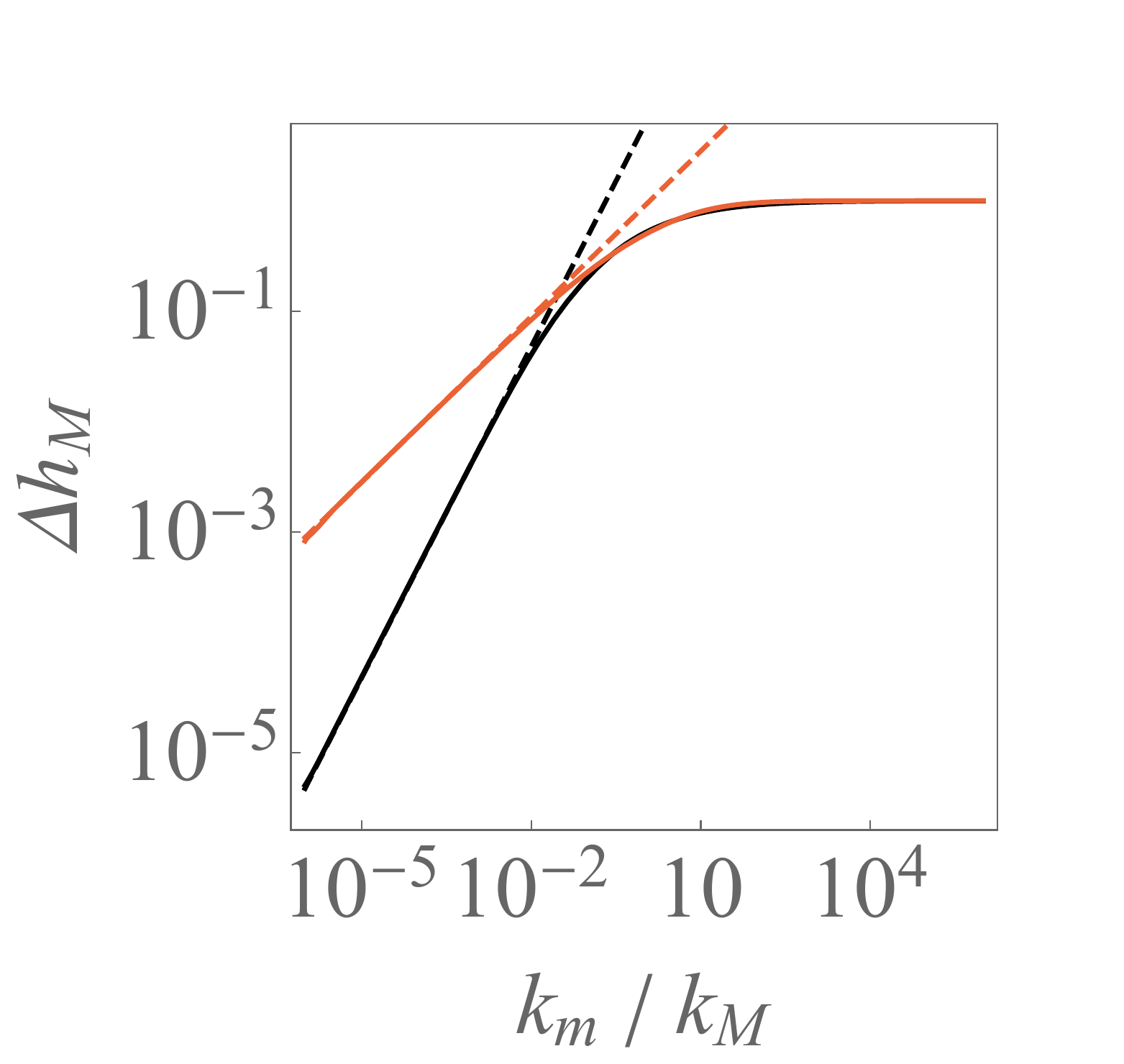}
		\end{minipage}
		\begin{minipage}{0.49\linewidth}
			\centering
			(b) \vspace{0.1cm} \\
			\includegraphics[height=\linewidth]{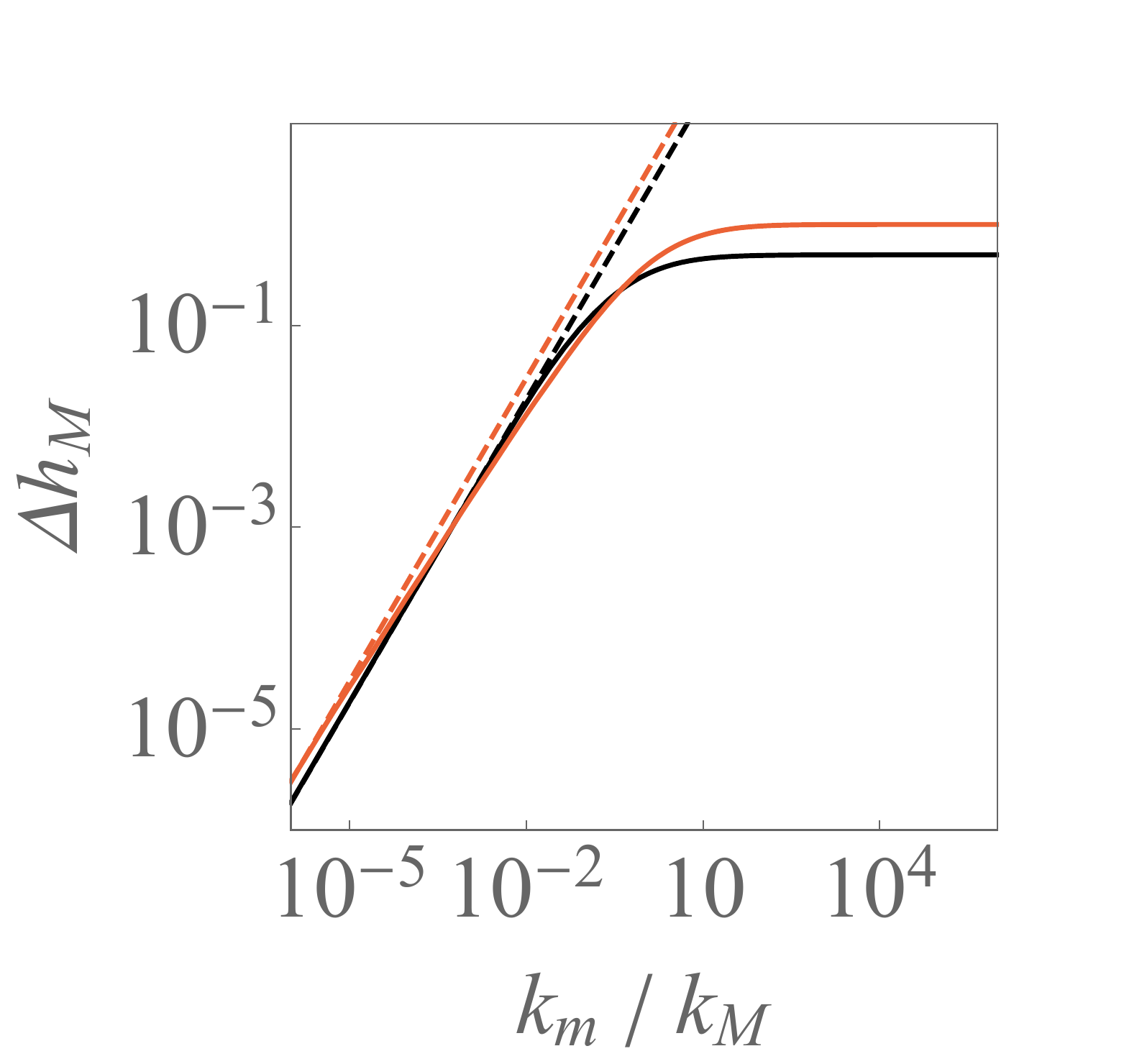}
		\end{minipage}
		\caption{Showing the asymptotic behavior of the EMT integrals $\Delta h_M$ as a function of
		$k_{m} / k_M$ for the TwK/GK (a) and TwK/H (b) models.
		In (a), the black and red curves emphasize the asymptotic behavior near the $J_G$ (with $M$ and
		$m$ representing the TwKL and GKL, respectively) and $J_{BG}$ (with $M$ and $m$ representing
		the GKL and TwKL, respectively).
		In (b), the black and red curves emphasize the asymptotic behavior near the $J_G$ (with $M$ and
		$m$ representing the TwKL and HL, respectively) and $J_{B}$ (with $M$ and $m$ representing
		the HL and TwKL, respectively).
		The dashed lines correspond to our asymptotic analytic predictions.
		}
		\label{fig:Integrals}
	\end{figure}

In Section~\ref{sec:Scaling} we have shown that $\Delta h_M \equiv 1-h_M \propto k_m / k_M$ near the $J_{G1}$,
$J_{G2}$ and $J_{B}$ points, and that $\Delta h_M \propto \sqrt{k_m / k_M}$ near the $J_{BG}$ point.
Figure~\ref{fig:Integrals} shows full numerical calculations of $\Delta h_M$ near the four jamming points and
confirms our analytical predictions.
In (a) we show $\Delta h_M$ as a function of $k_m / k_M$ for the TwK/GK model near $J_G$ (black, with $M$ 
and $m$ representing the twisted Kagome and generalized Kagome lattice, respectively) and near $J_{BG}$
(red, with $M$ and $m$ representing the generalized Kagome and twisted Kagome lattice, respectively).
Note that $\Delta h_M \propto \sqrt{k_m/k_M}$ near $J_{BG}$.
In (b) we show $\Delta h_M$ as a function of $k_m / k_M$ for the TwK/H model near $J_G$ (black, with $M$
and $m$ representing the twisted Kagome and honeycomb lattices, respectively) and near $J_{B}$ (red, with
$M$ and $m$ representing the honeycomb lattices and twisted Kagome lattice, respectively).
We have used $\alpha = \pi/12$ in both plots, and the dashed lines correspond to our asymptotic analytic
predictions.
We have used Eq.~\eqref{eq:cMdef} to calculate $c_M$ near the $J_{G1}$ and $J_{G2}$ points, and a numerical
fit to calculate $c_M$ near the $J_{BG}$ and $J_{B}$ points (see Table~\ref{tab:JammingPoints}).

\begin{figure}[!ht]
	\centering
	\begin{minipage}{0.48\linewidth}
		\centering
		(a) \vspace{0.1cm} \\
		\includegraphics[height=0.75\linewidth]{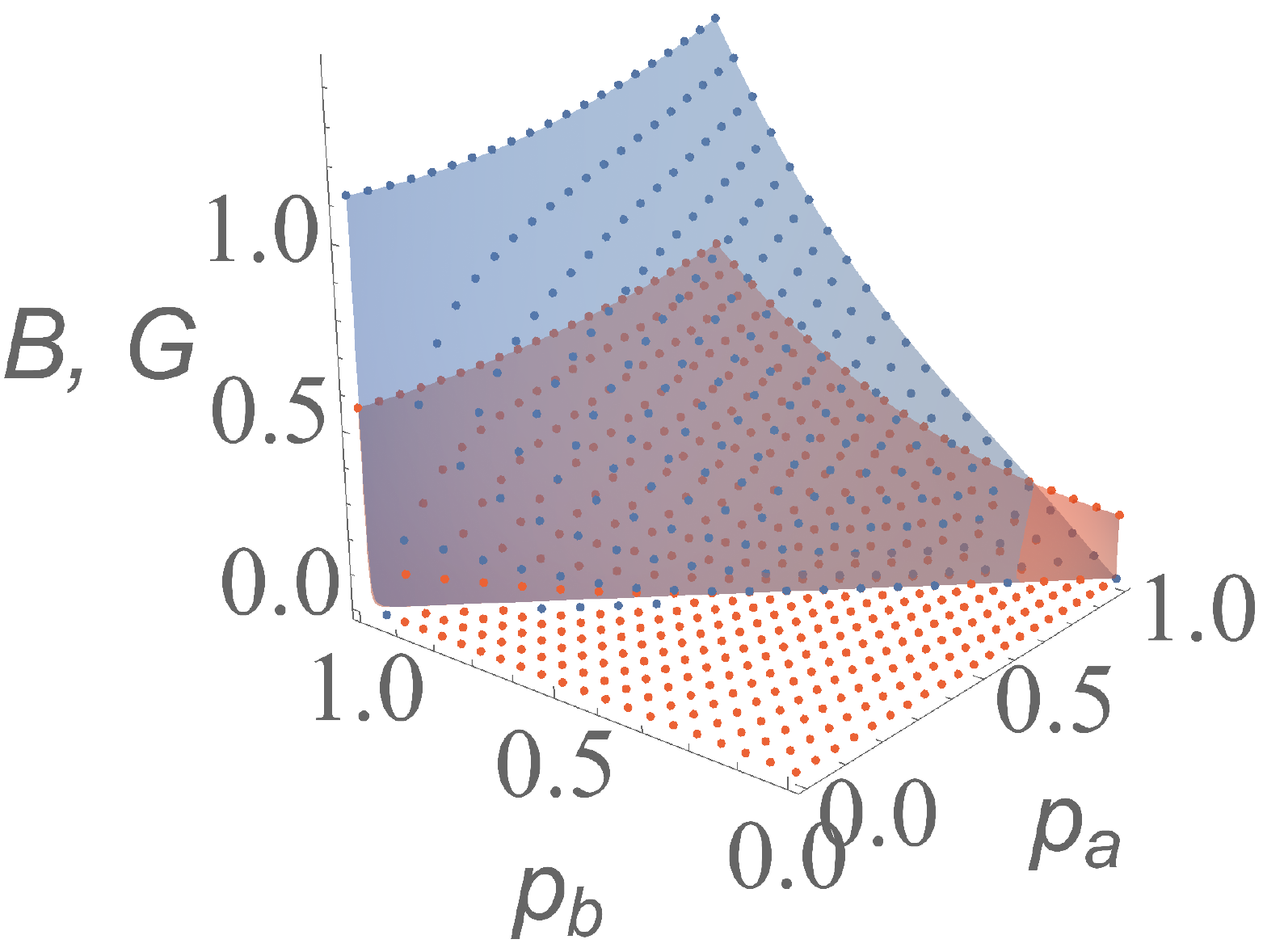}
	\end{minipage}
	\begin{minipage}{0.48\linewidth}
		\centering
		(b) \vspace{0.1cm} \\
		\includegraphics[height=0.75\linewidth]{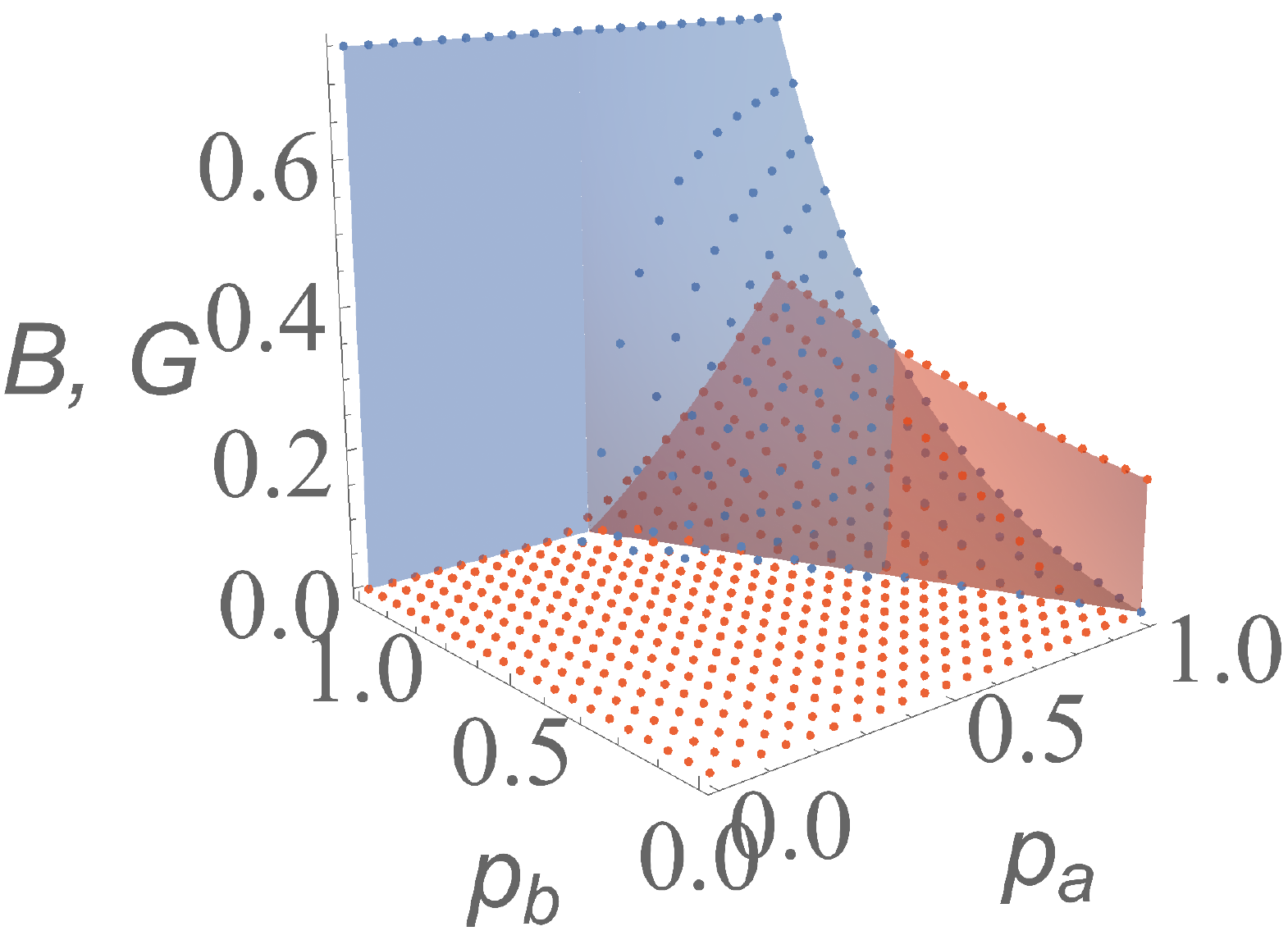}
	\end{minipage}
	\caption{Bulk (blue, upper surface on left side of each plot) and shear (red) moduli as a function of $p_a$ and $p_b$ for the TwK/GK (a) and TwK/H (b) models. The dots and surfaces correspond to numerical simulations and full solutions of the EMT equations, respectively.
		\label{Fig:3DPlots}}
\end{figure}

Finally, Figure~\ref{Fig:3DPlots} shows three-dimensional plots of $B$ (blue) and $G$ (red) as a function of
$p_a$ and $p_b$ for the TwK/GK (a) and TwK/H (b) models.
The dots and surfaces represent results from simulations and EMT, respectively.
As it should be anticipated (see Figure~\ref{Fig:PhaseDiagrams}), the agreement between EMT and simulations
is best near the shear-jamming points $J_G$.


\begin{thebibliography}{39}%
\makeatletter
\providecommand \@ifxundefined [1]{%
 \@ifx{#1\undefined}
}%
\providecommand \@ifnum [1]{%
 \ifnum #1\expandafter \@firstoftwo
 \else \expandafter \@secondoftwo
 \fi
}%
\providecommand \@ifx [1]{%
 \ifx #1\expandafter \@firstoftwo
 \else \expandafter \@secondoftwo
 \fi
}%
\providecommand \natexlab [1]{#1}%
\providecommand \enquote  [1]{``#1''}%
\providecommand \bibnamefont  [1]{#1}%
\providecommand \bibfnamefont [1]{#1}%
\providecommand \citenamefont [1]{#1}%
\providecommand \href@noop [0]{\@secondoftwo}%
\providecommand \href [0]{\begingroup \@sanitize@url \@href}%
\providecommand \@href[1]{\@@startlink{#1}\@@href}%
\providecommand \@@href[1]{\endgroup#1\@@endlink}%
\providecommand \@sanitize@url [0]{\catcode `\\12\catcode `\$12\catcode
  `\&12\catcode `\#12\catcode `\^12\catcode `\_12\catcode `\%12\relax}%
\providecommand \@@startlink[1]{}%
\providecommand \@@endlink[0]{}%
\providecommand \url  [0]{\begingroup\@sanitize@url \@url }%
\providecommand \@url [1]{\endgroup\@href {#1}{\urlprefix }}%
\providecommand \urlprefix  [0]{URL }%
\providecommand \Eprint [0]{\href }%
\providecommand \doibase [0]{http://dx.doi.org/}%
\providecommand \selectlanguage [0]{\@gobble}%
\providecommand \bibinfo  [0]{\@secondoftwo}%
\providecommand \bibfield  [0]{\@secondoftwo}%
\providecommand \translation [1]{[#1]}%
\providecommand \BibitemOpen [0]{}%
\providecommand \bibitemStop [0]{}%
\providecommand \bibitemNoStop [0]{.\EOS\space}%
\providecommand \EOS [0]{\spacefactor3000\relax}%
\providecommand \BibitemShut  [1]{\csname bibitem#1\endcsname}%
\let\auto@bib@innerbib\@empty
\bibitem [{\citenamefont {Ashcroft}\ and\ \citenamefont
  {Mermin}(1976)}]{AshcroftMer1976}%
  \BibitemOpen
  \bibfield  {author} {\bibinfo {author} {\bibfnamefont {N.~W.}\ \bibnamefont
  {Ashcroft}}\ and\ \bibinfo {author} {\bibfnamefont {N.~D.}\ \bibnamefont
  {Mermin}},\ }\href@noop {} {\emph {\bibinfo {title} {Solid state physics}}}\
  (\bibinfo  {publisher} {Hold, Rinehart, and Winston, New York},\ \bibinfo
  {year} {1976})\BibitemShut {NoStop}%
\bibitem [{\citenamefont {Thorpe}(1983)}]{Thorpe1983}%
  \BibitemOpen
  \bibfield  {author} {\bibinfo {author} {\bibfnamefont {M.}~\bibnamefont
  {Thorpe}},\ }\href {\doibase https://doi.org/10.1016/0022-3093(83)90424-6}
  {\bibfield  {journal} {\bibinfo  {journal} {Journal of Non-Crystalline
  Solids}\ }\textbf {\bibinfo {volume} {57}},\ \bibinfo {pages} {355 }
  (\bibinfo {year} {1983})}\BibitemShut {NoStop}%
\bibitem [{\citenamefont {Feng}\ \emph {et~al.}(1985)\citenamefont {Feng},
  \citenamefont {Thorpe},\ and\ \citenamefont {Garboczi}}]{FengGar1985}%
  \BibitemOpen
  \bibfield  {author} {\bibinfo {author} {\bibfnamefont {S.}~\bibnamefont
  {Feng}}, \bibinfo {author} {\bibfnamefont {M.~F.}\ \bibnamefont {Thorpe}}, \
  and\ \bibinfo {author} {\bibfnamefont {E.}~\bibnamefont {Garboczi}},\ }\href
  {<Go to ISI>://A1985TZ38500031} {\bibfield  {journal} {\bibinfo  {journal}
  {Physical Review B}\ }\textbf {\bibinfo {volume} {31}},\ \bibinfo {pages}
  {276} (\bibinfo {year} {1985})}\BibitemShut {NoStop}%
\bibitem [{\citenamefont {Souslov}\ \emph {et~al.}(2009)\citenamefont
  {Souslov}, \citenamefont {Liu},\ and\ \citenamefont
  {Lubensky}}]{SouslovLub2009}%
  \BibitemOpen
  \bibfield  {author} {\bibinfo {author} {\bibfnamefont {A.}~\bibnamefont
  {Souslov}}, \bibinfo {author} {\bibfnamefont {A.~J.}\ \bibnamefont {Liu}}, \
  and\ \bibinfo {author} {\bibfnamefont {T.~C.}\ \bibnamefont {Lubensky}},\
  }\href {\doibase 10.1103/PhysRevLett.103.205503} {\bibfield  {journal}
  {\bibinfo  {journal} {Phys. Rev. Lett.}\ }\textbf {\bibinfo {volume} {103}},\
  \bibinfo {pages} {205503} (\bibinfo {year} {2009})}\BibitemShut {NoStop}%
\bibitem [{\citenamefont {Binder}\ and\ \citenamefont
  {Kob}(2011)}]{Binder2011}%
  \BibitemOpen
  \bibfield  {author} {\bibinfo {author} {\bibfnamefont {K.}~\bibnamefont
  {Binder}}\ and\ \bibinfo {author} {\bibfnamefont {W.}~\bibnamefont {Kob}},\
  }\href@noop {} {\emph {\bibinfo {title} {Glassy materials and disordered
  solids: An introduction to their statistical mechanics}}}\ (\bibinfo
  {publisher} {World scientific},\ \bibinfo {year} {2011})\BibitemShut
  {NoStop}%
\bibitem [{\citenamefont {Lubensky}\ \emph {et~al.}(2015)\citenamefont
  {Lubensky}, \citenamefont {Kane}, \citenamefont {Mao}, \citenamefont
  {Souslov},\ and\ \citenamefont {Sun}}]{LubenskyKai2015}%
  \BibitemOpen
  \bibfield  {author} {\bibinfo {author} {\bibfnamefont {T.~C.}\ \bibnamefont
  {Lubensky}}, \bibinfo {author} {\bibfnamefont {C.~L.}\ \bibnamefont {Kane}},
  \bibinfo {author} {\bibfnamefont {X.}~\bibnamefont {Mao}}, \bibinfo {author}
  {\bibfnamefont {A.}~\bibnamefont {Souslov}}, \ and\ \bibinfo {author}
  {\bibfnamefont {K.}~\bibnamefont {Sun}},\ }\href {\doibase
  10.1088/0034-4885/78/7/073901} {\bibfield  {journal} {\bibinfo  {journal}
  {Reports on progress in physics}\ }\textbf {\bibinfo {volume} {78}},\
  \bibinfo {pages} {073901} (\bibinfo {year} {2015})}\BibitemShut {NoStop}%
\bibitem [{\citenamefont {Mao}\ and\ \citenamefont
  {Lubensky}(2018)}]{MaoLub2018}%
  \BibitemOpen
  \bibfield  {author} {\bibinfo {author} {\bibfnamefont {X.}~\bibnamefont
  {Mao}}\ and\ \bibinfo {author} {\bibfnamefont {T.~C.}\ \bibnamefont
  {Lubensky}},\ }\href {\doibase 10.1146/annurev-conmatphys-033117-054235}
  {\bibfield  {journal} {\bibinfo  {journal} {Annual Review of Condensed Matter
  Physics}\ }\textbf {\bibinfo {volume} {9}},\ \bibinfo {pages} {413} (\bibinfo
  {year} {2018})},\ \Eprint
  {http://arxiv.org/abs/https://doi.org/10.1146/annurev-conmatphys-033117-054235}
  {https://doi.org/10.1146/annurev-conmatphys-033117-054235} \BibitemShut
  {NoStop}%
\bibitem [{\citenamefont {Maxwell}(1864)}]{Maxwell1864}%
  \BibitemOpen
  \bibfield  {author} {\bibinfo {author} {\bibfnamefont {J.~C.}\ \bibnamefont
  {Maxwell}},\ }\href@noop {} {\bibfield  {journal} {\bibinfo  {journal} {Phil.
  Mag.}\ }\textbf {\bibinfo {volume} {27}},\ \bibinfo {pages} {598} (\bibinfo
  {year} {1864})}\BibitemShut {NoStop}%
\bibitem [{\citenamefont {Schwartz}\ \emph {et~al.}(1985)\citenamefont
  {Schwartz}, \citenamefont {Feng}, \citenamefont {Thorpe},\ and\ \citenamefont
  {Sen}}]{SchwartzSen1985}%
  \BibitemOpen
  \bibfield  {author} {\bibinfo {author} {\bibfnamefont {L.~M.}\ \bibnamefont
  {Schwartz}}, \bibinfo {author} {\bibfnamefont {S.}~\bibnamefont {Feng}},
  \bibinfo {author} {\bibfnamefont {M.~F.}\ \bibnamefont {Thorpe}}, \ and\
  \bibinfo {author} {\bibfnamefont {P.~N.}\ \bibnamefont {Sen}},\ }\href {<Go
  to ISI>://A1985ARX0400033} {\bibfield  {journal} {\bibinfo  {journal}
  {Physical Review B}\ }\textbf {\bibinfo {volume} {32}},\ \bibinfo {pages}
  {4607} (\bibinfo {year} {1985})}\BibitemShut {NoStop}%
\bibitem [{\citenamefont {Guyon}\ \emph {et~al.}(1990)\citenamefont {Guyon},
  \citenamefont {Roux}, \citenamefont {Hansen}, \citenamefont {Bideau},
  \citenamefont {Troadec},\ and\ \citenamefont {Crapo}}]{GuyonCr1990}%
  \BibitemOpen
  \bibfield  {author} {\bibinfo {author} {\bibfnamefont {E.}~\bibnamefont
  {Guyon}}, \bibinfo {author} {\bibfnamefont {S.}~\bibnamefont {Roux}},
  \bibinfo {author} {\bibfnamefont {A.}~\bibnamefont {Hansen}}, \bibinfo
  {author} {\bibfnamefont {D.}~\bibnamefont {Bideau}}, \bibinfo {author}
  {\bibfnamefont {J.~P.}\ \bibnamefont {Troadec}}, \ and\ \bibinfo {author}
  {\bibfnamefont {H.}~\bibnamefont {Crapo}},\ }\href {\doibase
  10.1088/0034-4885/53/4/001} {\bibfield  {journal} {\bibinfo  {journal}
  {Reports on Progress in Physics}\ }\textbf {\bibinfo {volume} {53}},\
  \bibinfo {pages} {373} (\bibinfo {year} {1990})}\BibitemShut {NoStop}%
\bibitem [{\citenamefont {Liu}\ and\ \citenamefont {Nagel}(2010)}]{LiuNa2010}%
  \BibitemOpen
  \bibfield  {author} {\bibinfo {author} {\bibfnamefont {A.~J.}\ \bibnamefont
  {Liu}}\ and\ \bibinfo {author} {\bibfnamefont {S.~R.}\ \bibnamefont
  {Nagel}},\ }\enquote {\bibinfo {title} {The jamming transition and the
  marginally jammed solid},}\ in\ \href {\doibase
  10.1146/annurev-conmatphys-070909-104045} {\emph {\bibinfo {booktitle}
  {Annual Review of Condensed Matter Physics, Vol 1}}},\ \bibinfo {series}
  {Annual Review of Condensed Matter Physics}, Vol.~\bibinfo {volume} {1}\
  (\bibinfo {year} {2010})\ pp.\ \bibinfo {pages} {347--369}\BibitemShut
  {NoStop}%
\bibitem [{\citenamefont {Behringer}\ and\ \citenamefont
  {Chakraborty}(2019)}]{BehringerChak2019}%
  \BibitemOpen
  \bibfield  {author} {\bibinfo {author} {\bibfnamefont {R.~P.}\ \bibnamefont
  {Behringer}}\ and\ \bibinfo {author} {\bibfnamefont {B.}~\bibnamefont
  {Chakraborty}},\ }\href {\doibase 10.1088/1361-6633/aadc3c} {\bibfield
  {journal} {\bibinfo  {journal} {Reports on Progress in Physics}\ }\textbf
  {\bibinfo {volume} {82}} (\bibinfo {year} {2019}),\
  10.1088/1361-6633/aadc3c}\BibitemShut {NoStop}%
\bibitem [{\citenamefont {O'Hern}\ \emph {et~al.}(2003)\citenamefont {O'Hern},
  \citenamefont {Silbert}, \citenamefont {Liu},\ and\ \citenamefont
  {Nagel}}]{OhernNa2003}%
  \BibitemOpen
  \bibfield  {author} {\bibinfo {author} {\bibfnamefont {C.~S.}\ \bibnamefont
  {O'Hern}}, \bibinfo {author} {\bibfnamefont {L.~E.}\ \bibnamefont {Silbert}},
  \bibinfo {author} {\bibfnamefont {A.~J.}\ \bibnamefont {Liu}}, \ and\
  \bibinfo {author} {\bibfnamefont {S.~R.}\ \bibnamefont {Nagel}},\ }\href {<Go
  to ISI>://000184582400020} {\bibfield  {journal} {\bibinfo  {journal}
  {Physical Review E}\ }\textbf {\bibinfo {volume} {68}},\ \bibinfo {pages}
  {011306} (\bibinfo {year} {2003})}\BibitemShut {NoStop}%
\bibitem [{\citenamefont {Goodrich}\ \emph {et~al.}(2015)\citenamefont
  {Goodrich}, \citenamefont {Liu},\ and\ \citenamefont
  {Nagel}}]{GoodrichNAG2015}%
  \BibitemOpen
  \bibfield  {author} {\bibinfo {author} {\bibfnamefont {C.~P.}\ \bibnamefont
  {Goodrich}}, \bibinfo {author} {\bibfnamefont {A.~J.}\ \bibnamefont {Liu}}, \
  and\ \bibinfo {author} {\bibfnamefont {S.~R.}\ \bibnamefont {Nagel}},\ }\href
  {\doibase 10.1103/PhysRevLett.114.225501} {\bibfield  {journal} {\bibinfo
  {journal} {Physical Review Letters}\ }\textbf {\bibinfo {volume} {114}}
  (\bibinfo {year} {2015}),\ 10.1103/PhysRevLett.114.225501}\BibitemShut
  {NoStop}%
\bibitem [{\citenamefont {Hexner}\ \emph
  {et~al.}(2018{\natexlab{a}})\citenamefont {Hexner}, \citenamefont {Liu},\
  and\ \citenamefont {Nagel}}]{HexnerNag2018a}%
  \BibitemOpen
  \bibfield  {author} {\bibinfo {author} {\bibfnamefont {D.}~\bibnamefont
  {Hexner}}, \bibinfo {author} {\bibfnamefont {A.~J.}\ \bibnamefont {Liu}}, \
  and\ \bibinfo {author} {\bibfnamefont {S.~R.}\ \bibnamefont {Nagel}},\ }\href
  {\doibase 10.1103/PhysRevE.97.063001} {\bibfield  {journal} {\bibinfo
  {journal} {Physical Review E}\ }\textbf {\bibinfo {volume} {97}} (\bibinfo
  {year} {2018}{\natexlab{a}}),\ 10.1103/PhysRevE.97.063001}\BibitemShut
  {NoStop}%
\bibitem [{\citenamefont {Hexner}\ \emph
  {et~al.}(2018{\natexlab{b}})\citenamefont {Hexner}, \citenamefont {Liu},\
  and\ \citenamefont {Nagel}}]{HexnerNag2018b}%
  \BibitemOpen
  \bibfield  {author} {\bibinfo {author} {\bibfnamefont {D.}~\bibnamefont
  {Hexner}}, \bibinfo {author} {\bibfnamefont {A.~J.}\ \bibnamefont {Liu}}, \
  and\ \bibinfo {author} {\bibfnamefont {S.~R.}\ \bibnamefont {Nagel}},\ }\href
  {\doibase 10.1039/c7sm01727h} {\bibfield  {journal} {\bibinfo  {journal}
  {Soft Matter}\ }\textbf {\bibinfo {volume} {14}},\ \bibinfo {pages} {312}
  (\bibinfo {year} {2018}{\natexlab{b}})}\BibitemShut {NoStop}%
\bibitem [{\citenamefont {Reid}\ \emph {et~al.}(2018)\citenamefont {Reid},
  \citenamefont {Pashine}, \citenamefont {Wozniak}, \citenamefont {Jaeger},
  \citenamefont {Liu}, \citenamefont {Nagel},\ and\ \citenamefont
  {de~Pablo}}]{ReidPab2018}%
  \BibitemOpen
  \bibfield  {author} {\bibinfo {author} {\bibfnamefont {D.~R.}\ \bibnamefont
  {Reid}}, \bibinfo {author} {\bibfnamefont {N.}~\bibnamefont {Pashine}},
  \bibinfo {author} {\bibfnamefont {J.~M.}\ \bibnamefont {Wozniak}}, \bibinfo
  {author} {\bibfnamefont {H.~M.}\ \bibnamefont {Jaeger}}, \bibinfo {author}
  {\bibfnamefont {A.~J.}\ \bibnamefont {Liu}}, \bibinfo {author} {\bibfnamefont
  {S.~R.}\ \bibnamefont {Nagel}}, \ and\ \bibinfo {author} {\bibfnamefont
  {J.~J.}\ \bibnamefont {de~Pablo}},\ }\href {\doibase 10.1073/pnas.1717442115}
  {\bibfield  {journal} {\bibinfo  {journal} {Proceedings of the National
  Academy of Sciences}\ }\textbf {\bibinfo {volume} {115}},\ \bibinfo {pages}
  {E1384} (\bibinfo {year} {2018})},\ \Eprint
  {http://arxiv.org/abs/https://www.pnas.org/content/115/7/E1384.full.pdf}
  {https://www.pnas.org/content/115/7/E1384.full.pdf} \BibitemShut {NoStop}%
\bibitem [{\citenamefont {Liarte}\ \emph {et~al.}(2019)\citenamefont {Liarte},
  \citenamefont {Mao}, \citenamefont {Stenull},\ and\ \citenamefont
  {Lubensky}}]{LiarteLub2019}%
  \BibitemOpen
  \bibfield  {author} {\bibinfo {author} {\bibfnamefont {D.~B.}\ \bibnamefont
  {Liarte}}, \bibinfo {author} {\bibfnamefont {X.}~\bibnamefont {Mao}},
  \bibinfo {author} {\bibfnamefont {O.}~\bibnamefont {Stenull}}, \ and\
  \bibinfo {author} {\bibfnamefont {T.~C.}\ \bibnamefont {Lubensky}},\ }\href
  {\doibase 10.1103/PhysRevLett.122.128006} {\bibfield  {journal} {\bibinfo
  {journal} {Phys. Rev. Lett.}\ }\textbf {\bibinfo {volume} {122}},\ \bibinfo
  {pages} {128006} (\bibinfo {year} {2019})}\BibitemShut {NoStop}%
\bibitem [{\citenamefont {Chaikin}\ and\ \citenamefont
  {Lubensky}(1995)}]{ChaikinLub1995}%
  \BibitemOpen
  \bibfield  {author} {\bibinfo {author} {\bibfnamefont {P.}~\bibnamefont
  {Chaikin}}\ and\ \bibinfo {author} {\bibfnamefont {T.}~\bibnamefont
  {Lubensky}},\ }\href@noop {} {\emph {\bibinfo {title} {Principles of
  Condensed Matter Physics}}}\ (\bibinfo  {publisher} {Cambridge Press},\
  \bibinfo {address} {Cambridge},\ \bibinfo {year} {1995})\BibitemShut
  {NoStop}%
\bibitem [{\citenamefont {Sun}\ \emph {et~al.}(2012)\citenamefont {Sun},
  \citenamefont {Souslov}, \citenamefont {Mao},\ and\ \citenamefont
  {Lubensky}}]{SunLub2012}%
  \BibitemOpen
  \bibfield  {author} {\bibinfo {author} {\bibfnamefont {K.}~\bibnamefont
  {Sun}}, \bibinfo {author} {\bibfnamefont {A.}~\bibnamefont {Souslov}},
  \bibinfo {author} {\bibfnamefont {X.~M.}\ \bibnamefont {Mao}}, \ and\
  \bibinfo {author} {\bibfnamefont {T.~C.}\ \bibnamefont {Lubensky}},\ }\href
  {\doibase 10.1073/pnas.1119941109} {\bibfield  {journal} {\bibinfo  {journal}
  {Proceedings of the National Academy of Sciences of the United States of
  America}\ }\textbf {\bibinfo {volume} {109}},\ \bibinfo {pages} {12369}
  (\bibinfo {year} {2012})}\BibitemShut {NoStop}%
\bibitem [{Note1()}]{Note1}%
  \BibitemOpen
  \bibinfo {note} {Here we assume the twist angle $\alpha \in (0,\pi
  /3)$}\BibitemShut {NoStop}%
\bibitem [{Note2()}]{Note2}%
  \BibitemOpen
  \bibinfo {note} {On the other hand, the combination of the TwKL with one
  honeycomb lattice ({\protect \it e.g.} the one formed by red bonds in
  Fig.~\ref {Fig:GK/Hfig}) does not have $C_3$ symmetry. We need at least three
  honeycomb lattices to have isotropic elasticity in the TwK/H
  model.}\BibitemShut {Stop}%
\bibitem [{\citenamefont {Bi}\ \emph {et~al.}(2011)\citenamefont {Bi},
  \citenamefont {Zhang}, \citenamefont {Chakraborty},\ and\ \citenamefont
  {Behringer}}]{BiBeh2011}%
  \BibitemOpen
  \bibfield  {author} {\bibinfo {author} {\bibfnamefont {D.}~\bibnamefont
  {Bi}}, \bibinfo {author} {\bibfnamefont {J.}~\bibnamefont {Zhang}}, \bibinfo
  {author} {\bibfnamefont {B.}~\bibnamefont {Chakraborty}}, \ and\ \bibinfo
  {author} {\bibfnamefont {R.~P.}\ \bibnamefont {Behringer}},\ }\href@noop {}
  {\bibfield  {journal} {\bibinfo  {journal} {Nature}\ }\textbf {\bibinfo
  {volume} {480}},\ \bibinfo {pages} {355} (\bibinfo {year}
  {2011})}\BibitemShut {NoStop}%
\bibitem [{\citenamefont {Baity-Jesi}\ \emph {et~al.}(2017)\citenamefont
  {Baity-Jesi}, \citenamefont {Goodrich}, \citenamefont {Liu}, \citenamefont
  {Nagel},\ and\ \citenamefont {Sethna}}]{JesiSet2017}%
  \BibitemOpen
  \bibfield  {author} {\bibinfo {author} {\bibfnamefont {M.}~\bibnamefont
  {Baity-Jesi}}, \bibinfo {author} {\bibfnamefont {C.~P.}\ \bibnamefont
  {Goodrich}}, \bibinfo {author} {\bibfnamefont {A.~J.}\ \bibnamefont {Liu}},
  \bibinfo {author} {\bibfnamefont {S.~R.}\ \bibnamefont {Nagel}}, \ and\
  \bibinfo {author} {\bibfnamefont {J.~P.}\ \bibnamefont {Sethna}},\ }\href
  {\doibase 10.1007/s10955-016-1703-9} {\bibfield  {journal} {\bibinfo
  {journal} {Journal of Statistical Physics}\ }\textbf {\bibinfo {volume}
  {167}},\ \bibinfo {pages} {735} (\bibinfo {year} {2017})}\BibitemShut
  {NoStop}%
\bibitem [{Note3()}]{Note3}%
  \BibitemOpen
  \bibinfo {note} {Our use of shear-jamming, which is defined by a
  discontinuous jump in $G$ at the jamming point, differs from that of
  Refs.~\cite {BiBeh2011,JesiSet2017,BehringerChak2019}, which refers to
  jamming induced by shear.}\BibitemShut {Stop}%
\bibitem [{\citenamefont {Fruchart}\ \emph {et~al.}(2020)\citenamefont
  {Fruchart}, \citenamefont {Zhou},\ and\ \citenamefont
  {Vitelli}}]{FruchartVit2020}%
  \BibitemOpen
  \bibfield  {author} {\bibinfo {author} {\bibfnamefont {M.}~\bibnamefont
  {Fruchart}}, \bibinfo {author} {\bibfnamefont {Y.}~\bibnamefont {Zhou}}, \
  and\ \bibinfo {author} {\bibfnamefont {V.}~\bibnamefont {Vitelli}},\
  }\href@noop {} {\bibfield  {journal} {\bibinfo  {journal} {Nature}\ }\textbf
  {\bibinfo {volume} {577}},\ \bibinfo {pages} {636} (\bibinfo {year}
  {2020})}\BibitemShut {NoStop}%
\bibitem [{\citenamefont {Jacobs}\ and\ \citenamefont
  {Thorpe}(1995)}]{JacobsTho1995}%
  \BibitemOpen
  \bibfield  {author} {\bibinfo {author} {\bibfnamefont {D.~J.}\ \bibnamefont
  {Jacobs}}\ and\ \bibinfo {author} {\bibfnamefont {M.~F.}\ \bibnamefont
  {Thorpe}},\ }\href {\doibase 10.1103/PhysRevLett.75.4051} {\bibfield
  {journal} {\bibinfo  {journal} {Phys. Rev. Lett.}\ }\textbf {\bibinfo
  {volume} {75}},\ \bibinfo {pages} {4051} (\bibinfo {year}
  {1995})}\BibitemShut {NoStop}%
\bibitem [{Note4()}]{Note4}%
  \BibitemOpen
  \bibinfo {note} {The meaning of the index $M$ (from `majority') will become
  clear when we introduce the concept of a `majority' lattice in Sec.~\ref
  {sec:Scaling}; Specific numbers for $\protect \mathcal {C}_M$ can be obtained
  for each jamming point using Eqs.~\protect \textup {\hbox {\mathsurround \z@
  \protect \normalfont (\ignorespaces \ref {eq:kalphan0}\unskip \@@italiccorr
  )}},~\protect \textup {\hbox {\mathsurround \z@ \protect \normalfont
  (\ignorespaces \ref {eq:kalphan1}\unskip \@@italiccorr )}} and Table~\ref
  {tab:JammingPoints}.}\BibitemShut {Stop}%
\bibitem [{Note5()}]{Note5}%
  \BibitemOpen
  \bibinfo {note} {More precisely, we modify Eq.~\protect \textup {\hbox
  {\mathsurround \z@ \protect \normalfont (\ignorespaces \ref {eq:K-K0}\unskip
  \@@italiccorr )}} in two ways to collapse simulation data. First, we consider
  a constant $\protect \mathcal {C}_M$ that is different from the numerical
  value found in EMT. Second, we change the definition of $\Delta p_{RP}$ from
  $p_a \protect \mathaccentV {tilde}07E{z}_a + p_b \protect \mathaccentV
  {tilde}07E{z}_b- jd$ to $\Delta p_{RP} = (k_1 p_a + k_2) \protect
  \mathaccentV {tilde}07E{z}_a + p_b \protect \mathaccentV {tilde}07E{z}_b-
  jd$, and choose $k_1$ and $k_2$ so that $\Delta p_{RP}\approx 0$ when
  $\protect \qopname \relax m{min}(B,G) < 10^{-10}$ in the simulation
  data}\BibitemShut {NoStop}%
\bibitem [{\citenamefont {Hanifpour}\ \emph {et~al.}(2018)\citenamefont
  {Hanifpour}, \citenamefont {Petersen}, \citenamefont {Alava},\ and\
  \citenamefont {Zapperi}}]{HanifpourZap2018}%
  \BibitemOpen
  \bibfield  {author} {\bibinfo {author} {\bibfnamefont {M.}~\bibnamefont
  {Hanifpour}}, \bibinfo {author} {\bibfnamefont {C.~F.}\ \bibnamefont
  {Petersen}}, \bibinfo {author} {\bibfnamefont {M.~J.}\ \bibnamefont {Alava}},
  \ and\ \bibinfo {author} {\bibfnamefont {S.}~\bibnamefont {Zapperi}},\
  }\href@noop {} {\bibfield  {journal} {\bibinfo  {journal} {The European
  Physical Journal B}\ }\textbf {\bibinfo {volume} {91}},\ \bibinfo {pages}
  {271} (\bibinfo {year} {2018})}\BibitemShut {NoStop}%
\bibitem [{Note6()}]{Note6}%
  \BibitemOpen
  \bibinfo {note} {Note that our matrix of spring constants $\protect \bm {k}$
  is not (in general) proportional to the identity matrix, so we cannot use the
  limiting form $(k/2V) \DOTSB \sum@ \slimits@ _\alpha (\protect \text
  {U}_\protect \text {aff} \cdot \protect \mathaccentV {hat}05E{\protect \text
  {t}}_\alpha )^2$ on the right of Eq.~(3.10) of Ref.~\cite
  {LubenskyKai2015}.}\BibitemShut {Stop}%
\bibitem [{\citenamefont {Press}\ \emph {et~al.}(1988)\citenamefont {Press},
  \citenamefont {Flannery}, \citenamefont {Teukolsky},\ and\ \citenamefont
  {Vetterling}}]{Recipies-C}%
  \BibitemOpen
  \bibfield  {author} {\bibinfo {author} {\bibfnamefont {W.}~\bibnamefont
  {Press}}, \bibinfo {author} {\bibfnamefont {B.}~\bibnamefont {Flannery}},
  \bibinfo {author} {\bibfnamefont {S.}~\bibnamefont {Teukolsky}}, \ and\
  \bibinfo {author} {\bibfnamefont {W.}~\bibnamefont {Vetterling}},\
  }\href@noop {} {\emph {\bibinfo {title} {Numerical Recipies in C}}}\
  (\bibinfo  {publisher} {Cambridge Press},\ \bibinfo {address} {Cambridge,
  U.K.},\ \bibinfo {year} {1988})\BibitemShut {NoStop}%
\bibitem [{\citenamefont {Mao}\ and\ \citenamefont
  {Lubensky}(2011)}]{MaoLub2011}%
  \BibitemOpen
  \bibfield  {author} {\bibinfo {author} {\bibfnamefont {X.}~\bibnamefont
  {Mao}}\ and\ \bibinfo {author} {\bibfnamefont {T.~C.}\ \bibnamefont
  {Lubensky}},\ }\href {\doibase 10.1103/PhysRevE.83.011111} {\bibfield
  {journal} {\bibinfo  {journal} {Phys. Rev. E}\ }\textbf {\bibinfo {volume}
  {83}},\ \bibinfo {pages} {011111} (\bibinfo {year} {2011})}\BibitemShut
  {NoStop}%
\bibitem [{\citenamefont {Mao}\ \emph {et~al.}(2013)\citenamefont {Mao},
  \citenamefont {Stenull},\ and\ \citenamefont {Lubensky}}]{MaoLub2013}%
  \BibitemOpen
  \bibfield  {author} {\bibinfo {author} {\bibfnamefont {X.~M.}\ \bibnamefont
  {Mao}}, \bibinfo {author} {\bibfnamefont {O.}~\bibnamefont {Stenull}}, \ and\
  \bibinfo {author} {\bibfnamefont {T.~C.}\ \bibnamefont {Lubensky}},\ }\href
  {\doibase 10.1103/PhysRevE.87.042602} {\bibfield  {journal} {\bibinfo
  {journal} {Physical Review E}\ }\textbf {\bibinfo {volume} {87}},\ \bibinfo
  {pages} {042602} (\bibinfo {year} {2013})}\BibitemShut {NoStop}%
\bibitem [{\citenamefont {Liarte}\ \emph {et~al.}(2016)\citenamefont {Liarte},
  \citenamefont {Stenull}, \citenamefont {Mao},\ and\ \citenamefont
  {Lubensky}}]{LiarteLub2016}%
  \BibitemOpen
  \bibfield  {author} {\bibinfo {author} {\bibfnamefont {D.~B.}\ \bibnamefont
  {Liarte}}, \bibinfo {author} {\bibfnamefont {O.}~\bibnamefont {Stenull}},
  \bibinfo {author} {\bibfnamefont {X.~M.}\ \bibnamefont {Mao}}, \ and\
  \bibinfo {author} {\bibfnamefont {T.~C.}\ \bibnamefont {Lubensky}},\ }\href
  {\doibase 10.1088/0953-8984/28/16/165402} {\bibfield  {journal} {\bibinfo
  {journal} {Journal of Physics-Condensed Matter}\ }\textbf {\bibinfo {volume}
  {28}},\ \bibinfo {pages} {165402} (\bibinfo {year} {2016})}\BibitemShut
  {NoStop}%
\bibitem [{\citenamefont {Stenull}\ and\ \citenamefont
  {Lubensky}(2019)}]{StenullLub2019}%
  \BibitemOpen
  \bibfield  {author} {\bibinfo {author} {\bibfnamefont {O.}~\bibnamefont
  {Stenull}}\ and\ \bibinfo {author} {\bibfnamefont {T.~C.}\ \bibnamefont
  {Lubensky}},\ }\href {\doibase 10.1103/PhysRevLett.122.248002} {\bibfield
  {journal} {\bibinfo  {journal} {Physical Review Letters}\ }\textbf {\bibinfo
  {volume} {122}} (\bibinfo {year} {2019}),\
  10.1103/PhysRevLett.122.248002}\BibitemShut {NoStop}%
\bibitem [{\citenamefont {Rens}\ and\ \citenamefont
  {Lerner}(2019)}]{RensLer2019}%
  \BibitemOpen
  \bibfield  {author} {\bibinfo {author} {\bibfnamefont {R.}~\bibnamefont
  {Rens}}\ and\ \bibinfo {author} {\bibfnamefont {E.}~\bibnamefont {Lerner}},\
  }\href {\doibase 10.1140/epje/i2019-11888-5} {\bibfield  {journal} {\bibinfo
  {journal} {The European Physical Journal E}\ }\textbf {\bibinfo {volume}
  {42}},\ \bibinfo {pages} {114} (\bibinfo {year} {2019})}\BibitemShut
  {NoStop}%
\bibitem [{\citenamefont {Calladine}(1978)}]{Calladine1978}%
  \BibitemOpen
  \bibfield  {author} {\bibinfo {author} {\bibfnamefont {C.}~\bibnamefont
  {Calladine}},\ }\href {\doibase https://doi.org/10.1016/0020-7683(78)90052-5}
  {\bibfield  {journal} {\bibinfo  {journal} {International Journal of Solids
  and Structures}\ }\textbf {\bibinfo {volume} {14}},\ \bibinfo {pages} {161 }
  (\bibinfo {year} {1978})}\BibitemShut {NoStop}%
\bibitem [{\citenamefont {Pellegrino}\ and\ \citenamefont
  {Calladine}(1986)}]{PellegrinoCal1986}%
  \BibitemOpen
  \bibfield  {author} {\bibinfo {author} {\bibfnamefont {S.}~\bibnamefont
  {Pellegrino}}\ and\ \bibinfo {author} {\bibfnamefont {C.}~\bibnamefont
  {Calladine}},\ }\href {\doibase 10.1016/0020-7683(86)90014-4} {\bibfield
  {journal} {\bibinfo  {journal} {International Journal of Solids and
  Structures}\ }\textbf {\bibinfo {volume} {22}},\ \bibinfo {pages} {409}
  (\bibinfo {year} {1986})}\BibitemShut {NoStop}%
\end{thebibliography}
\end{document}